\documentclass[aps,prd,twocolumn,preprintnumbers,
superscriptaddress,nofootinbib]{revtex4-2}
\usepackage{dcolumn}
\usepackage{amsmath}
\usepackage{amssymb}
\usepackage{amsthm}
\usepackage{mathtools}
\usepackage{mathrsfs}
\usepackage{bm}
\usepackage{slashed} 
\usepackage{graphicx}
\usepackage{multirow}
\usepackage{tikz}
\usepackage{natbib}
\usepackage[caption=false]{subfig}
\usepackage{relsize}	
\usepackage{array}
\usepackage{float}
\usepackage{color}
\usepackage{xcolor}
\usepackage{soul}
\usepackage{verbatim} 
\usepackage{siunitx}
\usepackage{hyperref}
\usepackage{CJKfntef}
\usepackage{ulem}
\hypersetup{colorlinks=true,linkcolor=purple,anchorcolor=blue,citecolor=blue, filecolor=blue,urlcolor=red,bookmarksnumbered=true,
	pdfview=FitB}

\newcommand{\ie}{\textit{i.e.}} 

\newcommand{\tab}[1]{Table~#1}

\newcommand{\eqs}[1]{Eqs.~(#1)}

\def\be{\begin{eqnarray}}
\def\ee{\end{eqnarray}}

\usepackage{xcolor}
\colorlet{darkred}{red!80!black}
\colorlet{darkgreen}{green!50!black}
\colorlet{darkblue}{blue!50!black}
\newcommand{\remove}[1] {\textcolor{darkred}{\st{#1}}}

\begin{document}

\author{Yuanqi~Lu}
\email{yuanqilu@impcas.ac.cn}
\affiliation{Institute of Modern Physics, Chinese Academy of Sciences, Lanzhou, Gansu, 730000, China}
\affiliation{School of Nuclear Physics, University of Chinese Academy of Sciences, Beijing, 100049, China}
\affiliation{CAS Key Laboratory of High Precision Nuclear Spectroscopy, Institute of Modern Physics, Chinese Academy of Sciences, Lanzhou 730000, China}

\author{Zhimin~Zhu}
\email{zhuzhimin@impcas.ac.cn}
\affiliation{Institute of Modern Physics, Chinese Academy of Sciences, Lanzhou, Gansu, 730000, China}
\affiliation{School of Nuclear Physics, University of Chinese Academy of Sciences, Beijing, 100049, China}
\affiliation{CAS Key Laboratory of High Precision Nuclear Spectroscopy, Institute of Modern Physics, Chinese Academy of Sciences, Lanzhou 730000, China}

\author{Jiangshan~Lan}
\email{jiangshanlan@impcas.ac.cn}
\affiliation{Institute of Modern Physics, Chinese Academy of Sciences, Lanzhou, Gansu, 730000, China}
\affiliation{School of Nuclear Physics, University of Chinese Academy of Sciences, Beijing, 100049, China}
\affiliation{CAS Key Laboratory of High Precision Nuclear Spectroscopy, Institute of Modern Physics, Chinese Academy of Sciences, Lanzhou 730000, China}

\author{Chandan~Mondal}
\email{mondal@impcas.ac.cn}
\affiliation{Institute of Modern Physics, Chinese Academy of Sciences, Lanzhou, Gansu, 730000, China}
\affiliation{School of Nuclear Physics, University of Chinese Academy of Sciences, Beijing, 100049, China}
\affiliation{CAS Key Laboratory of High Precision Nuclear Spectroscopy, Institute of Modern Physics, Chinese Academy of Sciences, Lanzhou 730000, China}

\author{Xingbo~Zhao}
\email{xbzhao@impcas.ac.cn}
\affiliation{Institute of Modern Physics, Chinese Academy of Sciences, Lanzhou, Gansu, 730000, China}
\affiliation{School of Nuclear Physics, University of Chinese Academy of Sciences, Beijing, 100049, China}
\affiliation{CAS Key Laboratory of High Precision Nuclear Spectroscopy, Institute of Modern Physics, Chinese Academy of Sciences, Lanzhou 730000, China}

\author{James~P.~Vary}
\email{jvary@iastate.edu}
\affiliation{Department of Physics and Astronomy, Iowa State University, Ames, IA 50011, USA}

\collaboration{BLFQ Collaboration}

\date{\today}

\title{Transverse structure of the kaon: A light-front Hamiltonian approach}

\begin{abstract}
We employ the Basis Light-Front Quantization (BLFQ) framework to compute the leading-twist (twist-2) and subleading-twist (twist-3) transverse-momentum-dependent parton distribution functions (TMDs) of the kaon. The light-front wave functions are obtained by diagonalizing a light-front QCD Hamiltonian that includes quark–antiquark ($|q\bar{q}\rangle$) and quark–antiquark–gluon ($|q\bar{q}g\rangle $) Fock components together with a three-dimensional confinement. Using the QCD equations of motion, the twist-3 TMDs are decomposed into twist-2 contributions and genuine twist-3 terms, the latter encoding quark–quark–gluon correlations beyond the probabilistic picture. These genuine twist-3 contributions arise from the interference between the $|q\bar{q}\rangle$ and $|q\bar{q}g\rangle$ sectors, which are usually neglected in the Wandzura–Wilczek approximation. This work provides the first theoretical predictions of kaon subleading-twist TMDs that explicitly account for Fock-sector interference. In addition, we present results for the kaon’s twist-2 and twist-3 collinear parton distribution functions (PDFs). The twist-2 PDFs are found to be in good agreement with the recent global analysis by the JAM Collaboration.
\end{abstract}
\maketitle

\section{Introduction}\label{sec: intro}
Understanding the internal structure of hadrons is a central goal of modern particle and nuclear physics~\cite{Accardi:2012qut,Bacchetta:2006tn,Belitsky:2002sm,Lai:2010vv,Pumplin:2002vw,Diehl:2003ny}. Direct access to this structure from the QCD first principles remains difficult due to the unresolved challenges of color confinement and chiral symmetry breaking. Nevertheless, the development of QCD factorization theorems~\cite{Collins:2011zzd,Diehl:2011yj,Collins:1996fb,Diehl:2003ny,Rogers:2015sqa,Sterman:1995fz,Collins:1985ue,Collins:1987pm,Collins:1998rz} allows one to separate experimental cross sections into short-distance perturbative contributions and long-distance nonperturbative components, with the latter encoding hadron structure~\cite{Collins:1987pm,Chernyak:1983ej,Muller:1994ses,Lampe:1998eu,Shuryak:1980tp}. A prime example is the extraction of collinear parton distribution functions (PDFs) from deep inelastic scattering (DIS) data~\cite{Hen:2016kwk,NNPDF:2017mvq}. PDFs describe the distribution of the longitudinal momentum fraction carried by hadronic constituents~\cite{Jaffe:1983hp,Collins:2011zzd}, thereby providing a one-dimensional view of hadron structure. 

Beyond the one-dimensional description offered by PDFs, the three-dimensional (3D) structure of hadrons can be explored through generalized parton distributions (GPDs)~\cite{Ji:1998pc,Polyakov:1999gs,Goeke:2001tz,Diehl:2003ny,Belitsky:2005qn} and transverse-momentum-dependent parton distribution functions (TMDs)~\cite{Collins:2011ca,Angeles-Martinez:2015sea,Diehl:2015uka,Bacchetta:2016ccz}. GPDs provide access to the spatial distribution and orbital motion of partons inside hadrons~\cite{Burkardt:2000za,Burkardt:2002hr,Diehl:2003ny,Liu:2022fvl}, and can be extracted experimentally from exclusive processes such as deeply virtual Compton scattering (DVCS) and deeply virtual meson production (DVMP)~\cite{Belitsky:2005qn,Diehl:2003ny,Ji:1996nm,Favart:2015umi,Boffi:2007yc}. TMDs, in contrast, encode information about the transverse momentum distribution of partons in addition to their longitudinal momentum fraction~\cite{Collins:2011zzd}. They can be extracted, particularly for quarks, from azimuthal-angle–dependent cross sections in semi-inclusive deep inelastic scattering (SIDIS) and Drell–Yan processes~\cite{Collins:2011zzd,Boer:1997nt,Brodsky:2002rv,Brodsky:2002cx}. Together, GPDs and TMDs provide complementary insights that enable a three-dimensional tomography of hadrons.

In theoretical descriptions of high-energy scattering processes, cross sections are expanded in powers of $1/Q$, where $Q$ is the scattering energy scale~\cite{Bacchetta_2019}. The leading term in this expansion is governed by leading-twist (twist-2) distributions, which in the parton model are interpreted as probability densities for finding partons inside a hadron~\cite{Jaffe:1983hp,Barone:2001sp,Collins:2011zzd,Aybat:2011zv}. Subleading contributions arise at twist-3, where the distributions go beyond the simple probabilistic picture and encode multiparton correlations within the hadron~\cite{Jaffe:1991kp,Jaffe:1991ra,Efremov:2002qh,AbdulKhalek:2021gbh}.

The main purpose of this work is to systematically explore the subleading-twist (twist-3) TMDs of the kaon, going beyond the Wandzura–Wilczek (WW) approximation~\cite{Bacchetta:2006tn,Wandzura:1977qf,Bacchetta:2019qkv,Freund_2003}. Twist-3 TMDs are not independent quantities; rather, through the QCD equations of motion, they can be decomposed into contributions from twist-2 distributions and genuine twist-3 terms~\cite{Bacchetta:2019qkv,Mulders:1995dh,Barone:2001sp}. The latter encode quark–quark–gluon correlations and arise from the interference between different light-front Fock sectors, specifically $|q\bar{q}\rangle$ and $|q\bar{q}g\rangle$, which are typically neglected within the WW approximation.
We compute the kaon TMDs using its light-front wave functions (LFWFs) obtained by diagonalizing the light-front QCD Hamiltonian. Our results provide new nonperturbative input on higher-twist dynamics, particularly the role of multi-parton interference, which will be relevant for future experimental studies at facilities such as the upcoming Electron–Ion Colliders in China (EicC) and the USA (EIC).

Twist-3 PDFs and TMDs contribute to a variety of observables in inclusive and semi-inclusive deep inelastic scattering (DIS and SIDIS). Although typically suppressed relative to twist-2, twist-3 effects can be significant in fixed-target kinematics. Indeed, one of the main goals of the Jefferson Lab 12 GeV program is to measure higher-twist spin-dependent azimuthal asymmetries in SIDIS~\cite{E12017,E12-06-112,E12-06-112B}, while the upcoming EICs \cite{AbdulKhalek:2021gbh,AbdulKhalek:2022hcn,Amoroso:2022eow,Anderle:2021wcy} will extend such studies to new kinematic regions\cite{Boer:2011fh,Accardi:2012qut}.

To date, twist-3 TMDs have been investigated within a variety of QCD-inspired models and perturbative light-front Hamiltonian approaches~\cite{Jakob:1997wg,Lu:2012gu,Mao:2013waa,Mao:2014aoa,Jaffe:1991ra,Signal:1996ct,Avakian:2010br,Lorce:2014hxa,Schweitzer:2003uy,Wakamatsu:2007nc,Wakamatsu:2003uu,Ohnishi:2003mf,Cebulla:2007ej,Balla:1997hf,Dressler:1999hc,Lorce:2016ugb,Pasquini:2018oyz,Burkardt:2001iy,Kundu:2001pk,Mukherjee:2009uy,Accardi:2009au,Choi:2025bxk}. More recently, unpolarized twist-2 and twist-3 time-reversal–even quark TMDs of the pion have been computed using the homogeneous Bethe–Salpeter integral equation~\cite{dePaula:2023ver} and the Basis Light-Front Quantization (BLFQ) framework~\cite{Zhu:2023lst}, while phenomenological extractions from pion–nucleus Drell–Yan data have been reported in Refs.~\cite{Cerutti:2022lmb,Barry:2023qqh}.


Although the kaon has received less attention than the pion, it represents an equally important probe of the underlying dynamics of the Standard Model. Kaons provide a rich source of information on the nature of fundamental interactions, including insights into the quark model of hadrons~\cite{Lan:2025fia}. Moreover, studies of kaon structure and dynamics serve as a valuable tool for exploring electroweak interactions and for advancing our understanding of CP violation~\cite{PhysRevLett.60.1695,BARR1993233,CPLEAR:1995epa}. In this context, investigating the partonic structure of the kaon through TMDs offers a unique opportunity to uncover its three-dimensional dynamics and to complement existing studies of the pion.

In this work, we present the twist-2 and twist-3  TMDs of the kaon obtained within the BLFQ framework~\cite{Vary:2009gt,Zhao:2014xaa}, which is a nonperturbative approach for solving relativistic many-body bound-state problems in quantum field theory~\cite{Vary:2009gt,Zhao:2014xaa,Wiecki:2014ola,Li:2015zda,Jia:2018ary,Qian:2020utg,Tang:2019gvn,Mondal:2019jdg,Xu:2021wwj,Liu:2022fvl,Nair:2022evk,Lan:2021wok,Adhikari:2021jrh,Mondal:2021czk,Hu:2020arv,Lan:2019img,Lan:2019rba,Lan:2019vui,Xu:2022abw,Peng:2022lte}. This framework has previously been applied successfully to study the leading-twist TMDs of spin-$\tfrac{1}{2}$ and spin-1 particles in QED, such as the electron~\cite{Hu:2020arv} and the photon~\cite{Nair:2022evk}, as well as for the proton~\cite{Hu:2022ctr}. More recently, the BLFQ framework has been extended to successfully investigate subleading twist TMDs of the pion~\cite{Zhu:2023lst} and the TMDs and GPDs of the proton~\cite{Zhang:2023xfe,Liu:2024umn,Yu:2024mxo}.

We stress that our computation of kaon TMDs is performed at the model scale, without incorporating the effects of TMD evolution~\cite{Collins:2011zzd,Collins:1981va,Catani:2000vq,Bozzi:2005wk,Bozzi:2008bb}. Within the nonperturbative BLFQ framework, the results should therefore be interpreted as representing the intrinsic, initial-scale structure of the kaon. In this sense, our calculation provides a physically motivated starting point for the kaon TMDs, while the inclusion of TMD evolution would be necessary to make direct comparisons with experimental data at higher scales. Such an analysis, however, lies beyond the scope of the present work.

Within the BLFQ framework~\cite{Vary:2009gt}, we employ the light-front QCD Hamiltonian~\cite{Lan:2021wok,Brodsky:1997de} and solve for its mass eigenvalues and eigenstates. The Hamiltonian is constructed with quark ($q$), antiquark ($\bar{q}$), and gluon ($g$) degrees of freedom, incorporating light-front QCD interactions relevant to the constituent $|q\bar{q}\rangle$ and $|q\bar{q}g\rangle$ Fock sectors of mesons, together with a complementary three-dimensional confinement potential~\cite{Li:2015zda}. We solve this Hamiltonian in the leading two Fock sectors and determine the model parameters by fitting the mass spectra of light mesons, including the kaon~\cite{Lan:2021wok}. The resulting LFWFs, obtained as Hamiltonian eigenstates, have already been successfully applied to describe a wide range of kaon observables, such as electromagnetic form factors, charge radius, decay constant, PDFs, and even cross sections for kaon-induced charmonium production, revealing consistency with available experimental measurements and other models~\cite{Qian2020,Jia:2018ary,lan2025strangemesonsdynamicalgluon}. In this work, we extend these studies to compute the kaon's subleading-twist TMDs and, for the first time, to predict the genuine twist-3 TMDs. In addition, we present results for the kaon’s twist-2 and twist-3 collinear PDFs. The twist-2 PDFs are found to be in good agreement with the recent global analysis by the JAM Collaboration~\cite{Barry:2025wjx}.

\section{Basis light-front quantization framework}
In light-front field theory, bound states are obtained by solving the light-front stationary Schr\"{o}dinger equation,  
\begin{equation}
P^+ P^- |\Psi\rangle = M^2 |\Psi\rangle ,\label{eigenequation}
\end{equation}
where $P^-=P^0-P^3$ is the light-front Hamiltonian, $P^+=P^0+P^3$ is the longitudinal momentum of the system, and $M^2$ is the invariant mass squared of the state. At fixed light-front time, $x^+ \equiv x^0 + x^3$, the meson state can be expanded in terms of its quark, antiquark, and gluon Fock components,  
\begin{equation}
|\Psi\rangle = \psi_{(q\bar{q})}|q\bar{q}\rangle 
+ \psi_{(q\bar{q}g)}|q\bar{q}g\rangle 
+ \psi_{(q\bar{q}q\bar{q})}|q\bar{q}q\bar{q}\rangle 
+ \cdots , 
\label{Fockexpansion}
\end{equation}
where $\psi_{(\cdots)}$ denotes the LFWF corresponding to a given Fock sector. For practical calculations, this infinite expansion must be truncated. In the present work, we retain only the lowest two sectors, namely the quark--antiquark ($\psi_{(q\bar{q})}$) and quark--antiquark--gluon ($\psi_{(q\bar{q}g)}$) components, which characterize the meson structure at the model scale.  

To address the eigenvalue problem, we introduce an effective light-front Hamiltonian of the form  
$P^- = P^-_{\mathrm{QCD}} + P^-_{\mathrm{C}}$,
where $P^-_{\mathrm{QCD}}$ represents the light-front QCD Hamiltonian describing the dynamics of the quark--antiquark ($|q\bar{q}\rangle$) and quark--antiquark--gluon ($|q\bar{q}g\rangle$) Fock sectors, while $P^-_{\mathrm{C}}$ denotes the model Hamiltonian implementing the confining interaction~\cite{Lan:2021wok}. 

In the light-front gauge $A^+=0$, the explicit form of the QCD Hamiltonian with a single dynamical gluon reads~\cite{Lan:2022blr,Zhu:2023lst,Brodsky_1998}:  
\begin{align}
    P_{\rm QCD}^-= &\int \mathrm{d}x^- \mathrm{d}^2 x^{\perp} \Big\{\frac{1}{2}\bar{\psi}\gamma^+\frac{m_{0}^2+(i\partial^\perp)^2}{i\partial^+}\psi\nonumber\\
    &+\frac{1}{2}A_a^i\left[m_g^2+(i\partial^\perp)^2\right] A^i_a +g\bar{\psi}\gamma_{\mu}T^aA_a^{\mu}\psi \nonumber\\
    &+ \frac{1}{2}g^2\bar{\psi}\gamma^+T^a\psi\frac{1}{(i\partial^+)^2}\bar{\psi}\gamma^+T^a\psi \Big\},
\end{align}
where $\psi$ and $A^\mu$ denote the quark and gluon fields, respectively, with $T^a = \lambda^a/2$ the SU(3) color generators. $\gamma^+=\gamma^0+\gamma^3$, represents the plus component of the Dirac matrices in light-front coordinates. The parameter $g$ is the QCD coupling constant, while $m_0$ and $m_g$ represent the bare quark mass and the model gluon mass, respectively. Although gluons are massless in QCD, a phenomenologically motivated gluon mass is introduced in our low-energy model to reproduce the meson mass spectra~\cite{Lan:2021wok}. To account for quark mass corrections from higher Fock sectors, we include a mass counterterm $\delta m_q = m_0 - m_q$ in the leading Fock sector, where $m_q$ is the renormalized quark mass. In addition, an independent quark mass $m_f$ is used in the vertex interaction, following Refs.~\cite{Burkardt:1998dd,Glazek:1992aq}.
%

The confining interaction in the leading $|q\bar{q}\rangle$ Fock sector consists of transverse and longitudinal contributions~\cite{Li:2015zda,Lan:2021wok},
\begin{equation}
  \begin{split}\label{eqn:Hc}
  &P_{\rm C}^-P^+=\kappa^4\left\{x(1-x)\vec{r}_{\perp}^{\,2}-\frac{\partial_{x}[x(1-x)\partial_{x}]}{(m_q+m_{\bar{q}})^2}\right\},
  \end{split}
\end{equation}
where $\kappa$ denotes the confinement strength and $\vec r_{\perp}=\sqrt{x(1-x)}(\vec r_{\perp q}-\vec r_{\perp \bar{q}})$ is the holographic variable~\cite{Brodsky:2014yha}.
In the $|q\bar{q}g\rangle$ sector, confinement is implemented through the introduction of a massive gluon together with a truncation of the BLFQ basis functions, as described below.

We employ the BLFQ framework~\cite{Vary_2010,Vary:2009gt,Zhao:2013cma,Wiecki:2014ola,Zhao:2014xaa,Li:2015zda,Lan:2021wok,Xu:2021wwj,Kaur:2024iwn}, a nonperturbative approach for solving the light-front eigenvalue problem, Eq.~(\ref{eigenequation}). Each Fock sector $|\cdots\rangle$ in Eq.~(\ref{Fockexpansion}) is constructed as a direct product of single-particle states $|\alpha\rangle=\otimes_i|\alpha_i\rangle$.

In BLFQ, the longitudinal degrees of freedom are represented using the Discretized Light-Cone Quantization (DLCQ) basis~\cite{Brodsky:1997de}. A fermion (boson) is confined to a one-dimensional box of length $2L$ with antiperiodic (periodic) boundary conditions, leading to longitudinal momenta
\begin{equation}
p^+ = \frac{2\pi}{L} k ,
\end{equation}
where $k$ is a half-integer for fermions and an integer for bosons. Zero modes for bosons are neglected.

The transverse degrees of freedom are described by two-dimensional harmonic-oscillator (2D-HO) basis functions $\Phi_{nm}(p_{\perp},b)$, with oscillator scale $b$. Here, $n$ and $m$ denote the radial and angular quantum numbers, respectively.

Each single-particle basis state is specified by four quantum numbers,
$|\alpha_i\rangle = |k_i, n_i, m_i, \lambda_i\rangle $,
where $\lambda_i$ denotes the light-front helicity. For Fock sectors that allow multiple color-singlet states (beyond those considered here), an additional label is required to distinguish among them.

Finally, the many-body basis states constructed in BLFQ have well-defined total angular momentum projection,
\begin{equation}
M_J = \sum_i (m_i + \lambda_i).
\end{equation}


For numerical calculations, the infinite basis of each Fock sector is truncated by introducing two truncation parameters $K$ and $N_{\rm max}$ in longitudinal and transverse directions, respectively. 
The parameter $K = \sum_i k_i$ limits the total longitudinal momentum quantum numbers.
Hence, for every parton indexed by $i$, the longitudinal momentum fraction is given by $x_i=p^+_i/P^+=k_i/K$. 
In the transverse direction, the parameter $N_{\mathrm{max}}$, set by $\Sigma_{i}(2n_i+|{m_{i}}|+1)\leq N_{\mathrm{max}}$, characterizes the truncation of the transverse center of mass motion~\cite{Wiecki:2014ola,Zhao:2014xaa}. 
The $N_{\mathrm{max}}$ truncation implicitly sets the infrared (IR) and ultraviolet (UV) cutoffs in momentum space: the IR cutoff $\lambda_{\mathrm{IR}}\simeq b/\sqrt{N_{\mathrm{max}}}$ and the UV cutoff $\Lambda_{\mathrm{UV}}\simeq b\sqrt{N_{\rm max}}$~\cite{Zhao:2014xaa}.

By diagonalizing the Hamiltonian matrix, we obtain the mass spectra $M^2$ and the corresponding LFWFs. The LFWF in the momentum space is schematically represented as 
\begin{align}
  \Psi^{M_J}_{\mathcal{N},\{\lambda_i\}}({\{x_i,p_{\perp i}\}})=\sum_{ \{n_i m_i\} }\psi^{M_J}_{\mathcal{N}}({\{\alpha}_i\})\prod_{i=1}^{\mathcal{N}}  \Phi_{n_i m_i}(p_{\perp i},b)\,,
\label{eqn:wf}
\end{align}
where $\psi^{M_J}_{\mathcal{N}=2}(\{\alpha_i\})$ and $\psi^{M_J}_{\mathcal{N}=3}(\{\alpha_i\})$ represent the eigenvector components corresponding to the $|q\bar{q}\rangle$ and $|q\bar{q}g\rangle$ Fock sectors, respectively. 
Using the truncation parameters $\{N_{\rm max}, K\} = \{14, 15\}$, the model parameters listed in Table~\ref{Tab:para} were determined to reproduce the mass spectra of strange mesons~\cite{Lan:2025fia}. The resulting LFWFs  provide a good quality description of a wide range of observables, including the kaon’s electromagnetic form factor and charge radius, decay constant, PDFs, as well as the kaon-nucleus-induced Drell-Yan cross section, which is
expected to be measured soon by COMPASS++/AMBER at CERN~\cite{Lan:2025fia}.



\begin{table}[htp]
  \caption{Model parameters ${m_q, m_{f_q}, m_g, b, \kappa, g}$ are taken from Ref.~\cite{Lan:2021wok}, while the strange quark masses $m_s$ and $m_{f_s}$ are fixed by fitting the experimental masses of $\phi(1020)$ and $f_1(1420)$~\cite{Lan:2025fia}. All values are in units of GeV, except for the coupling $g$.}
  \vspace{0.15cm}
  \label{Tab:para}
  \centering
  \begin{tabular}{cccccccc}
    \hline\hline
         ~$m_q$ ~&~ ${m}_{f_q}$ ~&~ $m_g$ ~&~ $b$ ~&~ $\kappa$ ~&~ $g$ ~&~ $m_s$ ~&~ $m_{f_s}$~ \\        
    \hline 
        ~0.39 ~&~ 5.69 ~&~ 0.60 ~&~ 0.29 ~&~ 0.65 ~&~ 1.92 ~&~ 0.55 ~&~ 7.13~ \\       
    \hline
  \end{tabular}
\end{table}

\section{Transverse-momentum-dependent parton distributions}


In the light-front formalism, quark and gluon TMDs are defined through the quark–quark and unintegrated gluon correlation functions, respectively~\cite{Mei_ner_2008,Chakrabarti:2023djs,Mulders:2000sh}:
\begin{gather}
    \begin{split}
        \Phi_{q}^{[\Gamma]}(x,k_{\perp})&=\frac{1}{2}\int\frac{\mathrm{d}z^-\mathrm{d}^2\vec{z}_{\perp}}{2(2\pi)^3}e^{ik\cdot z}\\
        &\times\langle P|\bar{\psi}(0)\Gamma\mathcal{W}(0,z)\psi(z)|P\rangle|_{z^+=0}\label{5},
    \end{split}\\
    \begin{split}
        \Phi^{g[ij]}(x,k_{\perp})&=\frac{1}{xP^+}\int\frac{\mathrm{d}z^-\mathrm{d}^2z_{\perp}}{2(2\pi)^2}e^{ik\cdot z}\\
        &\times\langle P|F^{+j}(0)\mathcal{W}(0,z)F^{+i}(z)|P\rangle|_{z^+=0}\label{6},
    \end{split}
\end{gather}
where $F^{\mu\nu}$ is the gluon field strength tensor. 
Here, $|P\rangle$ denotes the light-front bound state of the target meson with mass $M$ and momenta $(P^+, P_\perp)$, where we choose $P_\perp=0~$\cite{Collins:1992kk}. The Dirac matrix $\Gamma$ specifies the Lorentz structure of the correlator $\Phi^{[\Gamma]}$ and determines its twist-$\tau$~\cite{Jaffe:1991kp}. The Wilson line $\mathcal{W}$ ensures gauge invariance of the bilocal quark field operators in the correlation function~\cite{Bacchetta:2020vty}.

For spin-0 mesons, two leading-twist (twist-2) quark TMDs appear: the unpolarized distribution $f^q_{1}(x,k_\perp)$ and the Boer–Mulders function $h_{1}^{\perp q}(x,k_\perp)$. These are obtained from the parameterization of the quark–quark correlator with $\Gamma \equiv \gamma^{+}$ and $\Gamma \equiv \sigma^{j+}\gamma_{5}$, respectively~\cite{Pasquini:2014ppa,Ahmady:2019yvo}:
\begin{align}
    \Phi_q^{[\gamma^+]}(x,k_{\perp})&=f_1^q(x,k_{\perp})\label{tw-2TMDs1},\\
    \Phi_q^{[i\sigma^{j^+}\gamma_5]}(x,k_{\perp})&=-\frac{\varepsilon_{T}^{ij}k_{\perp}^i}{M}h_1^{\perp q}(x,k_{\perp})\label{tw-2TMDs2}.
    \end{align}
where $k_{\perp}$ is the transverse momentum of a struck parton, $\epsilon_{T}^{ij}$ is an antisymmetric tensor, $\varepsilon^{11}_T=\varepsilon^{22}_T=0$ and $\varepsilon^{12}_T=-\varepsilon^{21}_T=1$.
Meanwhile, the unpolarized twist-2 gluon TMDs are defined through the correlator as~\cite{PhysRevD.76.034002}
\begin{gather}
    \Phi^g(x,k_{\perp})=\delta^{ij}\Phi^{g[ij]}(x,k_{\perp})=f^g_1(x,k_{\perp})\label{9}.
\end{gather}
 
The subleading twist (twist-3) quark TMDs are defined as follows~\cite{Mei_ner_2008}:
\begin{align}
    \Phi_q^{[1]}(x,k_{\perp})&=\frac{M}{P^+}e^q(x,k_{\perp}),\label{tw-3TMDs-e}\\
    \Phi_q^{[\gamma^{j}]}(x,k_{\perp})&=\frac{M}{P^+}\frac{k^j_{\perp}}{M}f^{\perp q}(x,k_{\perp}),\label{tw-3TMDs-f}\\
    \Phi_q^{[\gamma^i\gamma_5]}(x,k_{\perp})&=-\frac{M}{P^+}\frac{\varepsilon_{T}^{ij}k_{\perp}^i}{M}g^{\perp q}(x,k_{\perp}),\label{tw-3TMDs-g}\\
    \Phi_q^{[i\varepsilon^{ij}\gamma_5]}(x,k_{\perp})&=-\frac{M}{P^+}\varepsilon_T^{ij}h^q(x,k_{\perp})\label{tw-3TMDs-h}.
    \end{align}
Note that the TMDs $h_{1}^{\perp q}(x,k_\perp)$, $g^{\perp q}(x,k_\perp)$, and $h^q(x,k_\perp)$ are time-reversal-odd (T-odd), while all others are time-reversal-even (T-even)~\cite{Meissner:2008ay}. Twist-3 TMD correlation functions are further suppressed by a factor of $M/P^+$. In the present work, we neglect the nontrivial effect of the Wilson line $\mathcal{W}$ by approximating it as the identity operator, $\mathcal{W} \approx \bf{1}$. Under this approximation, only the T-even TMDs remain~\cite{Lorce:2016ugb}.

We decompose the quark field $\psi$ into a “good” component $\psi_+$ and a “bad” component $\psi_-$, defined as $\psi_{\pm}=\tfrac{1}{4}\gamma^{\mp}\gamma^{\pm}\psi$ in light-front field theory~\cite{Kogut:1969xa}. The bad component $\psi_-$ can be represented by $\psi_+$ and the gauge fields as
\begin{equation}
\psi_-(z)=\frac{\gamma^+}{2i\partial^+}\left[i(\partial_j-igA_{\perp j}(z))\gamma_j+m_q\right]\psi_+(z).
\end{equation}
Note that $m_q$ describes the quark mass and $j=1,2$. This constraint equation follows from the equations of motion in the light-cone gauge, $A^+=0$~\cite{Brodsky_1998}.

At leading twist, Eq.~\eqref{tw-2TMDs1}, the relevant Dirac matrices project the correlator onto $\bar{\psi}_+\psi_+$. In contrast, for subleading twist, Eq.~\eqref{tw-3TMDs-e}, the Dirac structures project onto $\bar{\psi}_+\psi_-+\bar{\psi}_-\psi_+$. Consequently, twist-3 TMDs cannot be interpreted as probability densities (or their differences); instead, they are expressed in terms of quark–quark and quark–quark–gluon matrix elements.


Following the EOM of quark fields~\cite{Efremov_2003}, the higher-twist operators can be decomposed into a combination of twist-2 and genuine higher-twist contributions. Accordingly, the twist-2 and the twist-3 TMDs are related through the following relations~\cite{Lorc__2016,Zhu:2023lst}:
\begin{equation}
\begin{aligned}
    xe^q(x,k_{\perp})&=x\tilde{e}^q(x,k_{\perp})+\frac{m_q}{M}f_1^q(x,k_{\perp}),\\
    xf^{\perp q}(x,k_{\perp})&=x\tilde{f}^{\perp q}(x,k_{\perp})+f_1^q(x,k_{\perp})\label{16},
    \end{aligned}
\end{equation}
where the tilde terms, commonly referred to as genuine twist-3 TMDs, are defined as~\cite{Lorc__2015,Zhu:2023lst}
\begin{gather}
\begin{split}
    &\tilde{e}^q(x,k_{\perp})=\frac{i}{Mx}\frac{g}{2}\int\frac{\mathrm{d}z^-\mathrm{d}^2\vec{z}_{\perp}}{2(2\pi)^3}e^{ik\cdot z}\\
    &\times\langle P|\bar{\psi}_+(0)\sigma^{j+}[A^j_{\perp}(z)-A^j_{\perp}(0)]\psi_+(z)|P\rangle|_{z^+=0}\label{17},
\end{split}\\
\begin{split}
    &2xk^j_{\perp}\tilde{f}^{\perp q}(x,k_{\perp})=g\int\frac{\mathrm{d}z^-\mathrm{d}^2\vec{z}_{\perp}}{2(2\pi)^3}e^{ik\cdot z}\\
    &\times\langle P|\bar{\psi}_+(0)\gamma^+[A^j_{\perp}(0)+A^j_{\perp}(z)]\psi_+(z)|P\rangle|_{z^+=0}\label{18}.
\end{split}
\end{gather}

Sum rules serve as a measure of the internal consistency and effectiveness of the model~\cite{Lorc__2016}. In this context, the twist-2 and twist-3 PDFs defined as the integrated TMDs, $j(x)=\int \mathrm{d}^2k_{\perp}j(x,k_{\perp})$, with $j(x,k_{\perp})$ denoting the TMDs and the PDFs are required to satisfy the following sum rules:
\begin{align}
  \int\mathrm{d}x f^q_1(x)&=N_q,\label{quarkNum}\\
  \sum_q\int\mathrm{d}x xf_1^q(x)&=1-\int\mathrm{d}x xf_1^g(x)\label{momSumRul},\\
  \sum_qm_q\int\mathrm{d}x e^q(x)&=\sigma_K,\label{scalarFF}\\
  \int\mathrm{d}x xe^q(x)&=\frac{m_q}{M}N_q.\label{firstmoment}
\end{align}
Equation~\eqref{quarkNum} expresses the number sum rule, where $N_q$ denotes the total number of valence quarks of a given flavor~\cite{Schwartz:2014sze}.
Equation~\eqref{momSumRul} represents the momentum sum rule, which requires that the quark, antiquark, and gluon constituents collectively carry the full light-front momentum of the meson.
It is important to note that, due to the explicit $k_\perp^j$ factor in Eq.~\eqref{tw-3TMDs-f}, no corresponding PDF or sum rule exists for the TMD $f^{\perp q}(x,k_\perp)$.

The zeroth moment of $e^q(x)$, given in Eq.~\eqref{scalarFF}, contains a singular contribution at $x=0$ and is related to the kaon’s scalar form factor at zero momentum transfer, $\sigma_{K}$\cite{JAFFE1992527,Efremov_2003}.
The second sum rule of $e^q(x)$, Eq.~\eqref{firstmoment}, can be reformulated as the first-moment sum rule of $\tilde{e}^q(x)$ by employing the twist relations in Eq.~\eqref{16} together with the number sum rule in Eq.~\eqref{quarkNum}.
\begin{equation}
    \int \mathrm{d}xx\tilde{e}^q(x)=0\label{23}.
\end{equation}
We will examine this sum rule within our model in the next section.

\begin{figure*}[htp]
    \centering
    \includegraphics[width=0.3\textwidth]{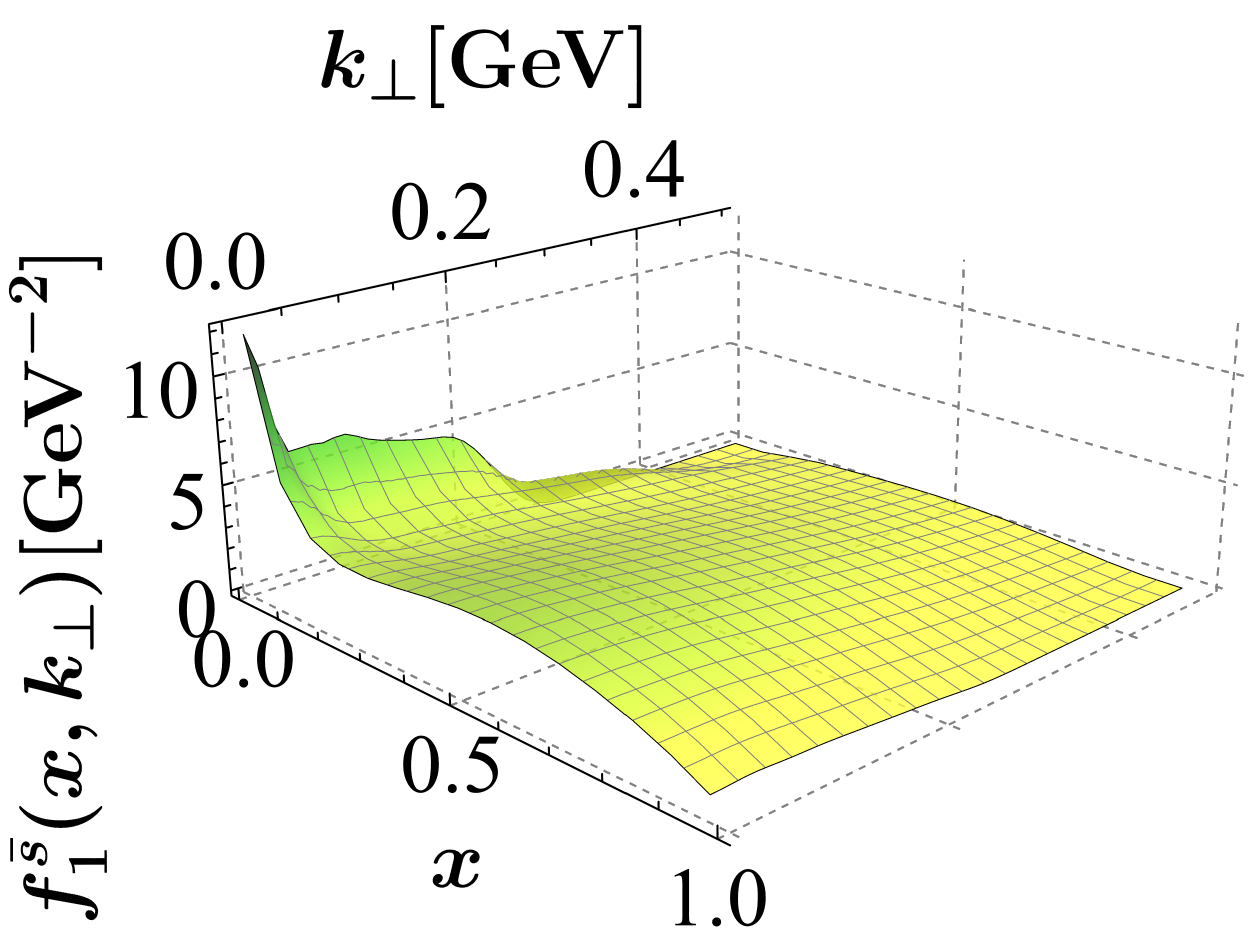}
    \includegraphics[width=0.3\textwidth]{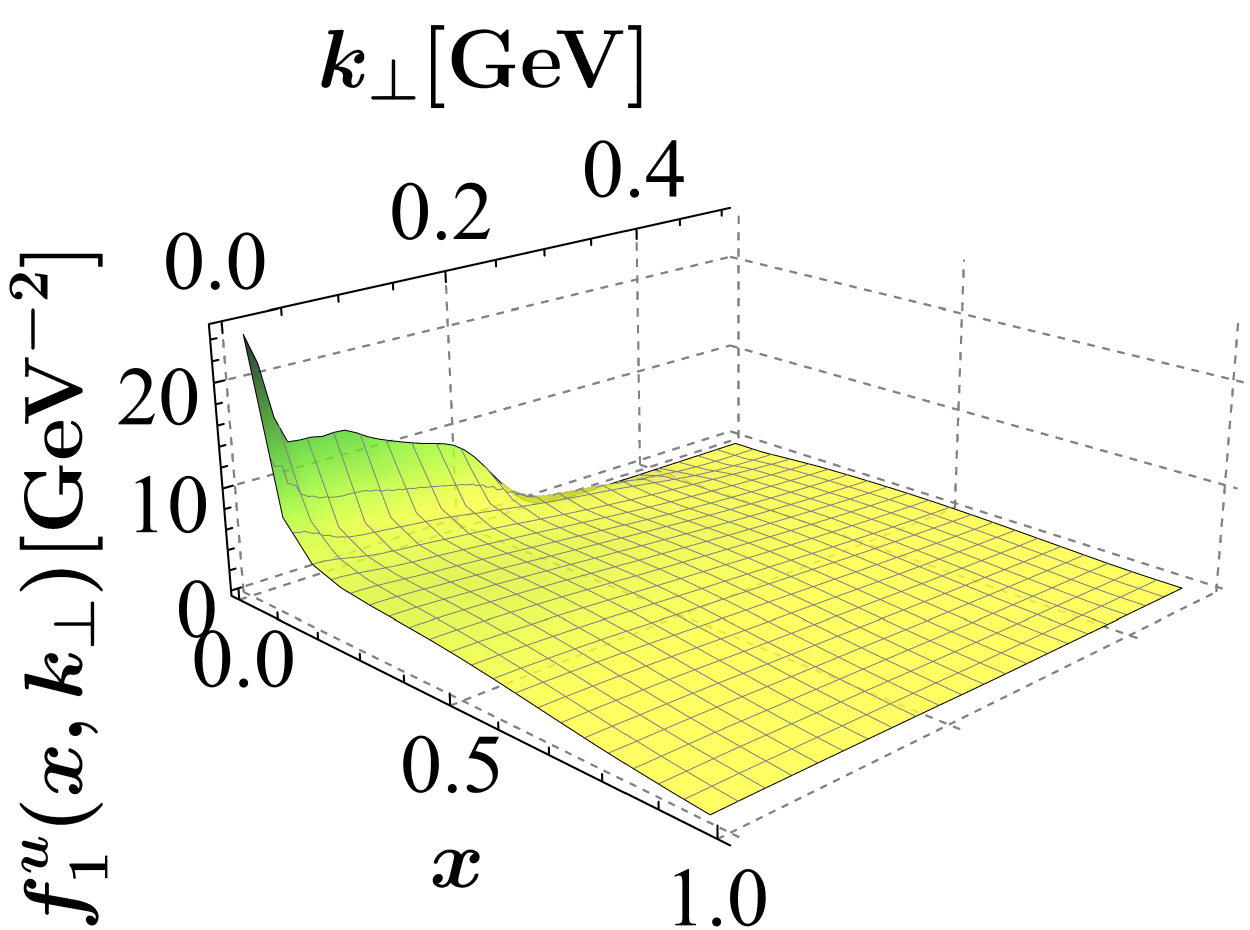}
    \includegraphics[width=0.3\textwidth]{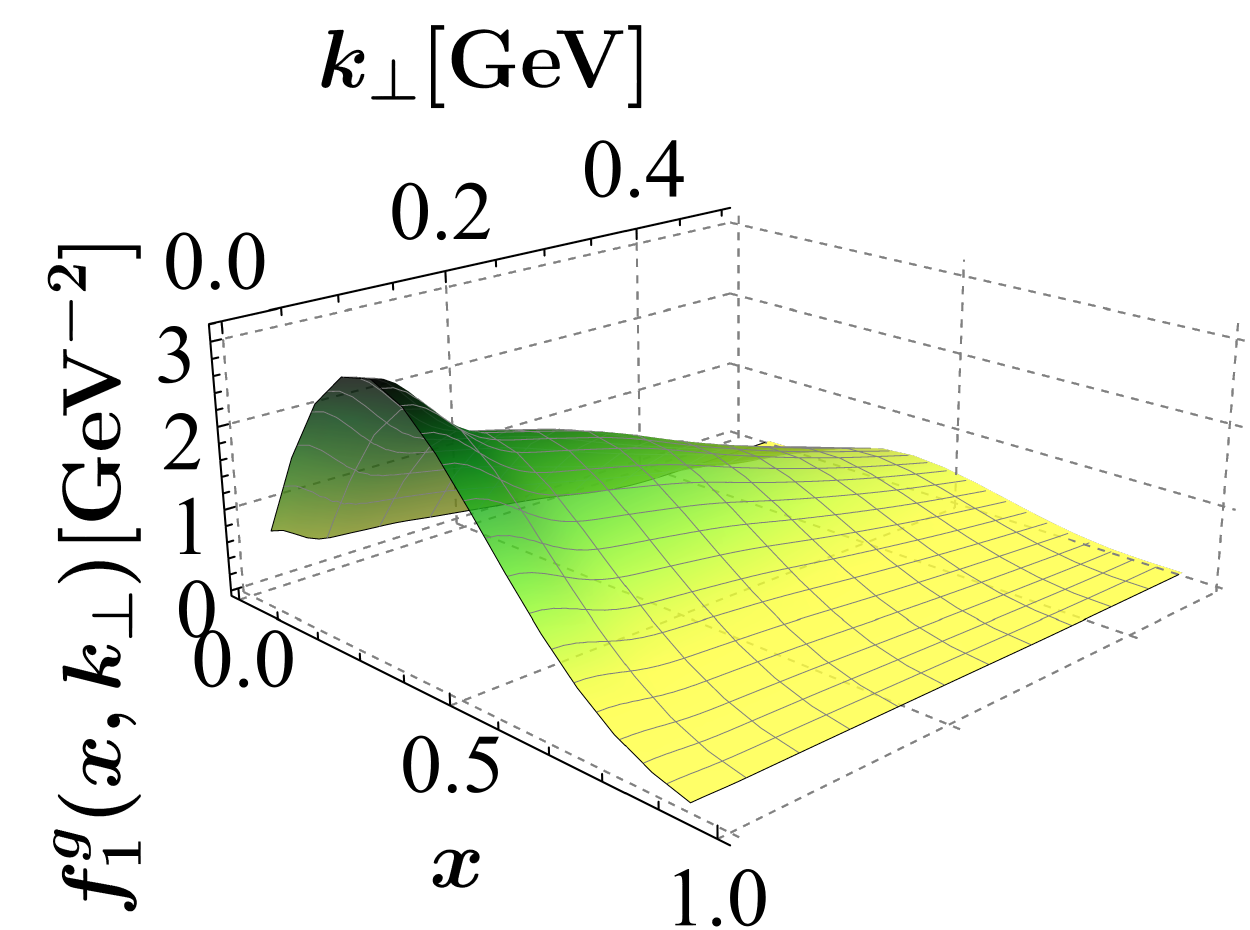}
\caption{3D plots of the BLFQ results for the twist-2 TMD
$f_1(x,k_\perp)$ of the kaon. The left, middle, and right panels correspond
to the $\bar{s}$ quark, $u$ quark, and gluon, respectively. As described in the text, all results are
obtained by averaging over BLFQ calculations with
$N_{\mathrm{max}}=\{12,14,16\}$ and $K=15$.}
    \label{Fig:1}
\end{figure*}

\begin{figure*}[htp]
    \centering
    \includegraphics[width=0.3\textwidth]{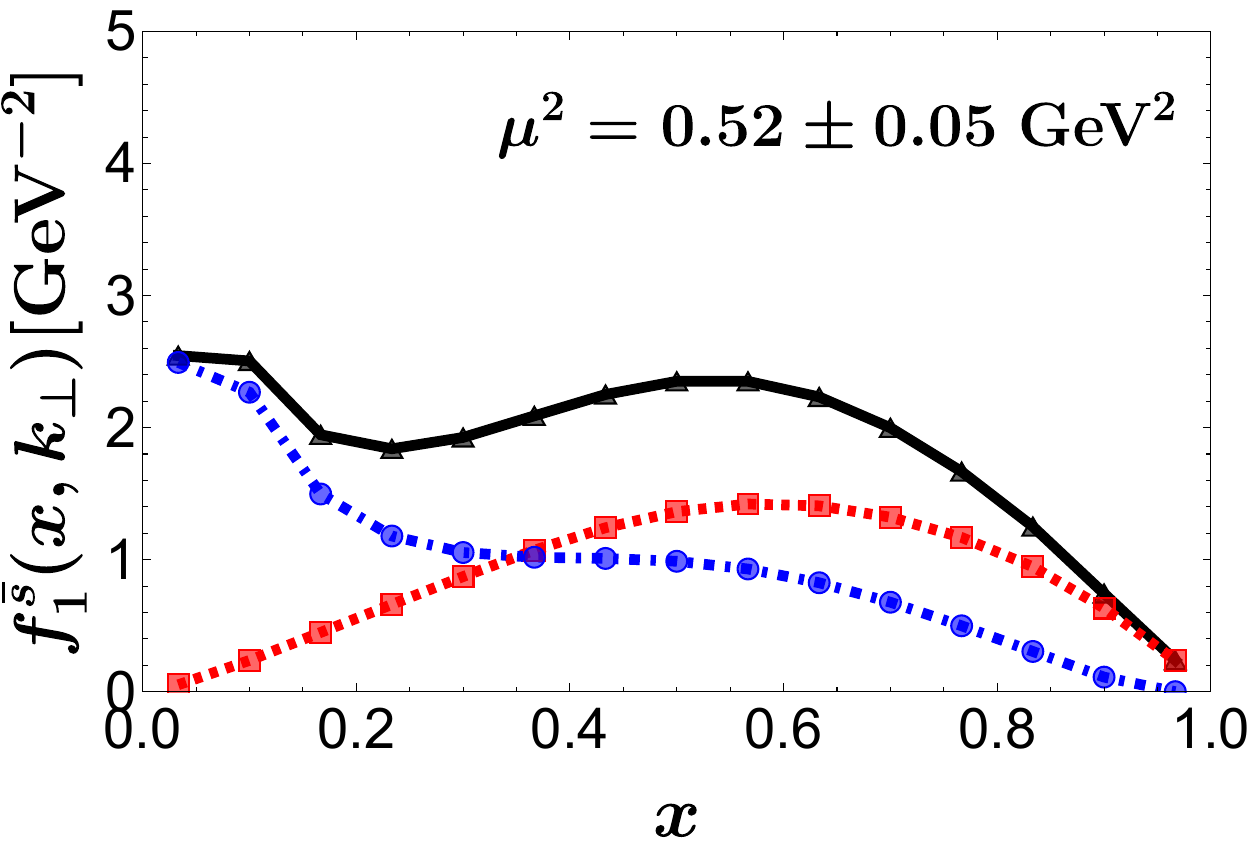}
    \includegraphics[width=0.3\textwidth]{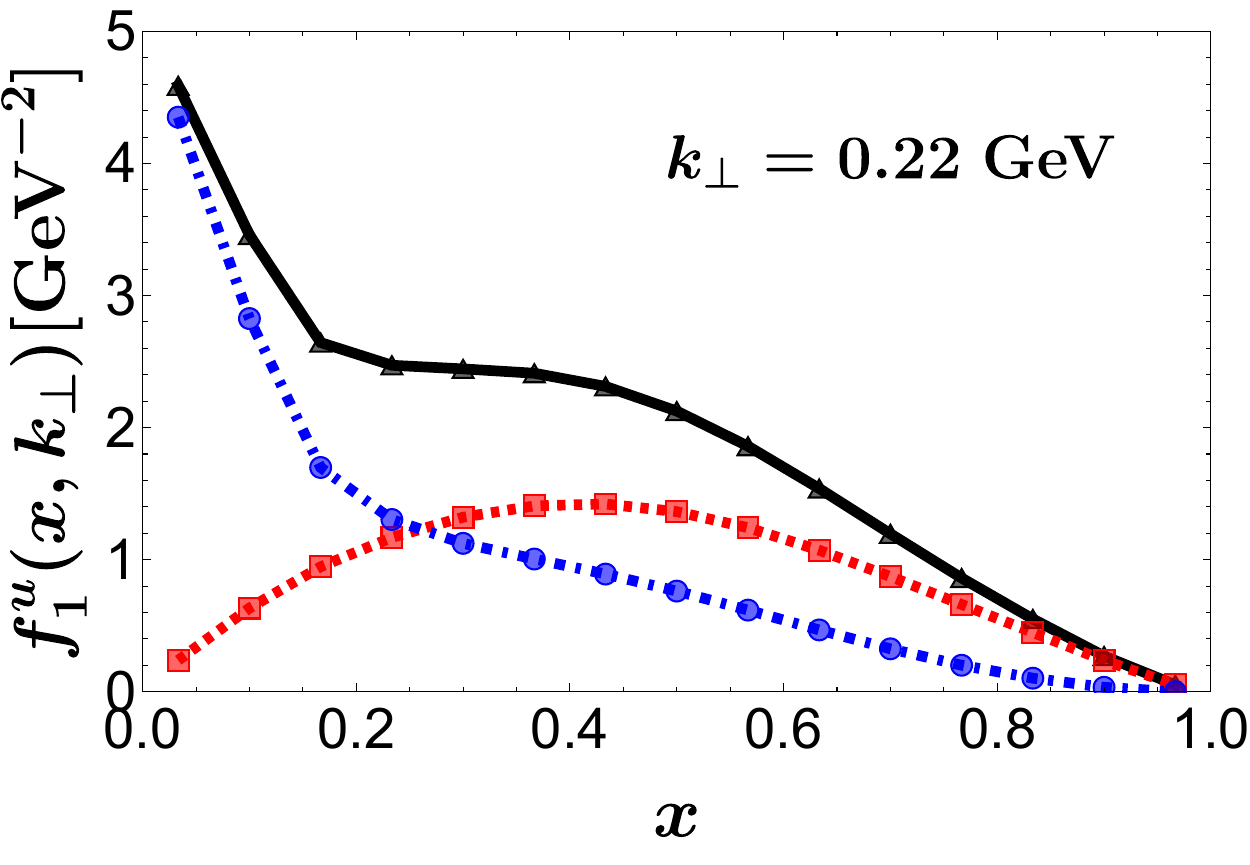}
    \includegraphics[width=0.3\textwidth]{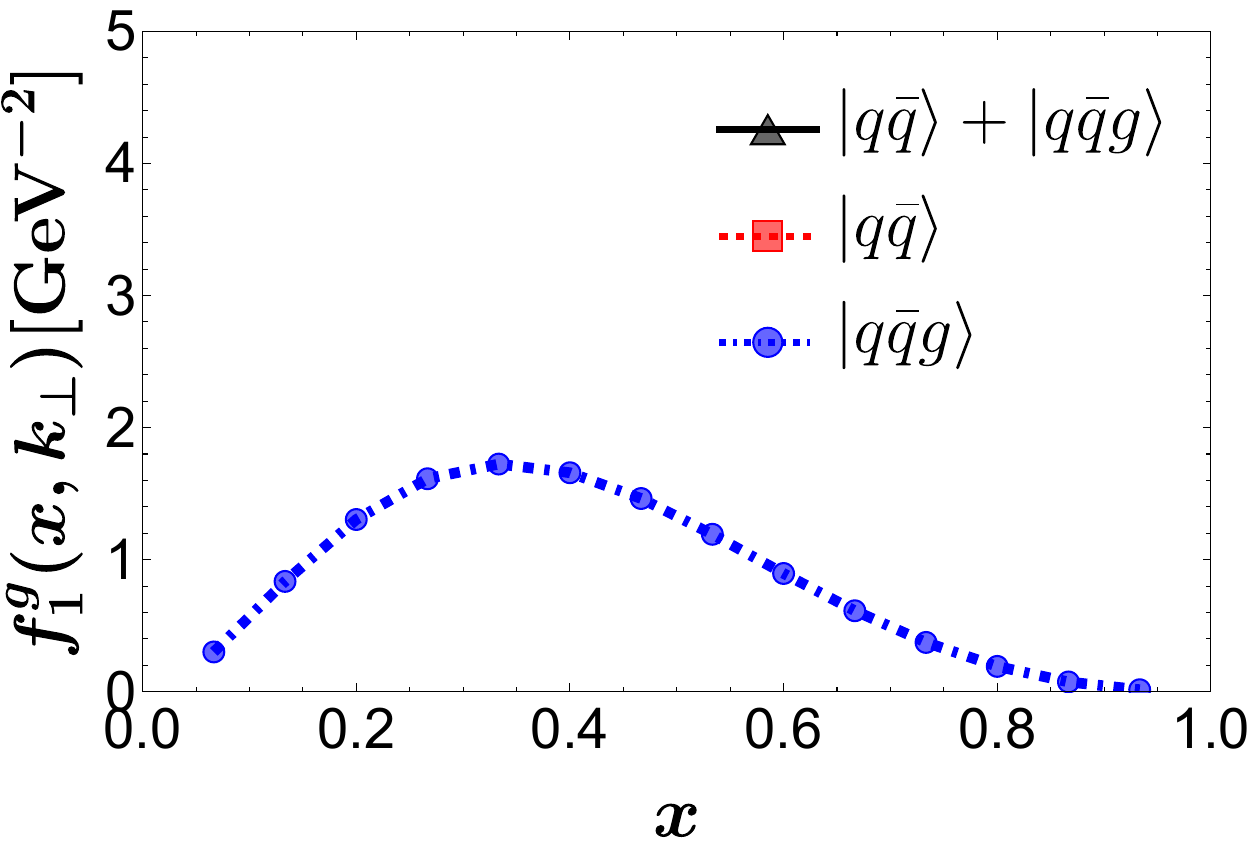}
    \includegraphics[width=0.3\textwidth]{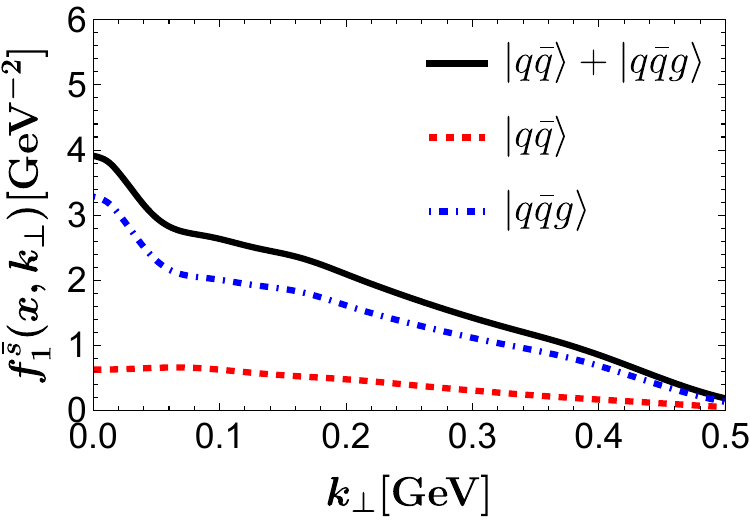}
    \includegraphics[width=0.3\textwidth]{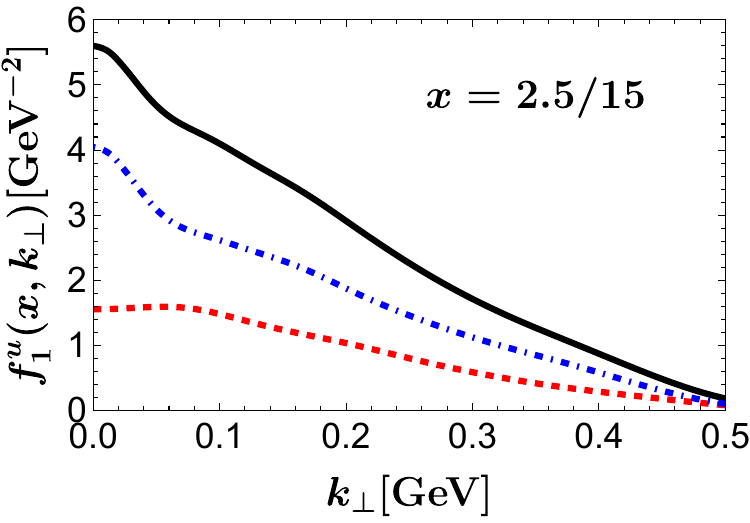}
    \includegraphics[width=0.3\textwidth]{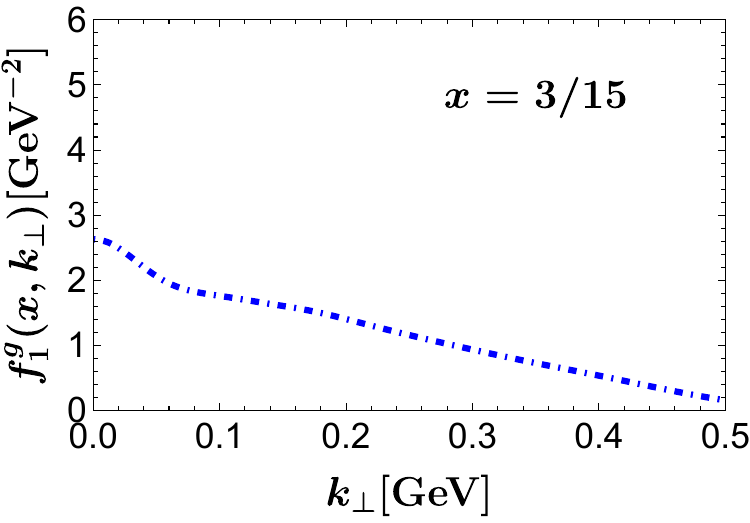}
\caption{Contributions from the $|q\bar{q}\rangle$ (red lines) and $|q\bar{q}g\rangle$ (blue lines) Fock sectors to the total (black lines) twist-2 TMD $f_1(x,k_\perp)$ of the kaon. The left, middle, and right panels correspond to the $\bar{s}$ quark, $u$ quark, and gluon, respectively. In the upper panels, the TMDs are shown as functions of $x$ at fixed $k_\perp = 0.22~\mathrm{GeV}$. In the lower panels, the TMDs are shown as functions of $k_\perp$ at fixed $x = 2.5/15$ for a quark and $x = 3/15$ for a gluon. As described in the text, all results are obtained by averaging over BLFQ calculations with $N_{\mathrm{max}}=\{12,14,16\}$ and $K=15$.}
    \label{Fig:2}
\end{figure*}

\subsection*{Overlap representations of TMDs}


In this subsection, we review the LFWF overlap representations for the leading-twist and subleading-twist TMDs of the kaon. In Fock space, the meson state is expressed as a superposition of all possible Fock components,
\begin{equation}
\begin{split}
    |P\rangle&=\sum_{n}\sum_{\lambda_1,\lambda_2,...,\lambda_n}\int\prod_1^n\frac{[\mathrm{d}x_i\mathrm{d}^2p_{\perp i}]}{2(2\pi)^3\sqrt{x_i}}\\
    &\times2(2\pi)^3\delta(1-\sum_1^nx_i)\delta^2(P_{\perp}-\sum_1^np_{\perp i})\\
    &\times\Psi_{n,\{\lambda_i\}}(\{x_i,k_{\perp i}\})|\{x_iP^+,p_{\perp i},\lambda_i\}\rangle\label{24},
\end{split}
\end{equation}
where $k_{\perp i}$ and $p_{\perp i}$ represent the relative transverse momentum and physical transverse momentum, respectively, \ie, $k_{\perp i} = p_{\perp i} - x_i P_\perp$. The corresponding orthonormal relation is
\begin{equation}
    \langle P^{\prime}|P\rangle=\sum_{n,n^{\prime}}2P^+(2\pi)^3\delta^3(P-P^{\prime})\delta_{nn^{\prime}}P_{n}\label{25},
\end{equation}
where $\delta^3(P-P^{\prime})=\delta(P^+-P^{\prime +})\delta^2(P_{\perp}-P^{\prime}_{\perp})$, and $P_n$ defines the probability of finding the $n$-parton state in the meson. Although the complete expansion includes all Fock states, in this work we restrict our attention to the two lowest components, the two-body LFWF $\psi_2$ for the $|q\bar{q}\rangle$ state and the three-body LFWF $\psi_3$ for the $|q\bar{q}g\rangle$ state. Using the model parameters listed in~\tab{I}, we find that the probabilities for the $|q\bar{q}\rangle$ and $|q\bar{q}g\rangle$ Fock sectors are 52.64\% and 47.36\%, respectively.

Using the meson state in Eq.~\eqref{25} together with the quantized quark and gluon fields in the Lepage–Brodsky prescription\cite{BRODSKY1998299,PhysRevD.22.2157}, the twist-2 quark TMD $f^q_1(x,k_{\perp})$ and the twist-3 quark TMDs $\tilde{e}^q(x,k_{\perp})$ and $\tilde{f}^q_{\perp}(x,k_{\perp})$ can be expressed in terms of overlaps of the LFWFs as~\cite{Zhu:2023lst}
\begin{gather}
\begin{split}
    &f^q_1(x,k_{\perp})=\sum_{\lambda_1,\lambda_2}\mathrm{d}[12]\Psi_{2,\lambda_1\lambda_2}^*\Psi_{2,\lambda_1\lambda_2}\delta^3(\tilde{p}^{\prime}_1-\tilde{k})\\
    &+\sum_{\lambda_1,\lambda_2,\lambda_g}\int \mathrm{d}[123]\Psi_{3,\lambda_1\lambda_2\lambda_g}^*\Psi_{3,\lambda_1\lambda_2\lambda_g}\delta^3(\tilde{p}_1-\tilde{k})\label{26},
\end{split}\\
\begin{split}
\tilde{e}^q(x,k_{\perp})&=-\frac{gC_F\sqrt{2}}{Mx}\sum_{\lambda_2,\lambda_g}\int \mathrm{d}[12]\mathrm{d}[123]\frac{2(2\pi)^3}{\sqrt{x_g}}\\
&\times(\Psi^*_{2,-\lambda_2}\Psi_{3,+\lambda_2\lambda_g}-\Psi_{2,+\lambda_2}\Psi^*_{3,-\lambda_2\lambda_g})\\
&\times[\delta^3(\tilde{p}^{\prime}_1-\tilde{k})-\delta^3(\tilde{p}_1-\tilde{k})]\delta^3(\tilde{p}^{\prime}_2-\tilde{p}_2)\label{27},
\end{split}\\
\begin{split}
    \tilde{f}^q_{\perp}(x,k_{\perp})&=\frac{gC_F\sqrt{2}}{2xk_{\perp}}\sum_{\lambda_1,\lambda_2,\lambda_g}\int \mathrm{d}[12]\mathrm{d}[123]\frac{2(2\pi)^3}{\sqrt{x_g}}(-)^{\frac{\lambda_g+1}{2}}\\
    &\times[\Psi^*_{2,\lambda_1\lambda_2}\Psi_{3,\lambda_1\lambda_2\lambda_g}\delta^3(\tilde{p}_1-\tilde{k})\\
    &+\Psi_{2,\lambda_1\lambda_2}\Psi^*_{3,\lambda_1\lambda_2\lambda_g}\delta^3(\tilde{p}^{\prime}_1-\tilde{k})]\delta^3(\tilde{p}^{\prime}_2-\tilde{p}_2)\label{28}.
\end{split}
\end{gather}
The twist-2 gluon TMD $f^g_1(x,k_{\perp})$ is expressed as 
\begin{align}
    f^g_1(x,k_{\perp})&=\sum_{\lambda_1,\lambda_2,\lambda_g}\int\mathrm{d}[123]\Psi^*_{3,\lambda_1\lambda_2\lambda_g}\Psi_{3,\lambda_1\lambda_2\lambda_g}\nonumber\\
    &\times\delta^3(\tilde{p_g}-\tilde{k}).\label{29}
\end{align}

The conventions we adopt are
\begin{equation}
\begin{aligned}
    \mathrm{d}[12]&\equiv\frac{\prod_1^2 \mathrm{d}x_i^{\prime}\mathrm{d}^2p_{\perp i}^{\prime}}{[2(2\pi)^3]^2}2(2\pi)^3\delta^3(\tilde{P}-\sum_1^2\tilde{p}^{\prime}_i)\label{30},\\
    \mathrm{d}[123]&\equiv\frac{\prod_1^3 \mathrm{d}x_i\mathrm{d}^2p_{\perp i}}{[2(2\pi)^3]^3}2(2\pi)^3\delta^3(\tilde{P}-\sum_1^3\tilde{p}_i),
    \end{aligned}
\end{equation}
where $\delta^3(\tilde{P}-\sum_1^n\tilde{p}_i)=\delta(1-\sum_1^nx_i)\delta^2(P_{\perp}-\sum_1^np_{\perp i})$. For simplicity, we omit the arguments $(\tilde{p}^{\prime}_1,\tilde{p}^{\prime}_2)$ from the quark-antiquark LFWFs and $(\tilde{p}_1,\tilde{p}_2,\tilde{p}_g)$ from the quark-antiquark-gluon LFWFs, where each momentum is denoted by $\tilde{k}\equiv(x,k_{\perp})$. 
The color factor, $C_F=2/\sqrt{3}$, derives from the quark-quark-gluon matrix element $\langle q\bar{q}|\Psi_bA^{\mu, a}T^a_{bc}\Psi_c|q\bar{q}g\rangle$ and its hermitian conjugate. 
As shown in~\eqs{\ref{26}}-(\ref{28}), the unpolarized TMD $f_1(x,k_{\perp})$ can be decomposed into the individual overlaps of the $|q\bar{q}\rangle$ and $|q\bar{q}g\rangle$ Fock sectors' LFWFs, while the genuine twist-3 TMDs $\tilde{e}(x,k_{\perp})$ and $\tilde{f}_{\perp}(x,k_{\perp})$ blend the $|q\bar{q}\rangle$ and the $|q\bar{q}g\rangle$ Fock sectors. 
%
This indicates only leading-twist TMDs and PDFs contribute to probabilistic interpretation. 
The twist-3 TMDs are interpreted as some interference of scattering between a coherent quark-gluon pair and a single quark~\cite{JAFFE1992527,PhysRevLett.67.552,Efremov_2003}.

\section{Numerical results}\label{sec:key_results}
Using the basis truncation parameters $\{N_{\text{max}},K\}=\{14,15\}$ summarized in Table~\ref{Tab:para}, we solve the light-front stationary Schr\"{o}dinger equation and obtain the two--Fock-sector LFWFs of the kaon, corresponding to a mass $M=0.485~\mathrm{GeV}$. 
The nonperturbative solutions obtained in single-particle coordinates Eq.~(\ref{eqn:wf}) are transformed to relative coordinates by factorizing out the center-of-mass motion~\cite{Wiecki:2014ola,Xu:2021wwj,Hu:2020arv}, yielding the boost-invariant LFWFs $\Psi_{n,\{\lambda_i\}}(x_i,p_{\perp i})$. These LFWFs are then employed in the overlap representations, Eqs.~(\ref{26})--(\ref{29}), to compute the quark and gluon TMDs and PDFs of the kaon.
%

The TMDs computed within BLFQ exhibit oscillations in the transverse momentum direction, reflecting the oscillatory behavior of the 2D-HO basis functions used in the transverse plane. The transverse basis states are determined by the transverse truncation parameter $N_{\mathrm{max}}$ and the 2D-HO scale $b$ together. These oscillatory artifacts, arising from the finite basis, can be reduced by averaging results across different $N_{\rm max}$ values. Following the procedure of Refs.~\cite{Hu:2020arv,Nair:2022evk,Zhu:2023lst}, we perform a two-step averaging at fixed $K$: 
first averaging the results obtained at $N_{\rm max}=n$ and $N_{\rm max}=n+2$,  
then averaging those at $N_{\rm max}=n+2$ and $N_{\rm max}=n+4$.
The final result is obtained by averaging these two intermediate averages, thus incorporating BLFQ outputs at $N_{\rm max}=\{n,\,n+2,\,n+4\}$.

\subsection{Twist-2 and twist-3 TMDs}
Unlike the pion, the kaon contains a strange quark as one of its valence constituents. Accordingly, we calculate two sets of results for different quark flavors within the $|q\bar q\rangle$ and $|q\bar q\rangle+|q\bar q g\rangle$ Fock sectors. In addition, we also evaluate the leading-twist TMDs and PDFs of the gluon in the $|q\bar q g\rangle$ sector.

In Fig.~\ref{Fig:1}, we show the three-dimensional profiles of the unpolarized valence-quark ($u$ and $\bar{s}$) and gluon leading-twist TMDs, $f_1(x,k_\perp)$, in the kaon. The results are obtained after averaging over $N_{\mathrm{max}}$ in the $|q\bar{q}\rangle+|q\bar{q}g\rangle$ Fock space and are presented as functions of the parton’s longitudinal momentum fraction $x$ and transverse momentum $k_\perp$. The left and middle panels display the total $u$- and $\bar{s}$-quark TMDs, which include contributions from both the leading and next-to-leading Fock sectors, while the right panel shows the gluon TMD.

We observe that the BLFQ results exhibit oscillatory behavior that depends on 
$N_{\mathrm{max}}$. The oscillation amplitude is more pronounced at small 
$k_\perp$ and low $x$, while the oscillation phase varies with changing 
$N_{\mathrm{max}}$. By averaging the results obtained at different 
$N_{\mathrm{max}}$ as described above, we effectively suppress these oscillations, leading to 
smoother distributions. Here, we present the results for the basis truncation 
$N_{\mathrm{max}}=\{12,14,16\}$ and $K=15$. In contrast, at large $k_\perp$ and 
for $x > 0.1$, the oscillation amplitude is already small, indicating better 
numerical stability of the results in that region. 

We find that the TMDs still exhibit small residual oscillations in the 
$k_\perp$ direction even after averaging. The TMDs drop to zero abruptly in the 
transverse plane at the ultraviolet (UV) cutoff $\sim b\sqrt{N_{\mathrm{max}}}$. 
Note that the UV cutoff increases with increasing $N_{\mathrm{max}}$, and 
consequently, the fall-off of the TMDs extends to higher transverse momenta for 
larger $N_{\mathrm{max}}$. This characteristic behavior of TMDs within our BLFQ 
framework has also been observed previously for physical spin-$\tfrac{1}{2}$ 
particles in QED and for light pseudoscalar particles in QCD, \ie, the electron 
and the pion, respectively, as reported in Refs.~\cite{Hu_2021,Zhu:2023lst}.

The quark TMD exhibits a pronounced peak at low $x$, highlighting the contribution of the next-to-leading Fock sector that includes a dynamical gluon, as shown in Figs.~\ref{Fig:1} and \ref{Fig:2}. 
In contrast, the leading (valence) Fock sector contributes predominantly at intermediate values of $x$ and features a broader distribution in transverse momentum $k_\perp$. 
This is illustrated in Fig.~\ref{Fig:2}, where we present two-dimensional slices of the total TMDs, which include contributions from both the leading and next-to-leading Fock sectors, along with their individual components: the leading ($|q\bar{q}\rangle$) and next-to-leading ($|q\bar{q}g\rangle$) sectors. 
Furthermore, the TMD from the valence sector decreases more slowly at large $k_\perp$ compared to the next-to-leading sector. 
This slower fall-off indicates that valence quarks are more likely to carry higher transverse momenta, reflecting stronger intrinsic transverse motion in the absence of explicit gluon degrees of freedom. 
Such momentum-space behavior is consistent with previous phenomenological studies~\cite{Kaur:2020vkq,Shi:2020pqe,Pasquini:2014ppa,Lorce:2016ugb,Shi:2018zqd,Ahmady:2019yvo} and with earlier BLFQ results for the pion~\cite{Lan:2025qio}. 
Finally, the strange-quark distribution in the kaon is significantly larger than that of the up quark at large $x$, indicating that the heavier quark carries a greater fraction of the longitudinal momentum, as expected.

\begin{figure*}[htp]
    \centering
    \includegraphics[width=0.4\textwidth]{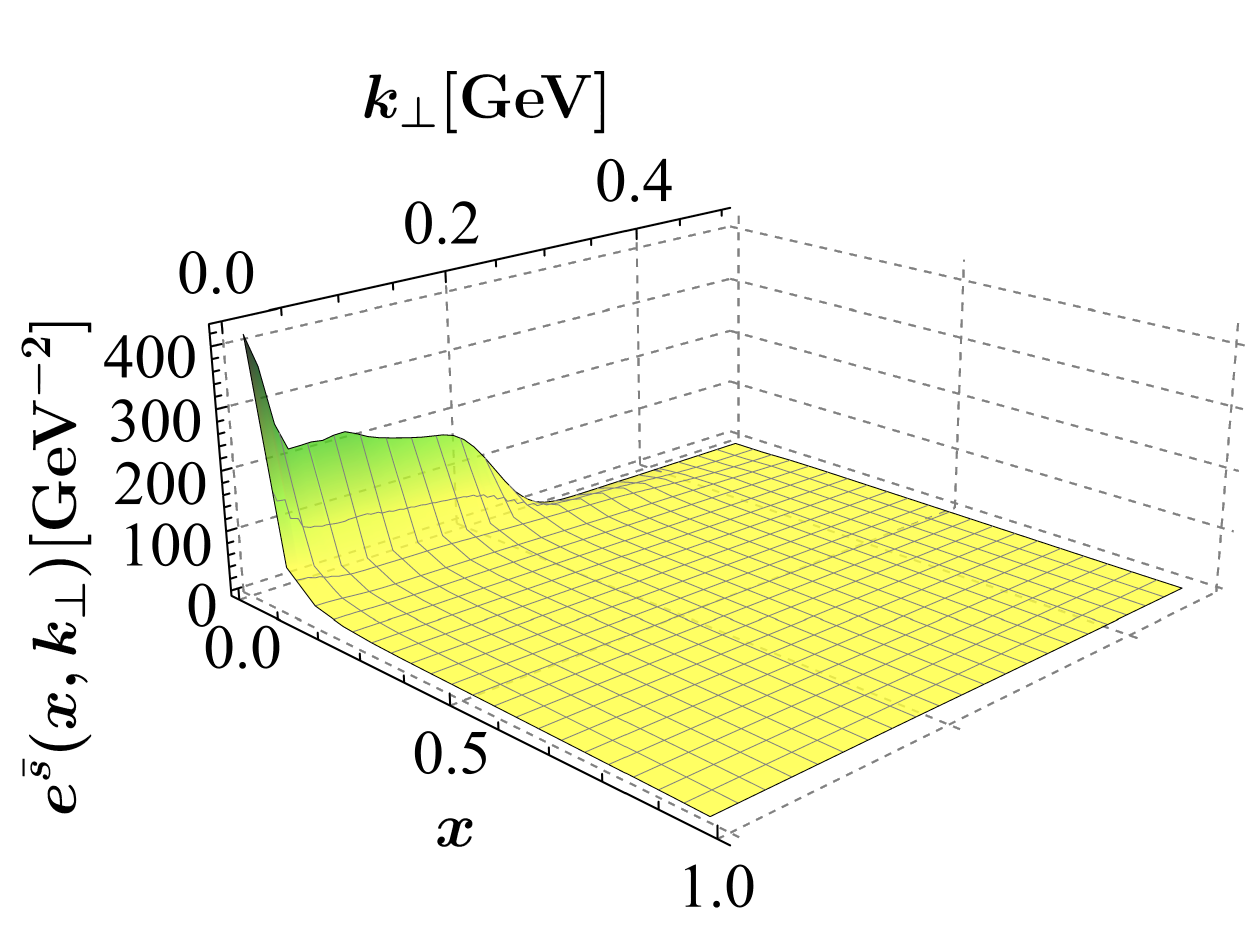}
    \includegraphics[width=0.4\textwidth]{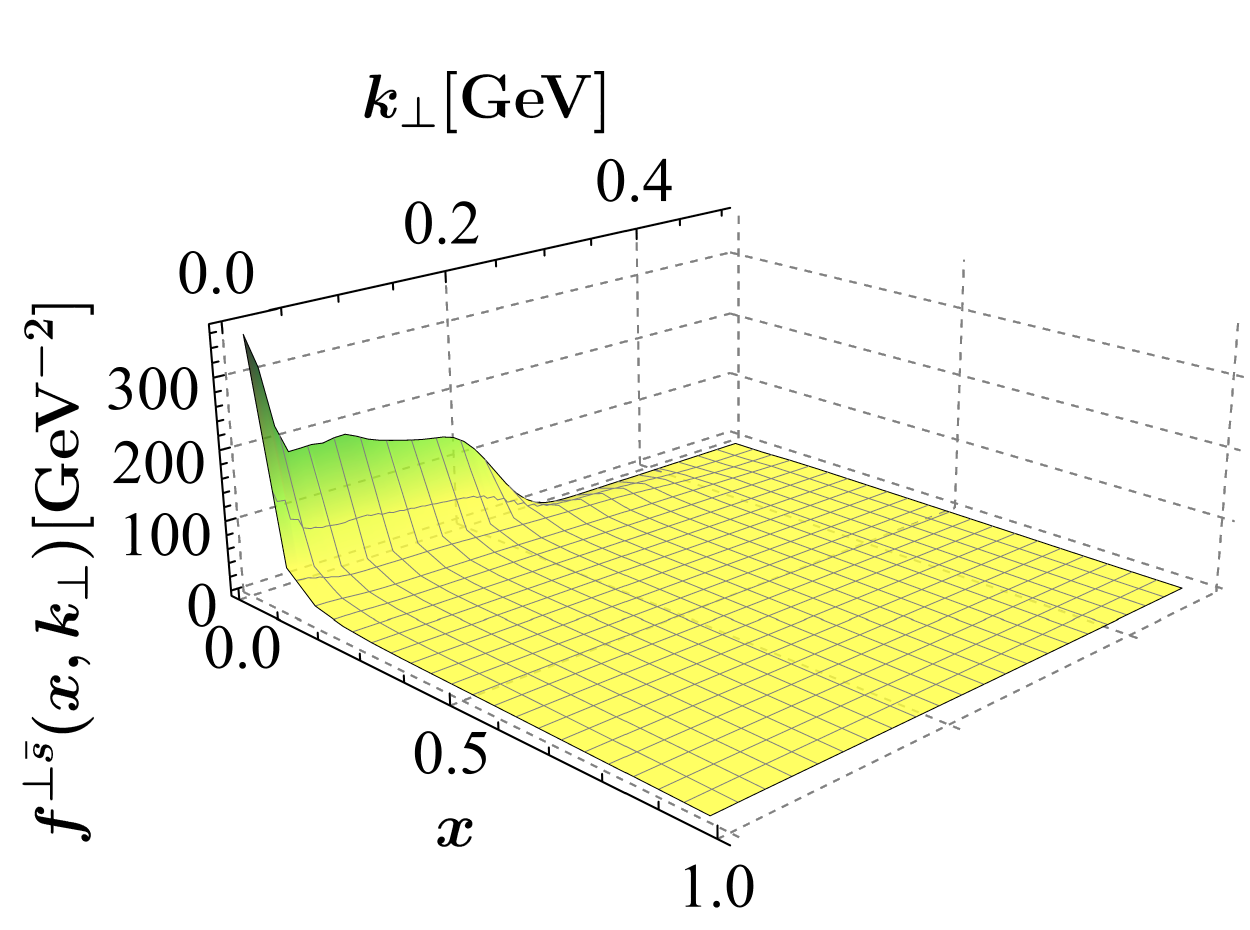}
    \includegraphics[width=0.4\textwidth]{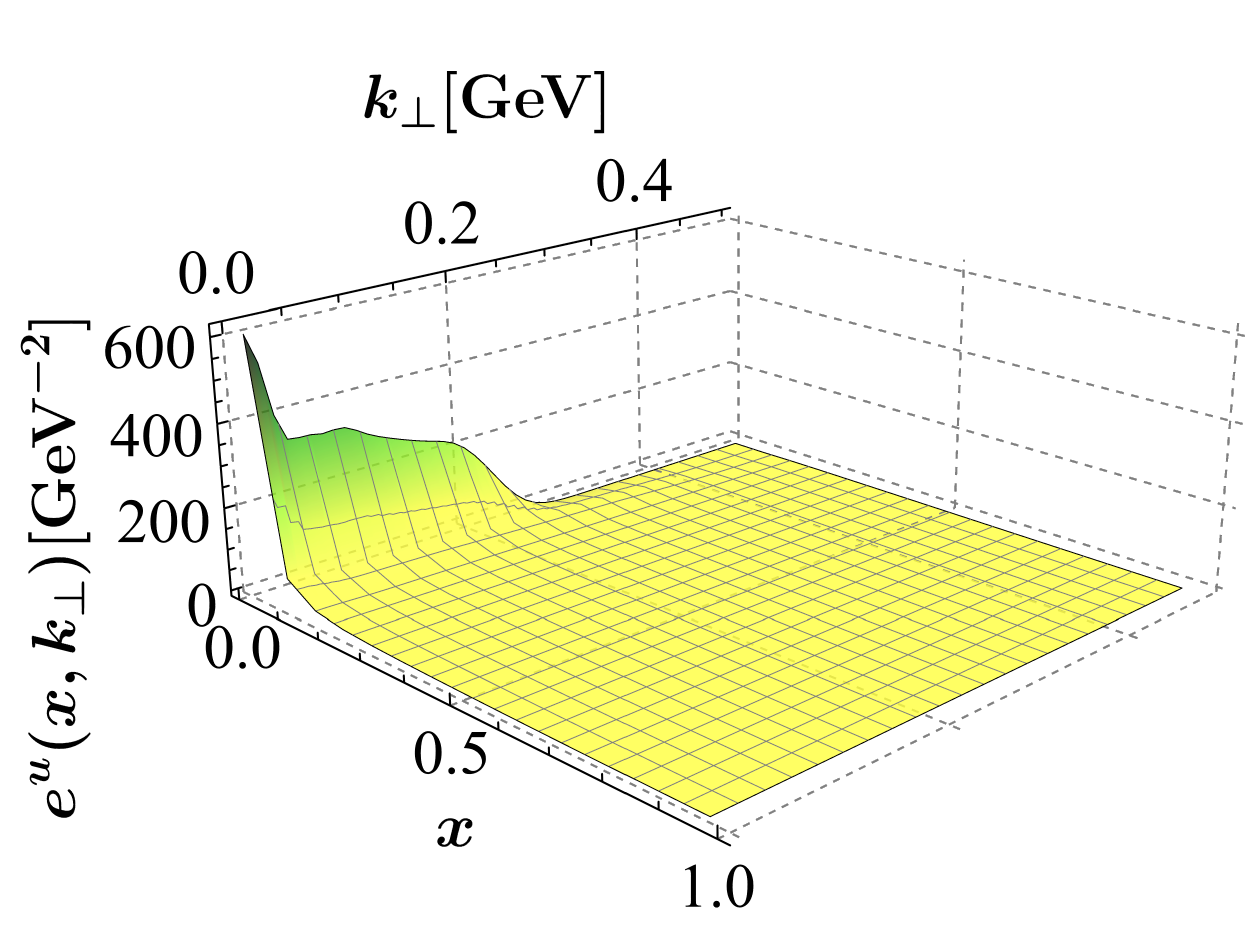}
    \includegraphics[width=0.4\textwidth]{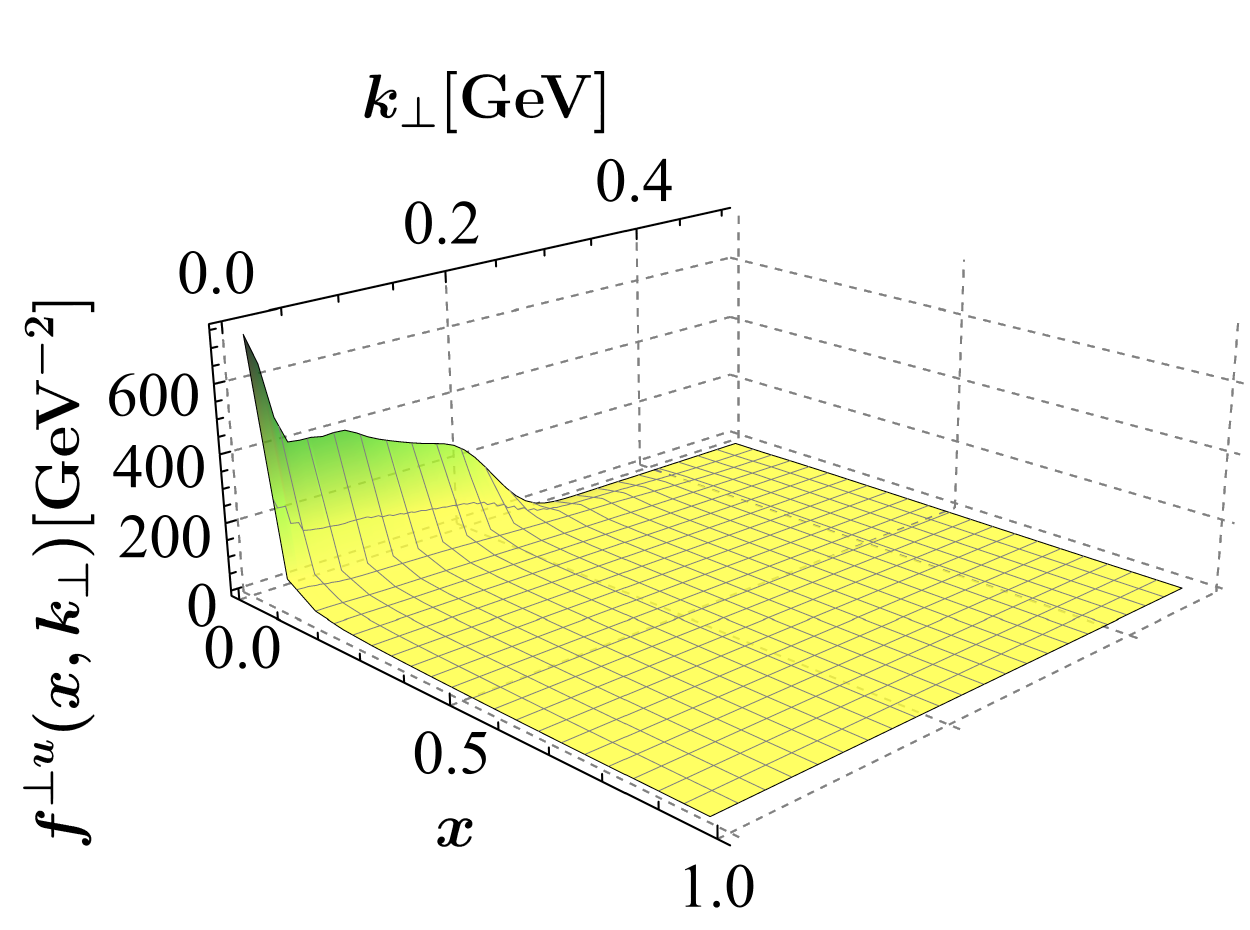}
    \caption{3D plots of the BLFQ results for the twist-3 TMDs $e(x,k_{\perp})$ (left panels) and $f^{\perp}(x,k_{\perp})$ (right panels). The upper panels correspond to the $\bar{s}$ quark, while the lower panels correspond to the $u$ quark. As described in the text, all results are obtained by averaging over BLFQ calculations with $N_{\mathrm{max}}=\{12,14,16\}$ and $K=15$. Note the differences in the vertical scales.}
    \label{tw-3TMDs}
\end{figure*}

The gluon TMD in the kaon is extracted from the $|q\bar{q}g\rangle$ Fock sector and is shown in the right panels of Figs.~\ref{Fig:1} and \ref{Fig:2}. 
The probability of finding an unpolarized gluon carrying a momentum fraction $x$ in the kaon peaks around $x \approx 0.33$. 
The distribution is narrower in $x$ than the quark TMD from the valence Fock sector, indicating a more localized gluon momentum structure along the longitudinal direction. 
We further observe that the probability of finding a gluon is smaller than that of finding a quark in the valence Fock sector at our model scale. 
Moreover, the overall shape of the gluon TMD is consistent with expectations for an effective massive gluon: it vanishes at both endpoints ($x \to 0$ and $x \to 1$) and is concentrated in the intermediate--$x$ region.

\begin{figure*}[htp]
    \centering
    \includegraphics[width=0.4\textwidth]{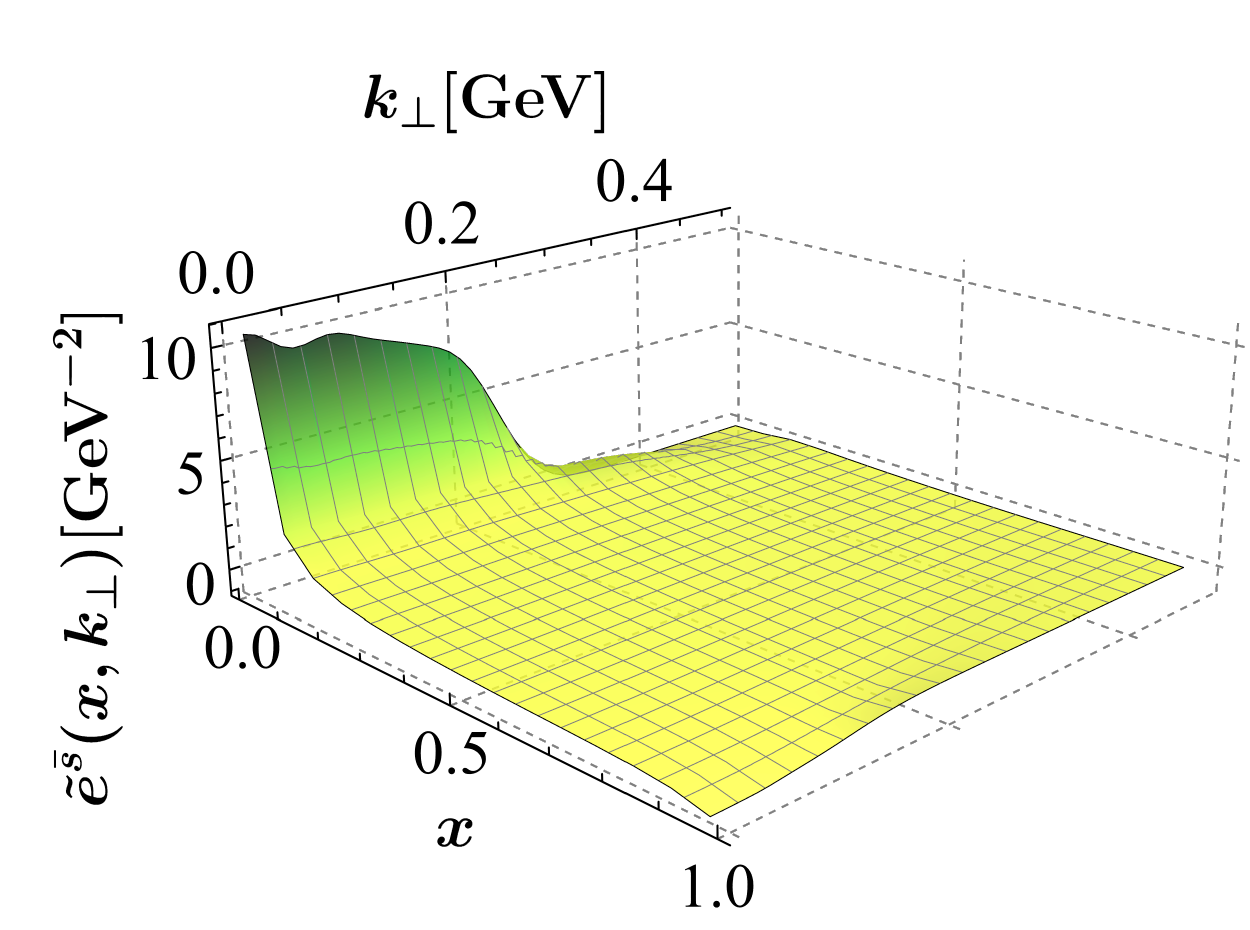}
    \includegraphics[width=0.4\textwidth]{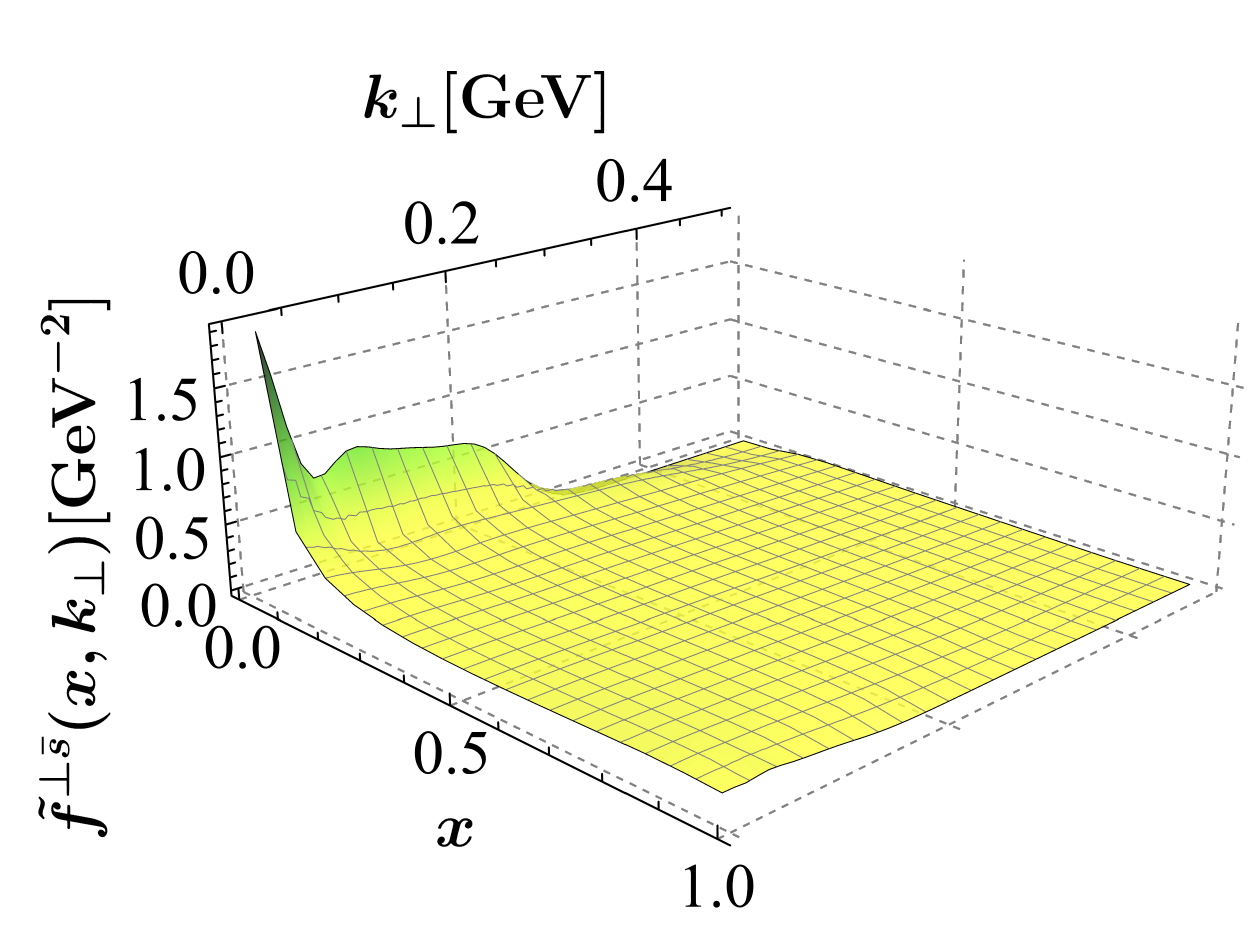}\\ 
    \includegraphics[width=0.4\textwidth]{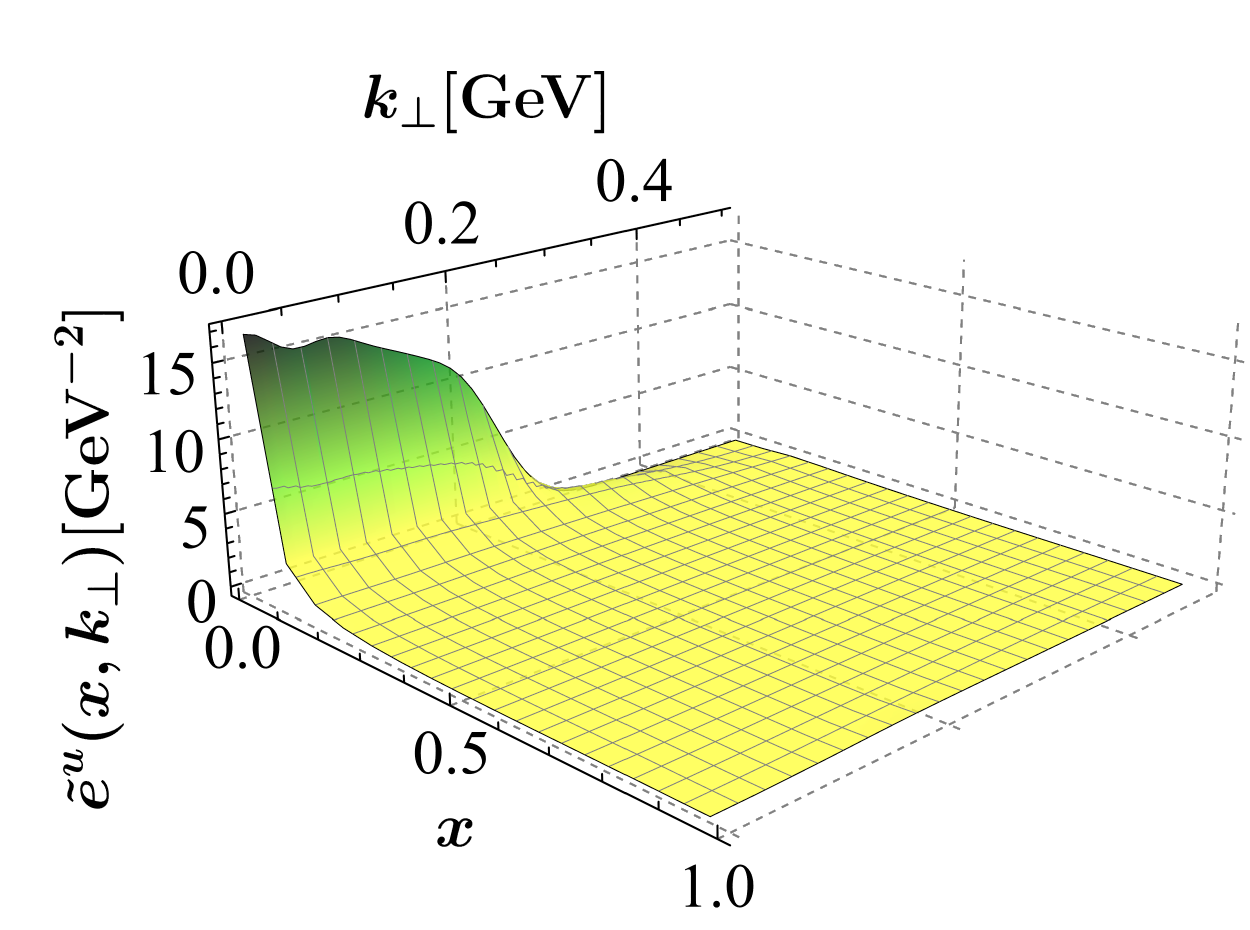}
    \includegraphics[width=0.4\textwidth]{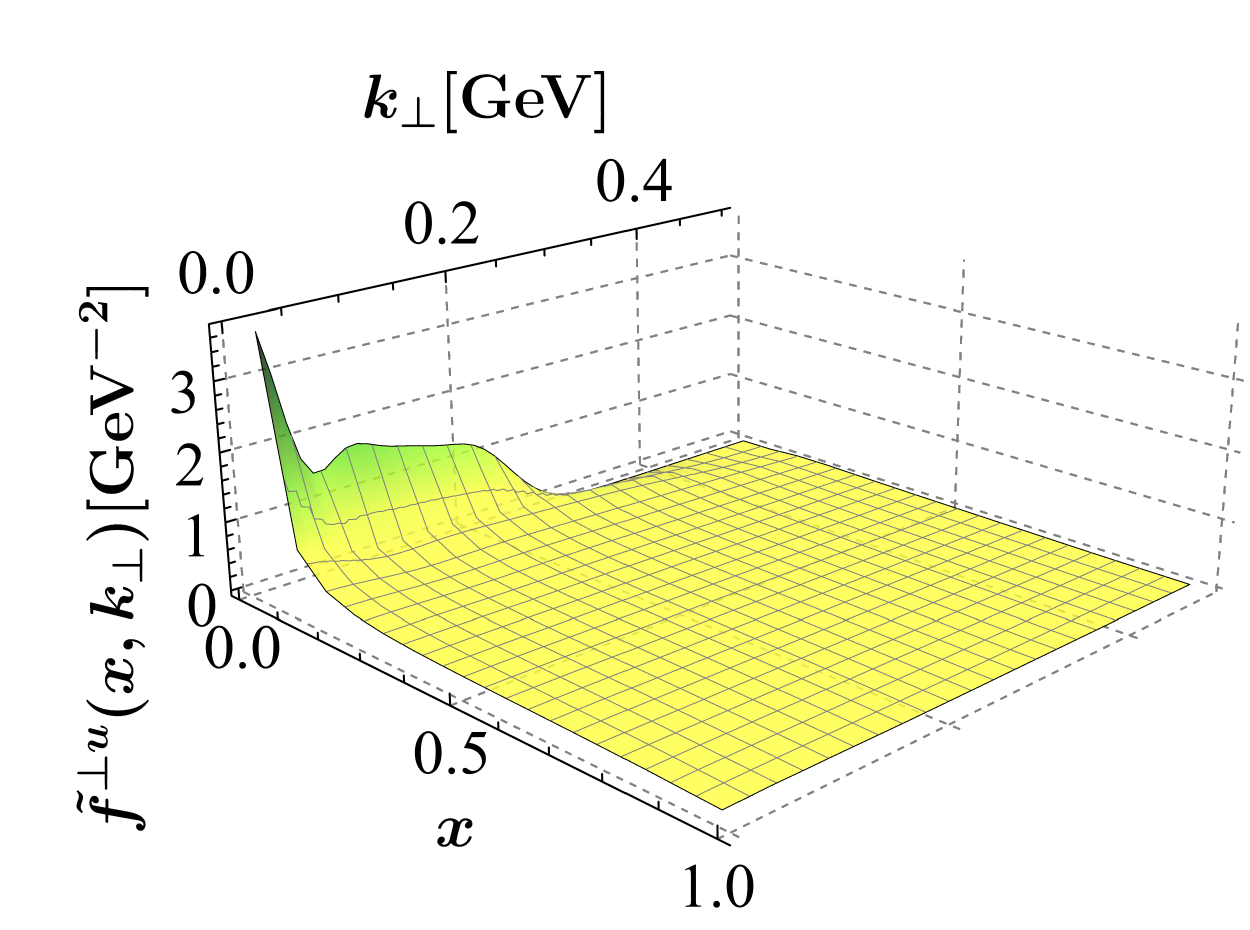}
    \caption{3D plots of the BLFQ results for the genuine twist-3 TMDs $\tilde{e}(x,k_{\perp})$ (left panels) and $\tilde{f}^{\perp}(x,k_{\perp})$ (right panels). The upper panels correspond to the $\bar{s}$ quark, while the lower panels correspond to the $u$ quark. As described in the text, all results are obtained by averaging over BLFQ calculations with $N_{\mathrm{max}}=\{12,14,16\}$ and $K=15$.}
    \label{genuine:tw-3TMDs}
\end{figure*}

\begin{figure*}[htp]
    \includegraphics[width=0.4\textwidth]{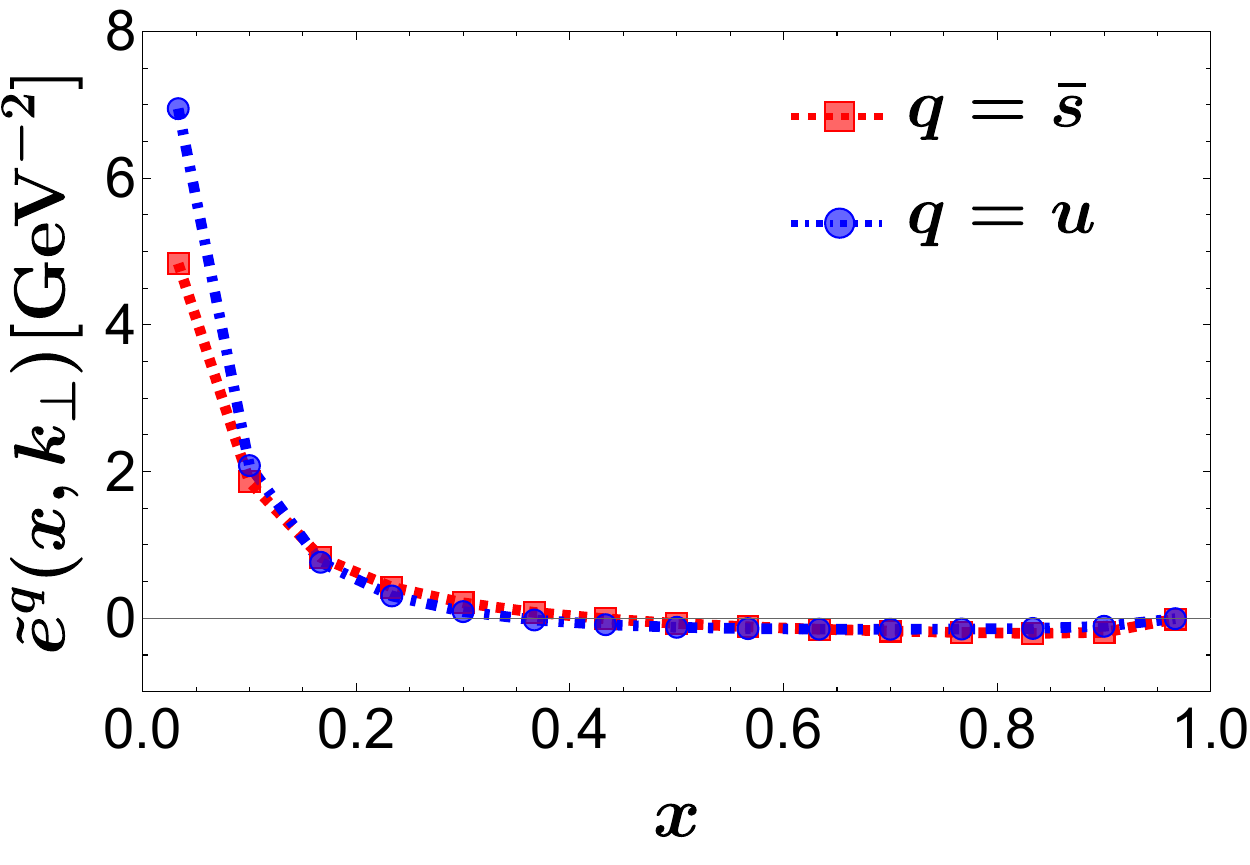}
    \includegraphics[width=0.4\textwidth]{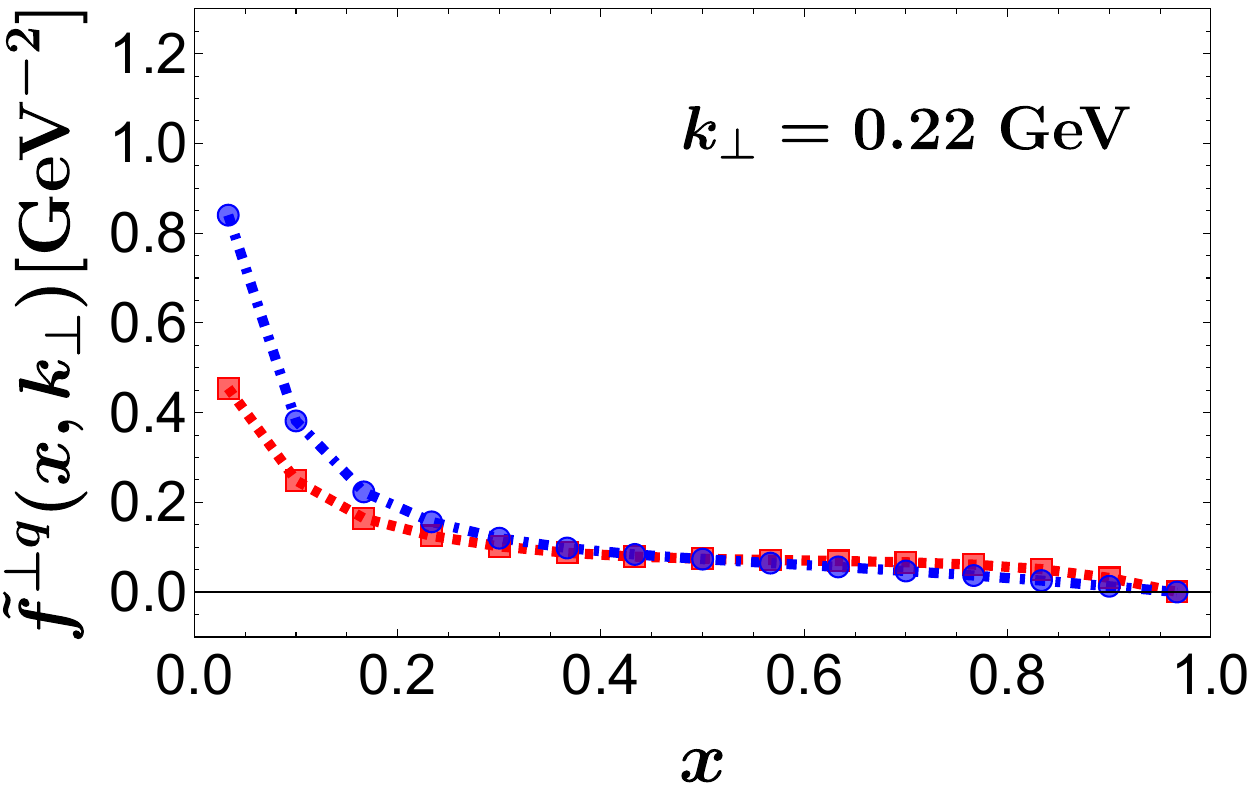}\\
    \includegraphics[width=0.4\textwidth]{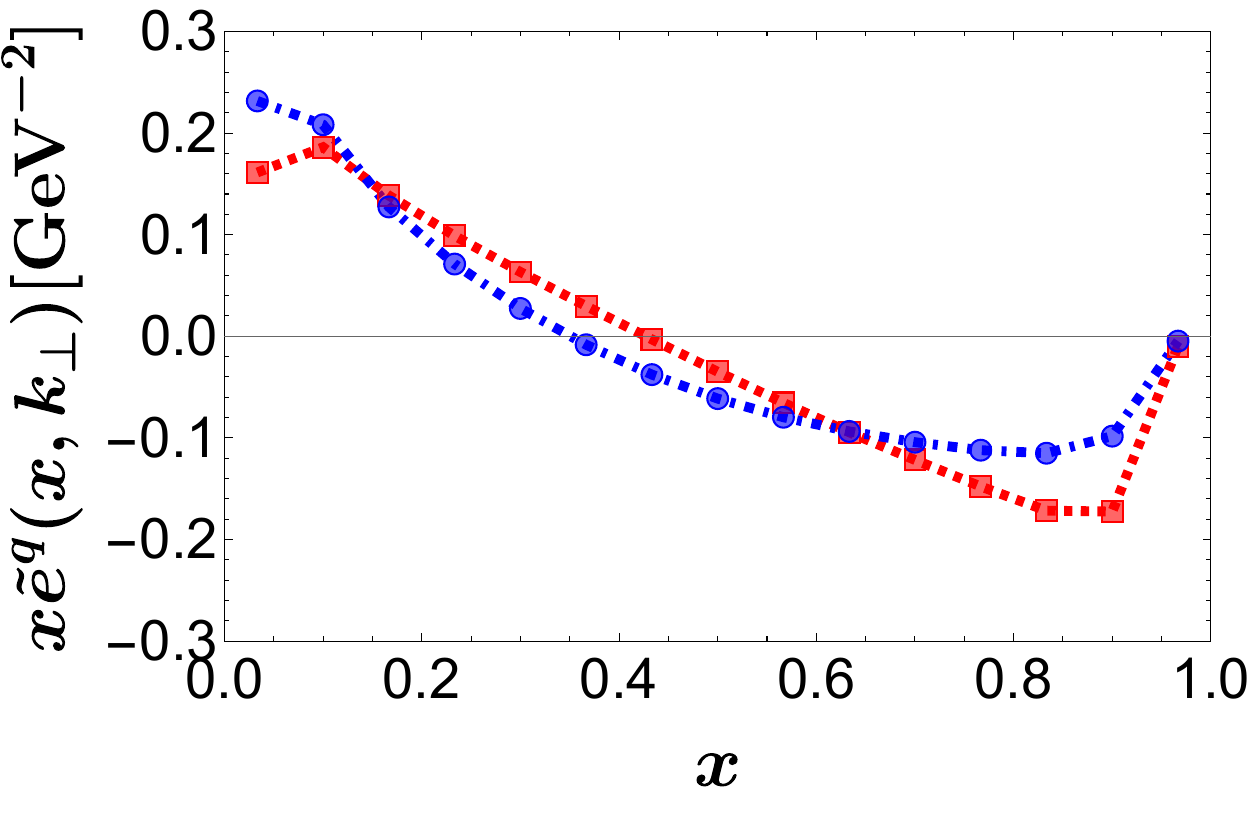}
    \includegraphics[width=0.4\textwidth]{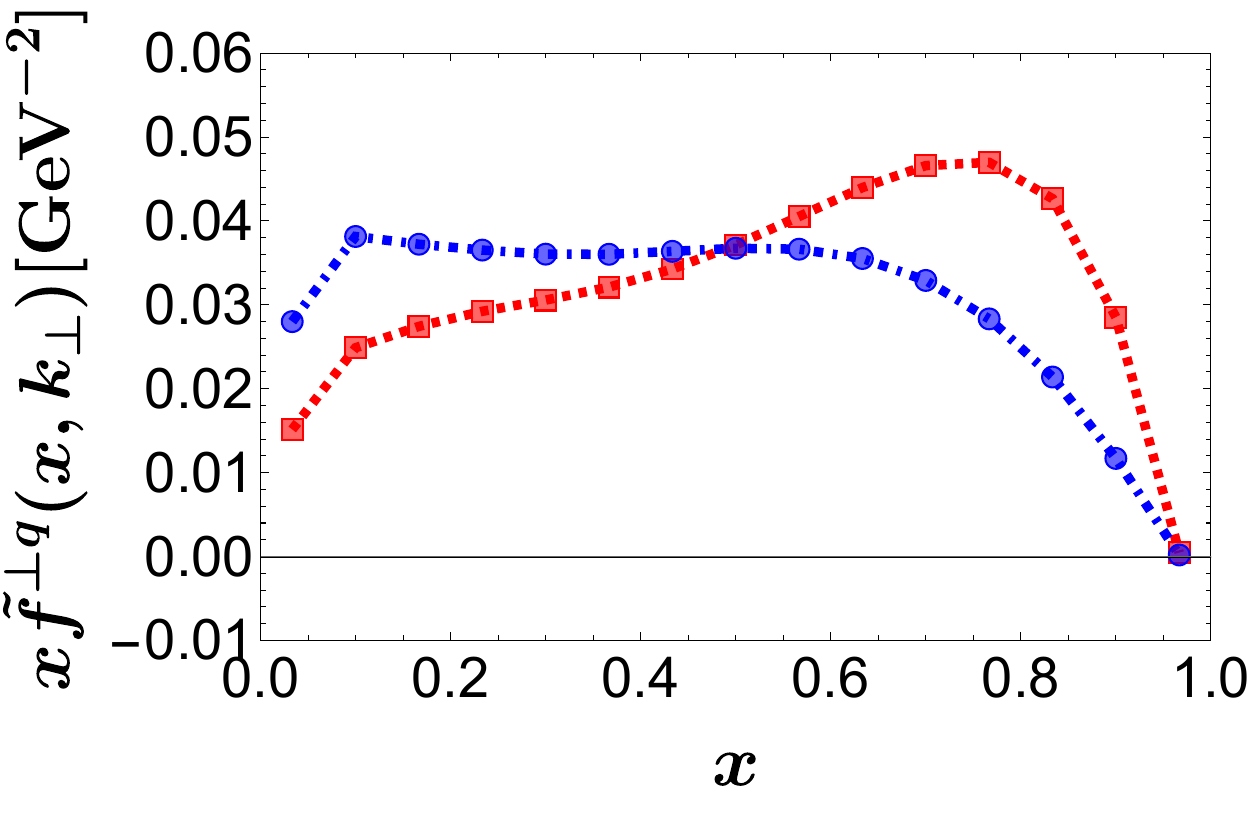}\\
    \includegraphics[width=0.4\textwidth]{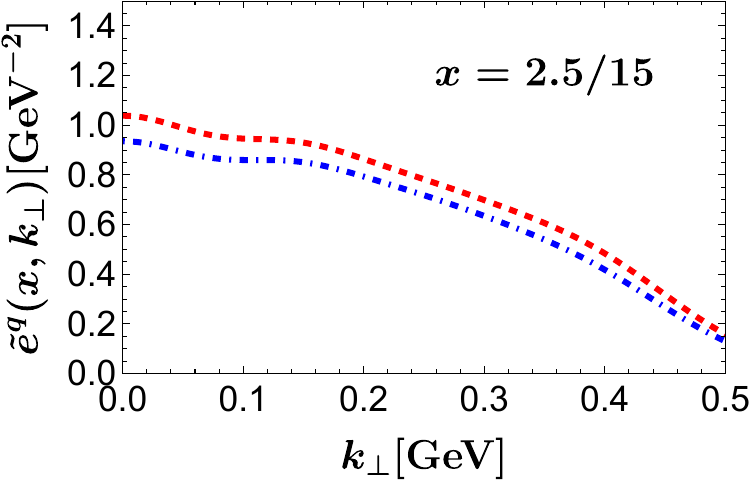}
    \includegraphics[width=0.4\textwidth]{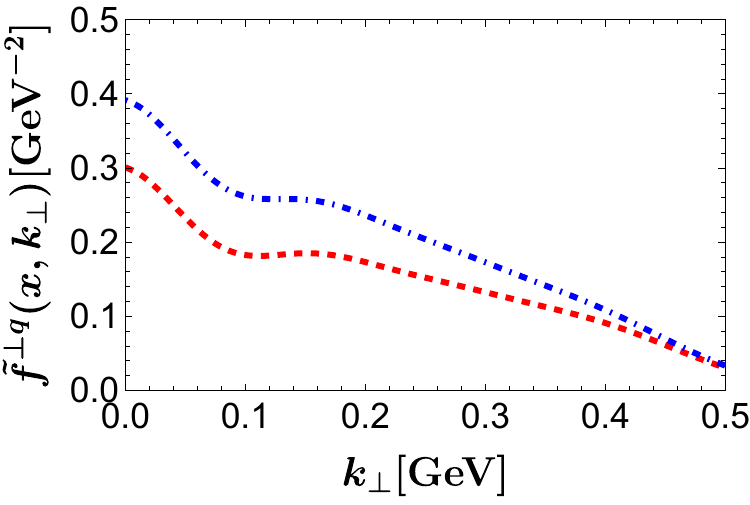}
\caption{2D slices of the 3D BLFQ results for the genuine twist-3 TMDs $\tilde{e}(x,k_\perp)$ (left panels) and $\tilde{f}^{\perp}(x,k_\perp)$ (right panels). In the upper panels, the TMDs are shown as functions of $x$ at fixed $k_\perp = 0.22~\mathrm{GeV}$. In the middle panel, the TMDs $x\tilde{e}(x,k_{\perp})$ and $x\tilde{f}(x,k_{\perp})$ are shown as functions of $x$ with the same fixed $k_{\perp}$. In the lower panels, the TMDs are shown as functions of $k_\perp$ at fixed $x = 2.5/15$. The red and blue curves correspond to the $\bar{s}$ and $u$ quarks, respectively. As described in the text, all results are obtained by averaging over BLFQ calculations with $N_{\mathrm{max}}=\{12,14,16\}$ and $K=15$.}

    \label{genuine:tw-3TMDs_2d}
\end{figure*}

\begin{figure*}[htp]
    \centering
    \includegraphics[width=0.335\textwidth]{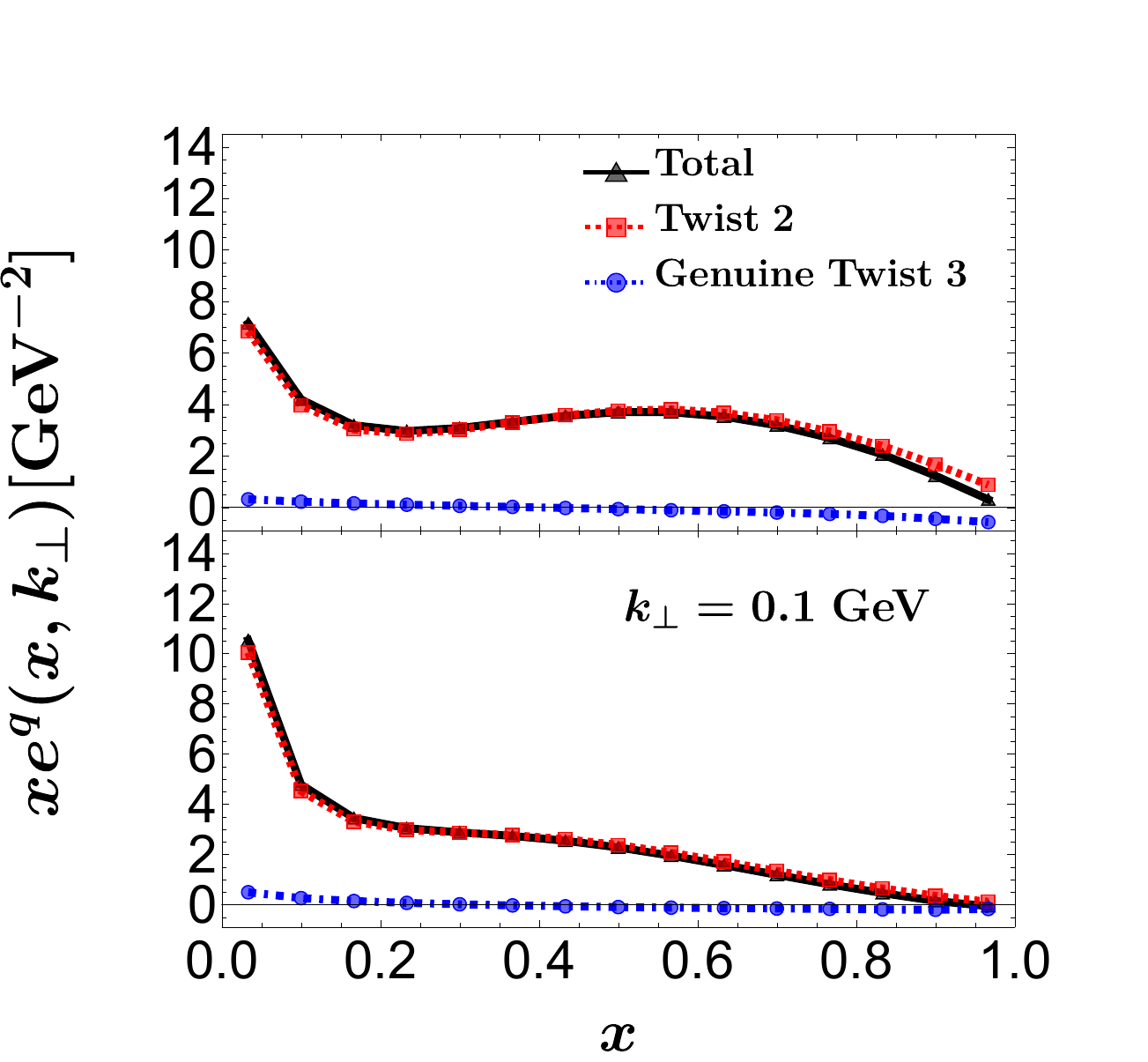}
    \includegraphics[width=0.32\textwidth]{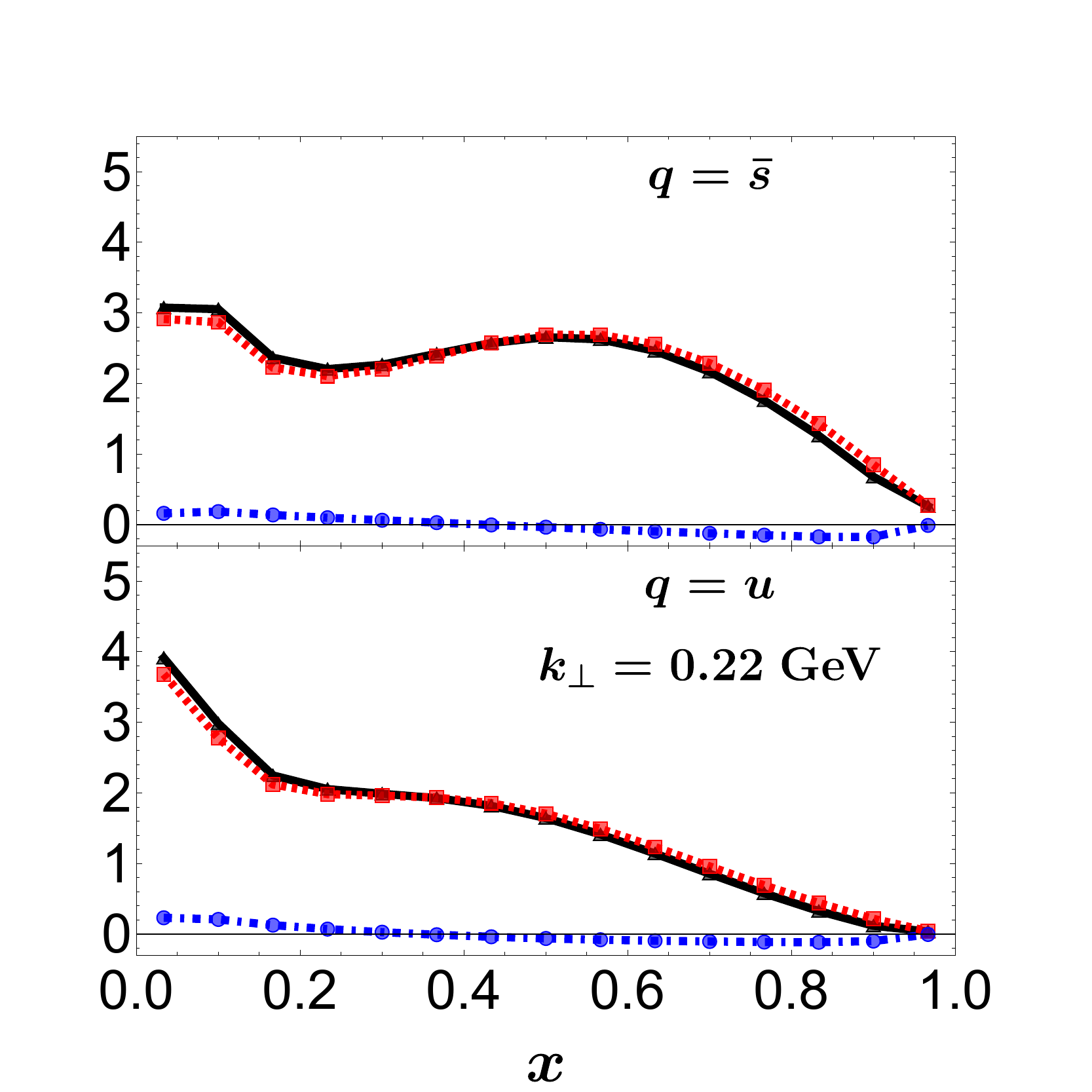}
    \includegraphics[width=0.32\textwidth]{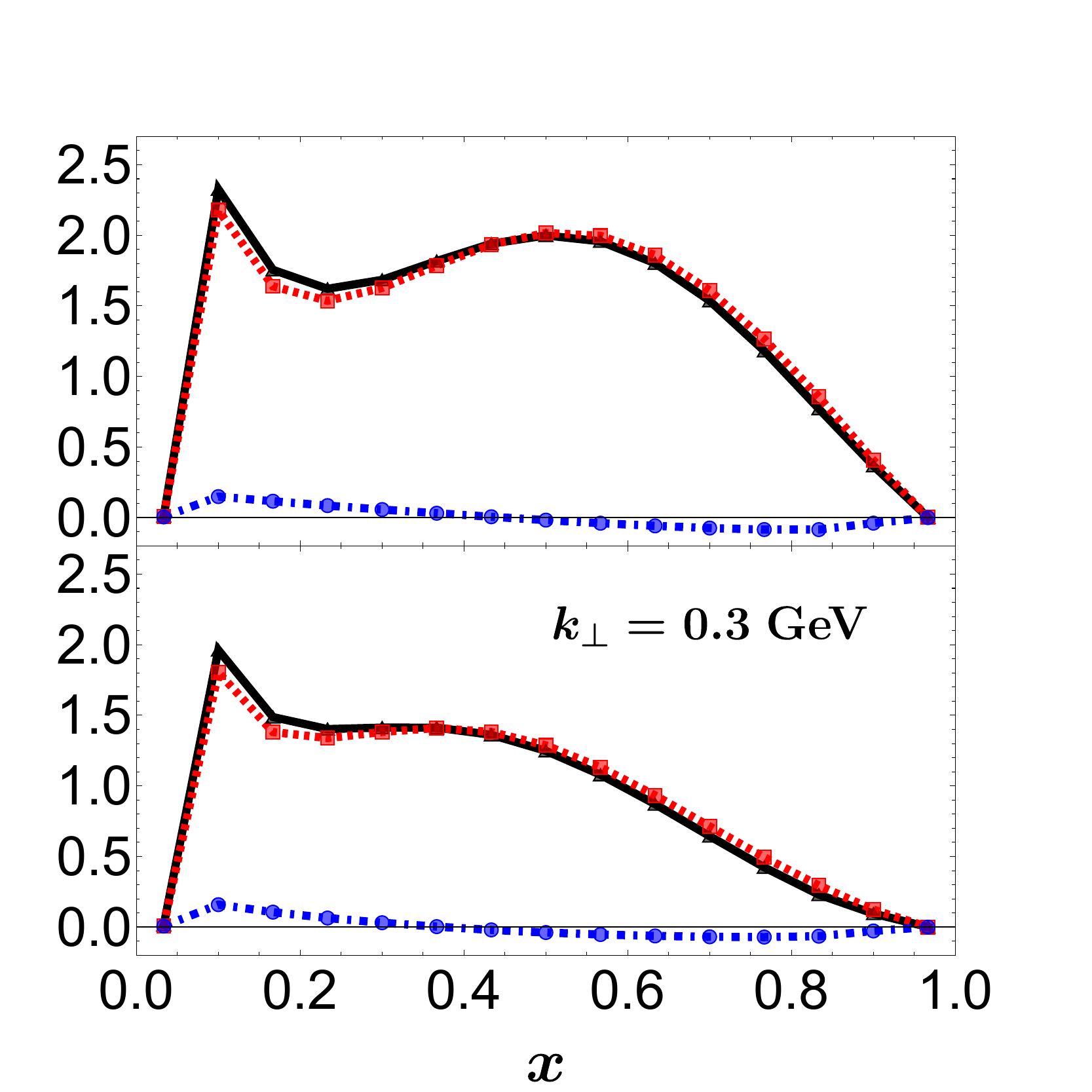}
    \includegraphics[width=0.335\textwidth]{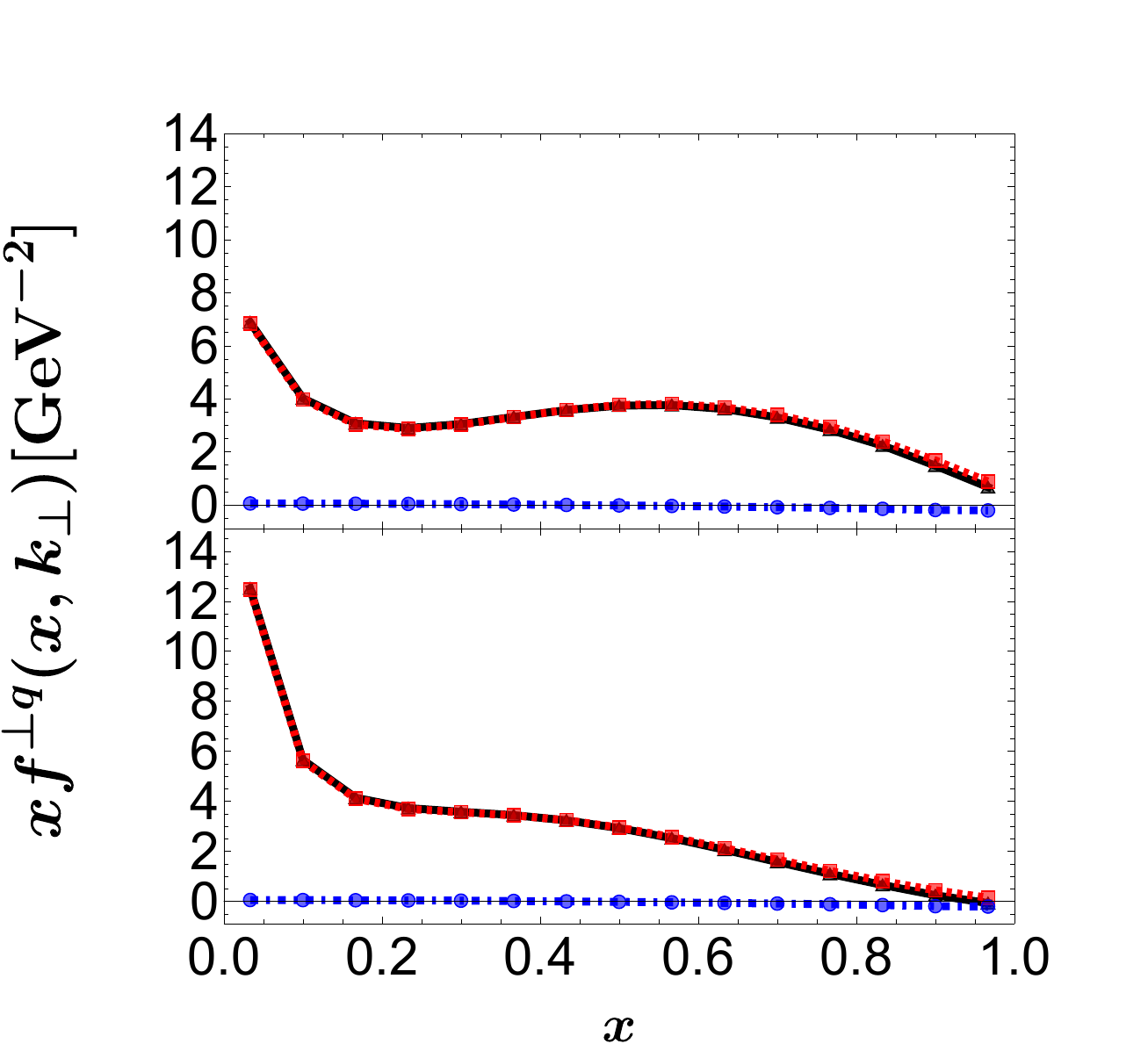}
    \includegraphics[width=0.32\textwidth]{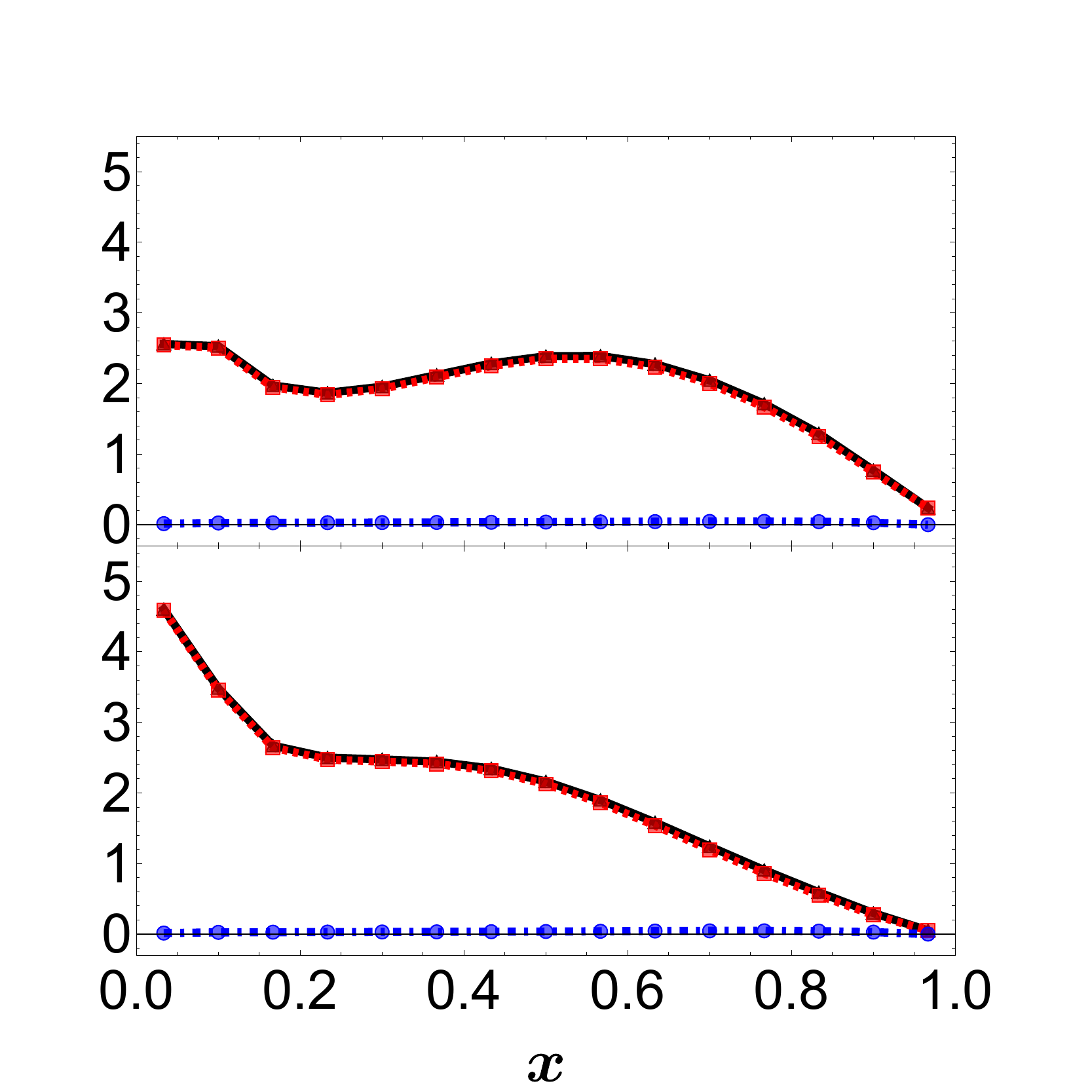}
    \includegraphics[width=0.32\textwidth]{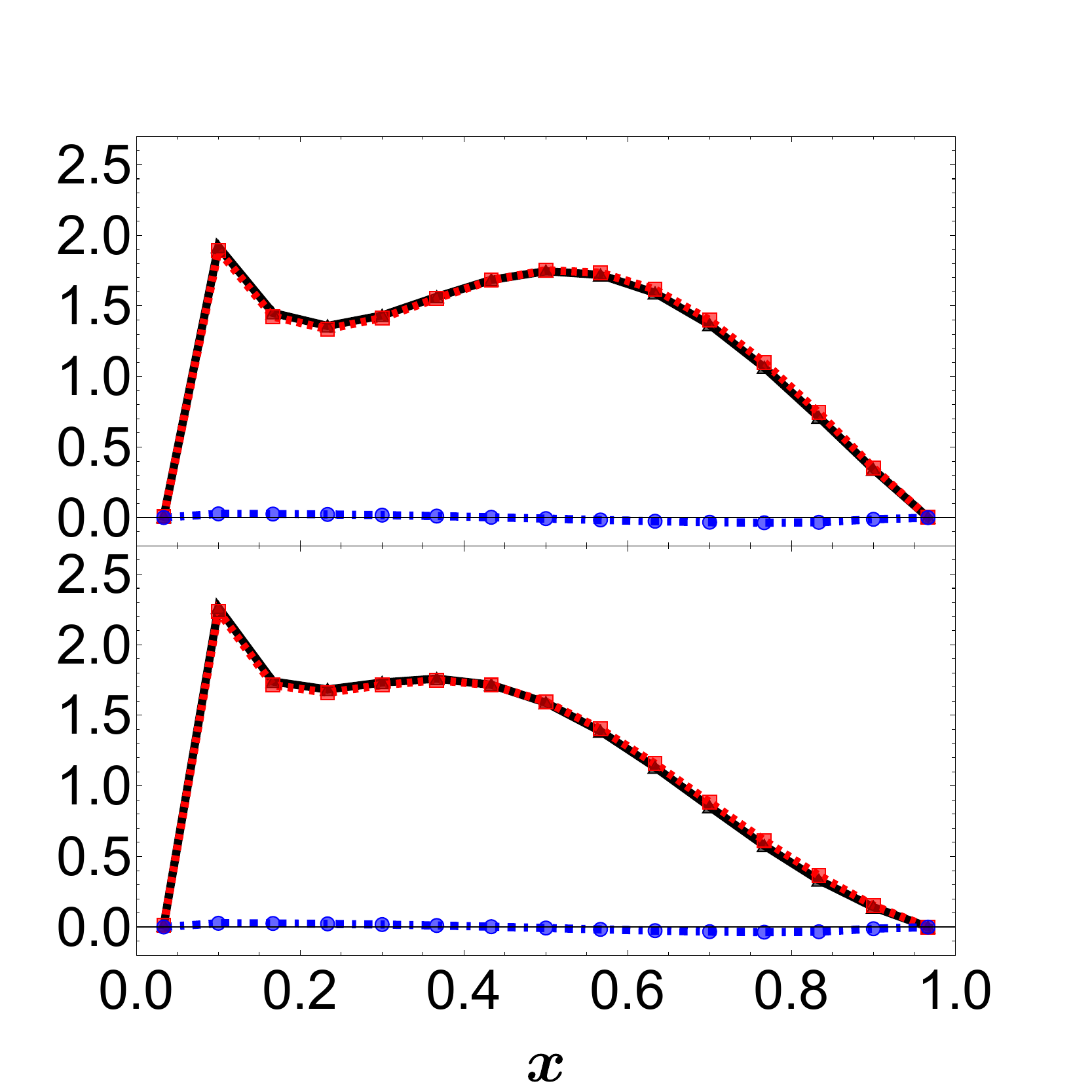}
\caption{The total twist-3 TMDs (black curves), the genuine twist-3 contributions (blue curves), and the twist-2 contributions (red curves) shown as functions of $x$ for three different values of $k_\perp$. The upper and lower panels correspond to the TMDs $e(x,k_\perp)$ and $f^{\perp}(x,k_\perp)$, respectively. As described in the text, the results are obtained in the BLFQ framework by averaging over truncation parameters $N_{\mathrm{max}}=\{12,14,16\}$ with $K=15$.}
    \label{tw-3TMDs_2D}
\end{figure*}

\begin{figure*}[htp]
    \centering
    \includegraphics[width=0.4\textwidth]{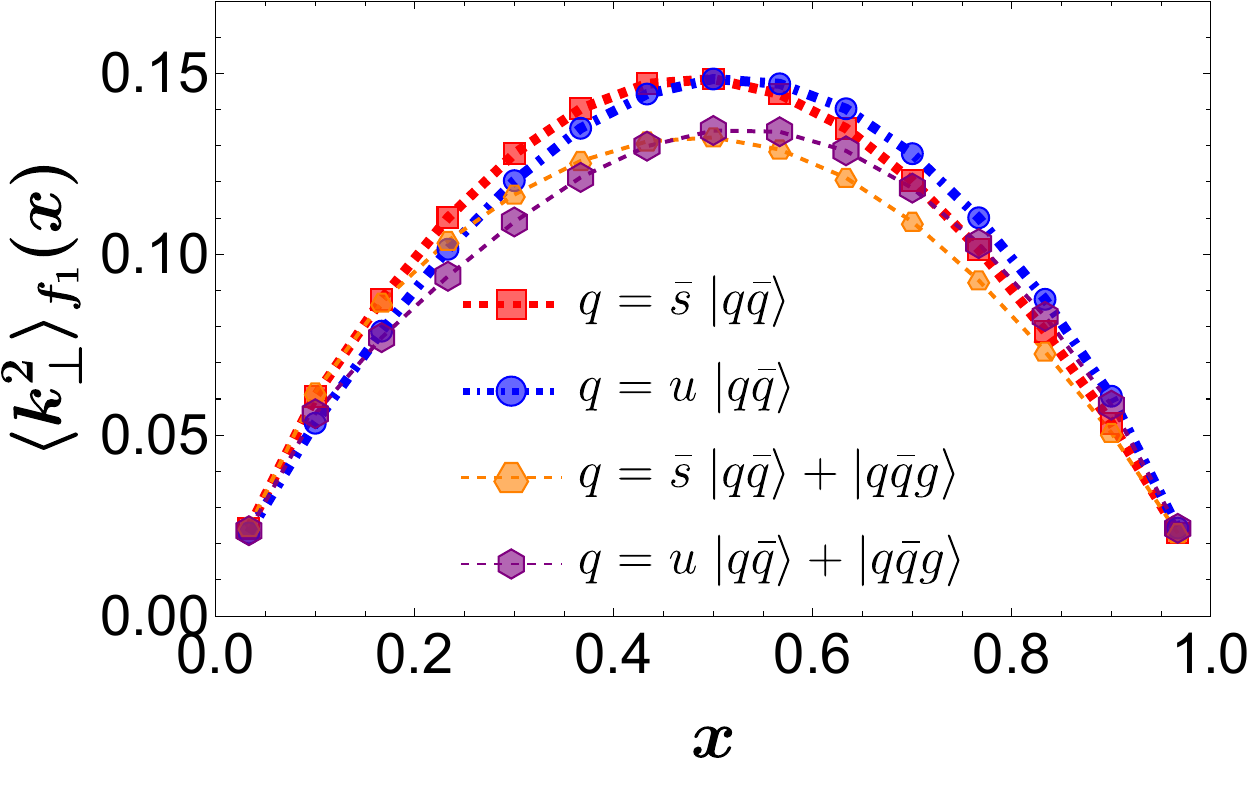}
    \includegraphics[width=0.4\textwidth]{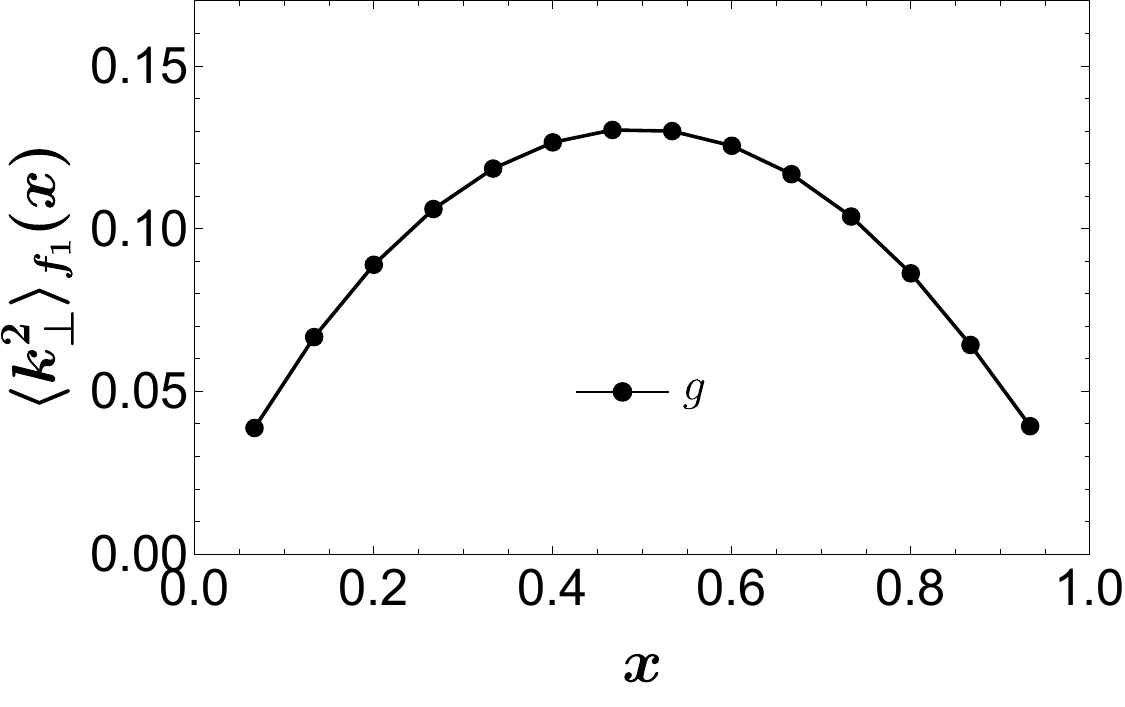}
\caption{The $x$-dependent mean squared transverse momentum of quarks (left panel) for the leading Fock sector $|q\bar{q}\rangle$ and for the combined Fock sectors $|q\bar{q}\rangle + |q\bar{q}g\rangle$, and of gluons (right panel). As described in the text, the results are obtained in the BLFQ framework by averaging over truncation parameters $N_{\mathrm{max}}=\{12,14,16\}$ with $K=15$.}
    \label{Fig:moments}
\end{figure*}

\begin{figure*}[htp]
    \centering
    \includegraphics[width=0.32\textwidth]{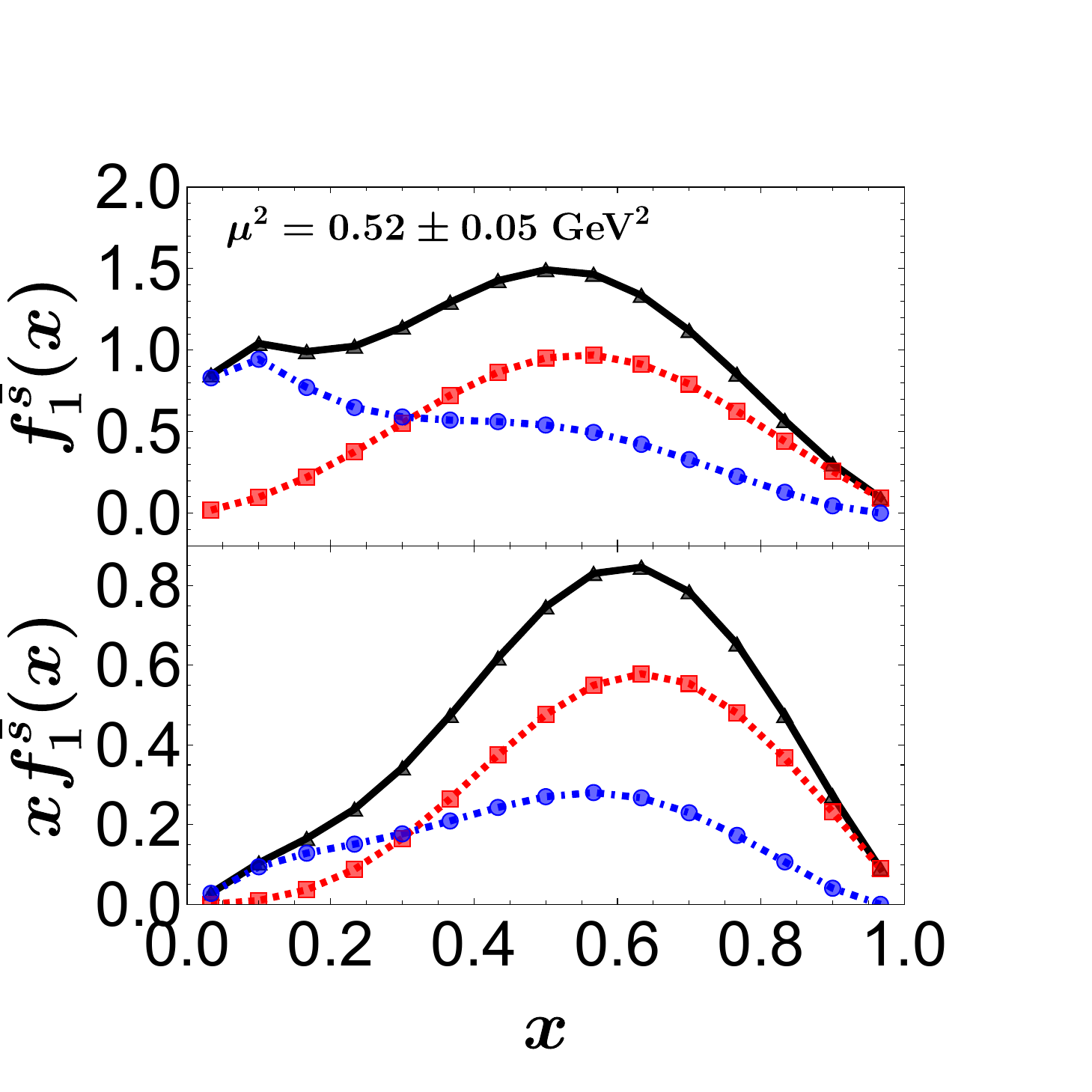}
    \includegraphics[width=0.32\textwidth]{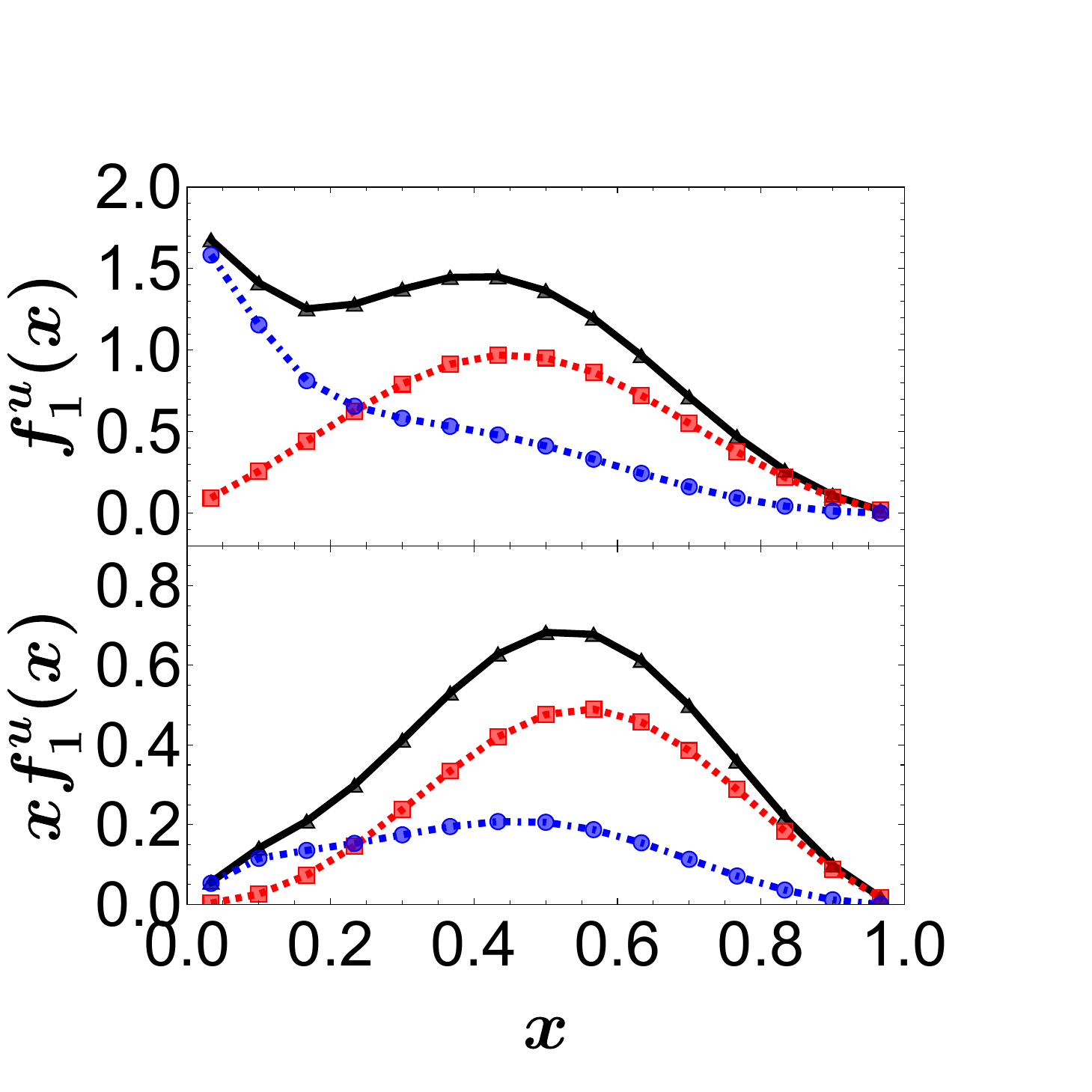}
    \includegraphics[width=0.32\textwidth]{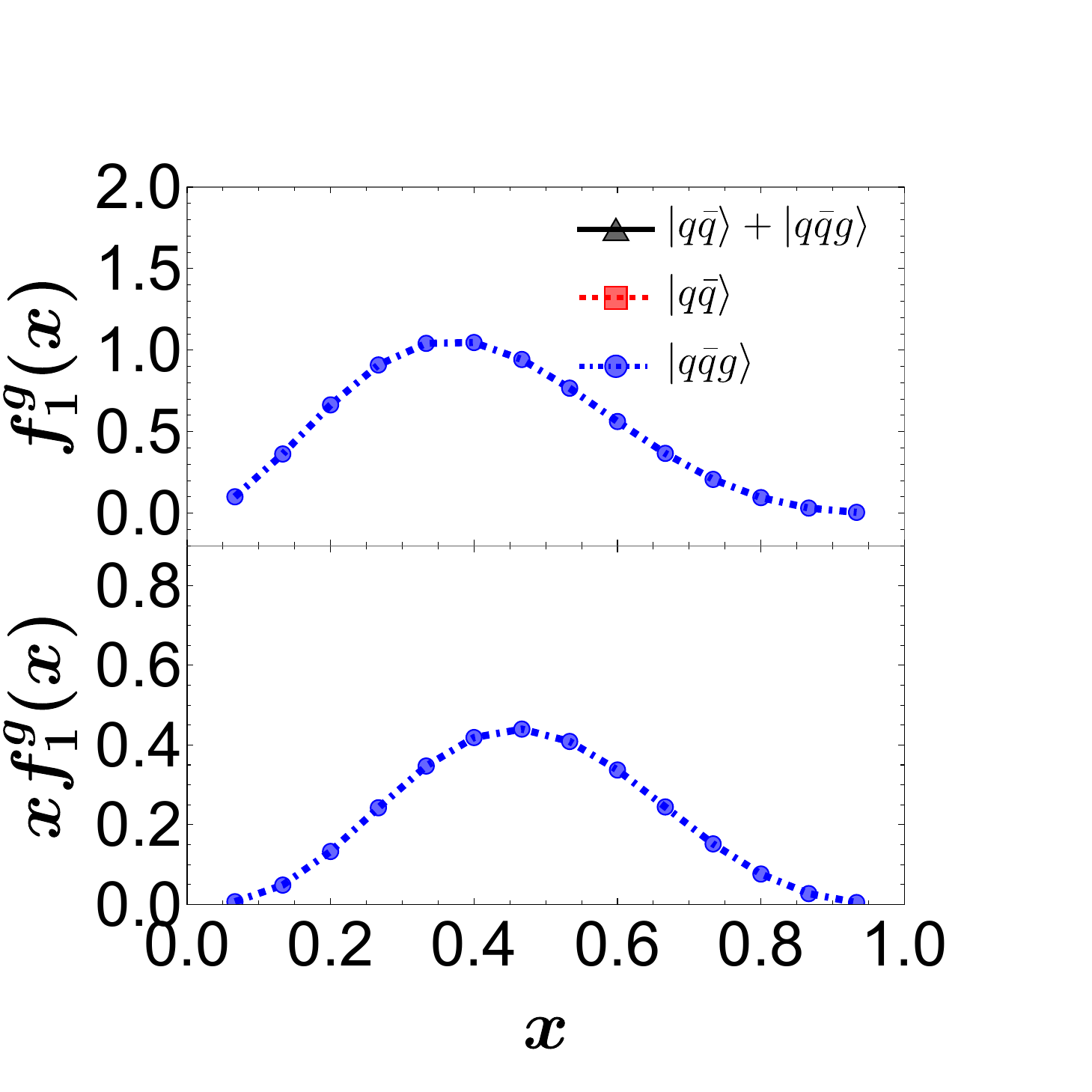}
\caption{Contributions from the $|q\bar{q}\rangle$ (red lines) and $|q\bar{q}g\rangle$ (blue lines) Fock sectors to the total (black lines) twist-2 PDFs $f_1(x)$ of the kaon. The left, middle, and right panels correspond to the $\bar{s}$ quark, $u$ quark, and gluon, respectively. As described in the text, all results are obtained by averaging over BLFQ calculations with $N_{\mathrm{max}}=\{12,14,16\}$ and $K=15$.}
    \label{Fig:t2_pdf}
\end{figure*}

\begin{figure*}[htp]
    \centering
    \includegraphics[width=0.4\textwidth]{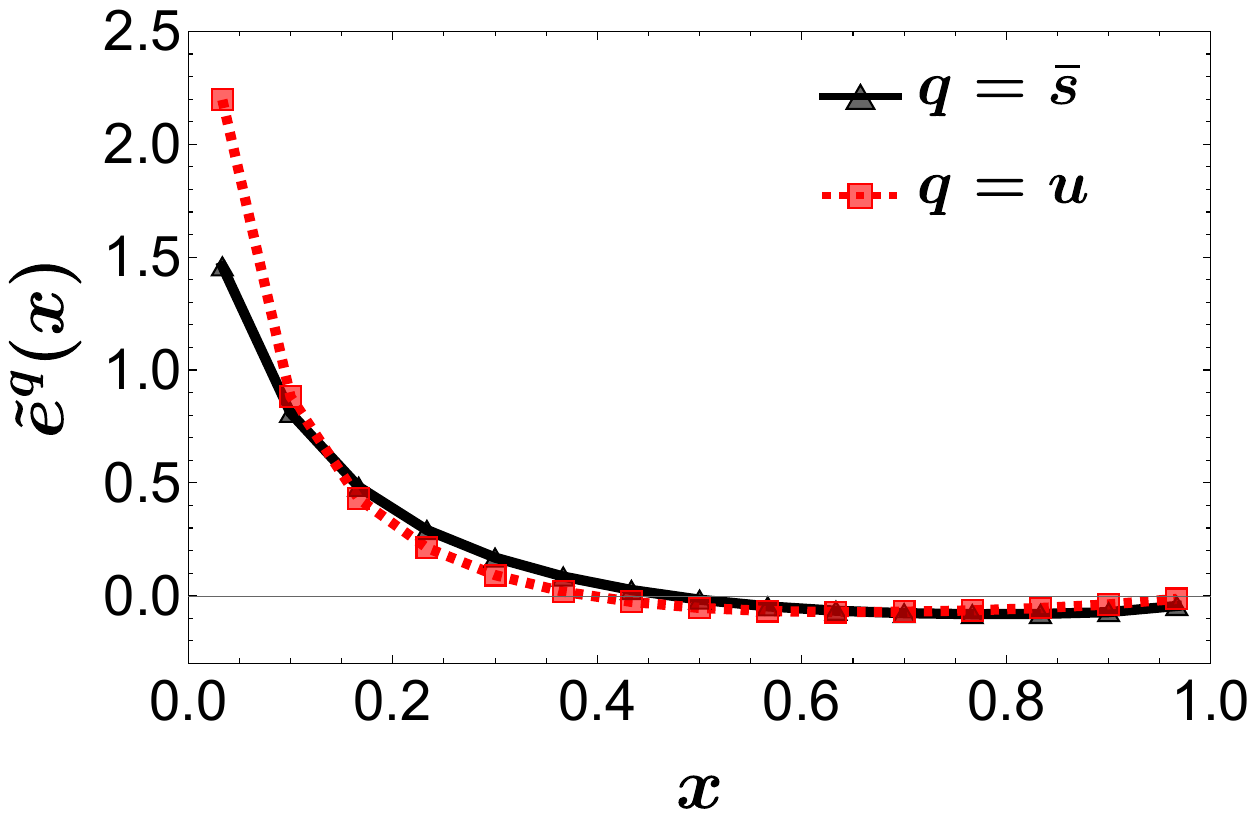}
    \includegraphics[width=0.4\textwidth]{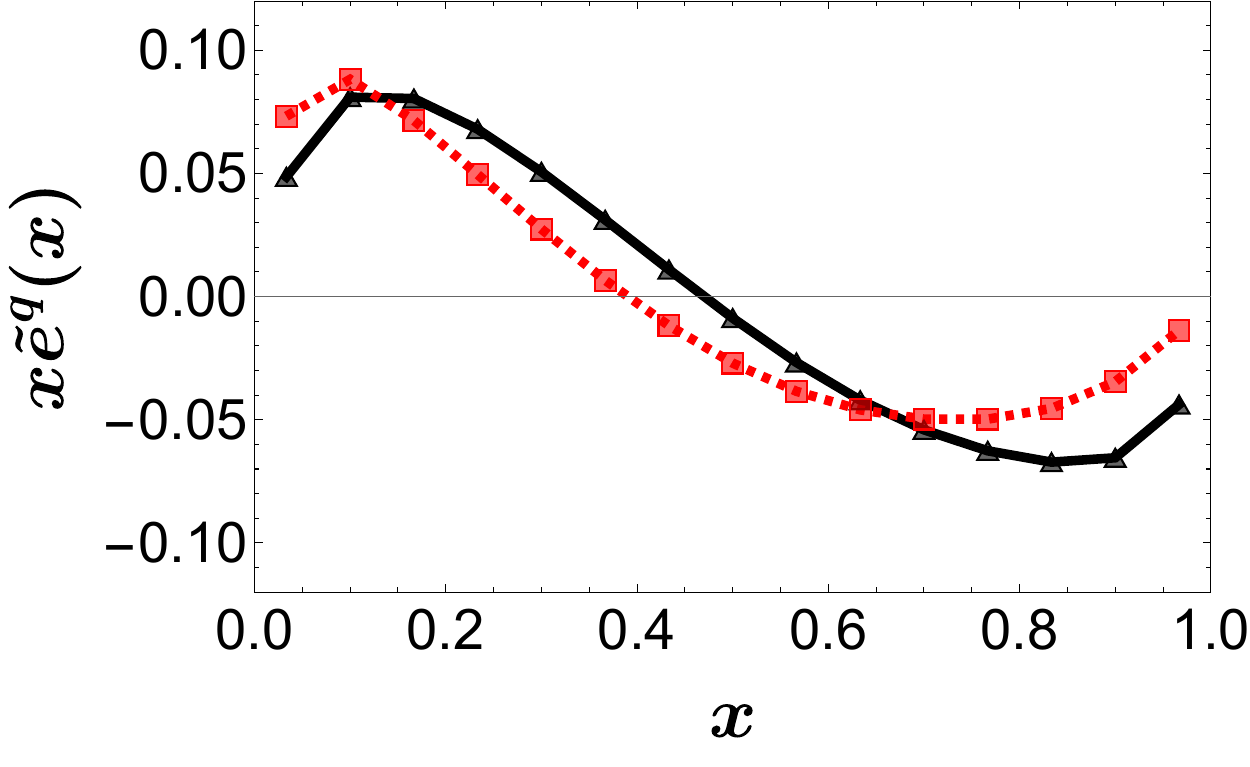}
    \includegraphics[width=0.4\textwidth]{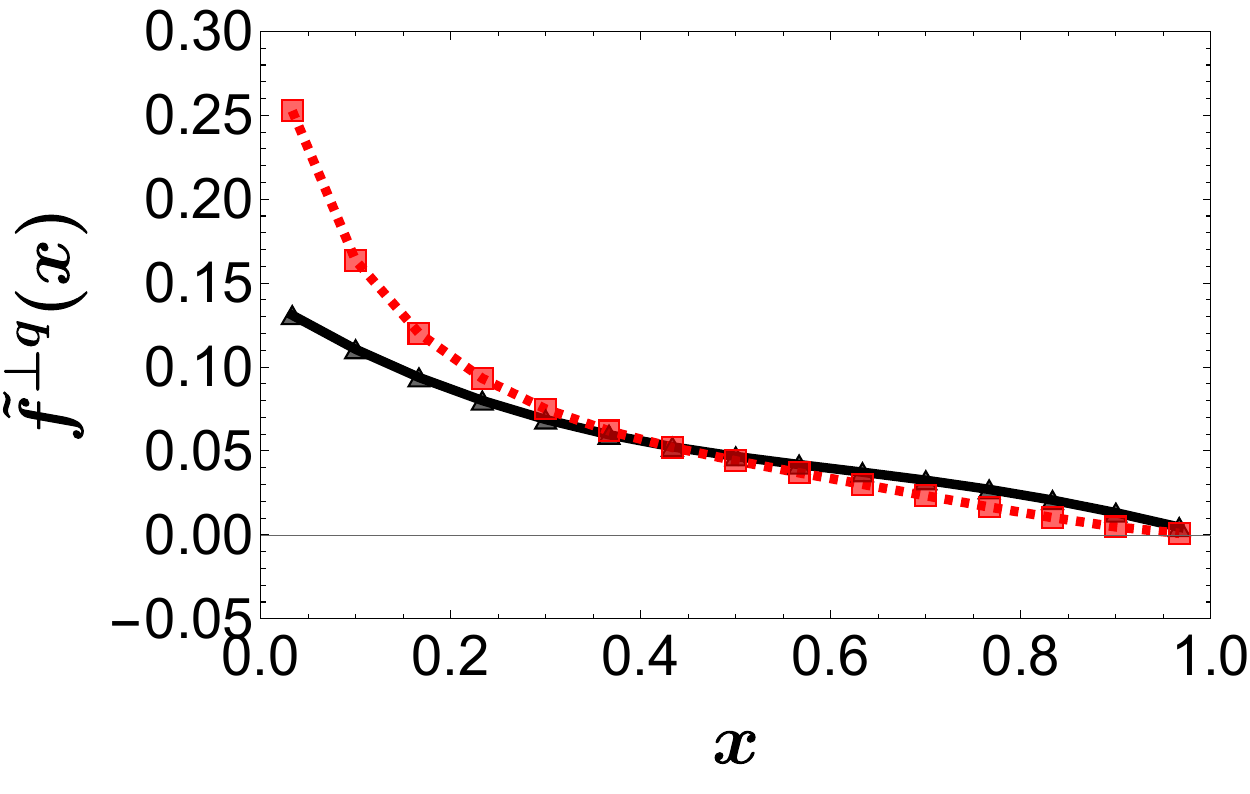}
    \includegraphics[width=0.4\textwidth]{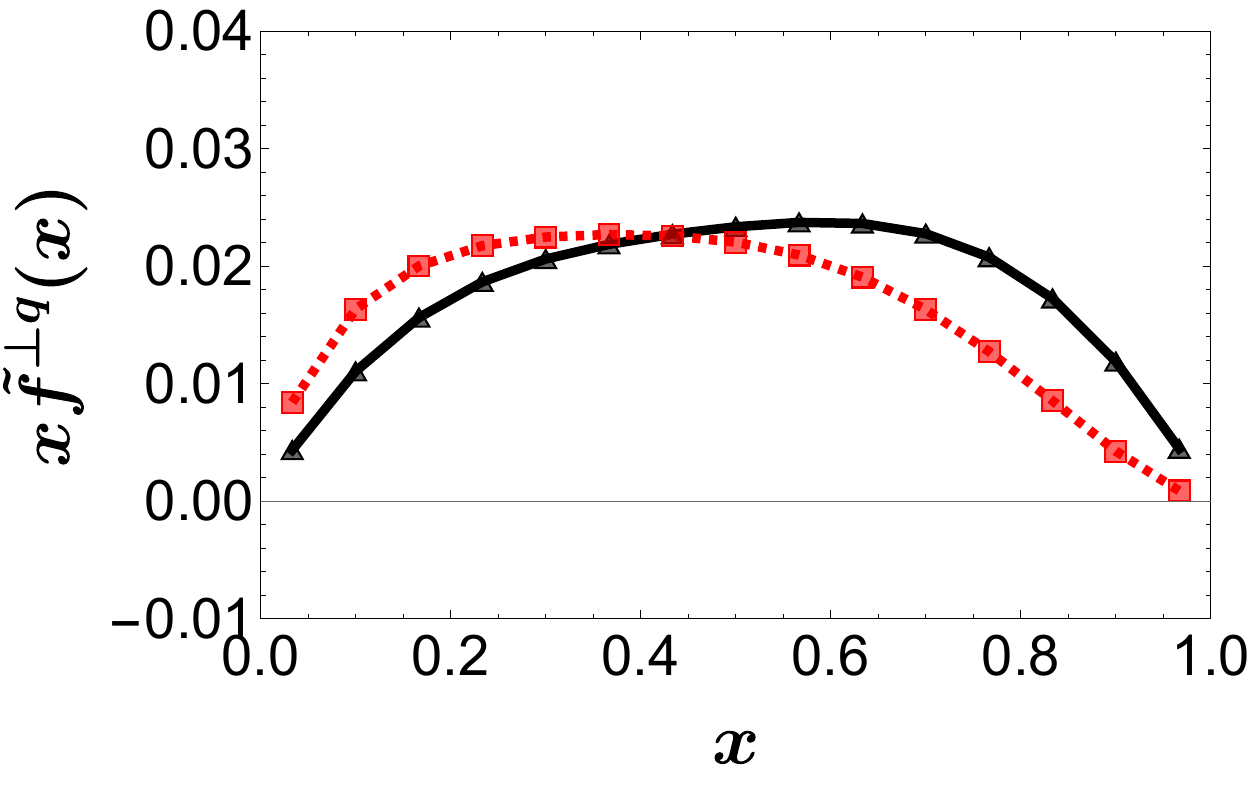}
\caption{The results of the genuine twist-3 PDFs $\tilde{e}$(x), $\tilde{f}^{\perp}(x)$ and $x\tilde{e}(x)$, $x\tilde{f}(x)$ with the different quarks $\bar{s}$ and $u$ of the kaon. As described in the text, all results are acquired by averaging through BLFQ approach with $N_{\mathrm{max}}=\{12,14,16\}$ and $K=15$.}
    \label{Fig:t4_pdf}
\end{figure*}

\begin{figure*}
    \centering
    \includegraphics[width=0.4\textwidth]{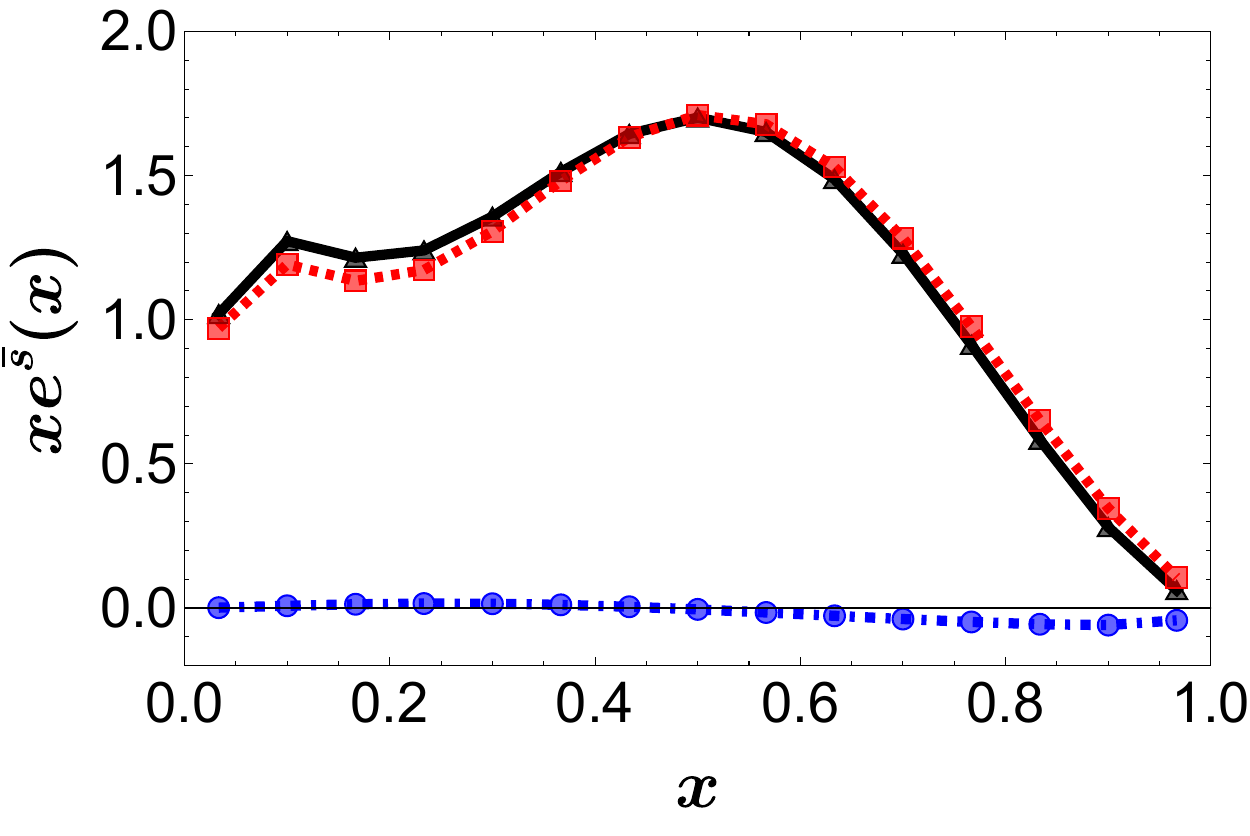}
    \includegraphics[width=0.4\textwidth]{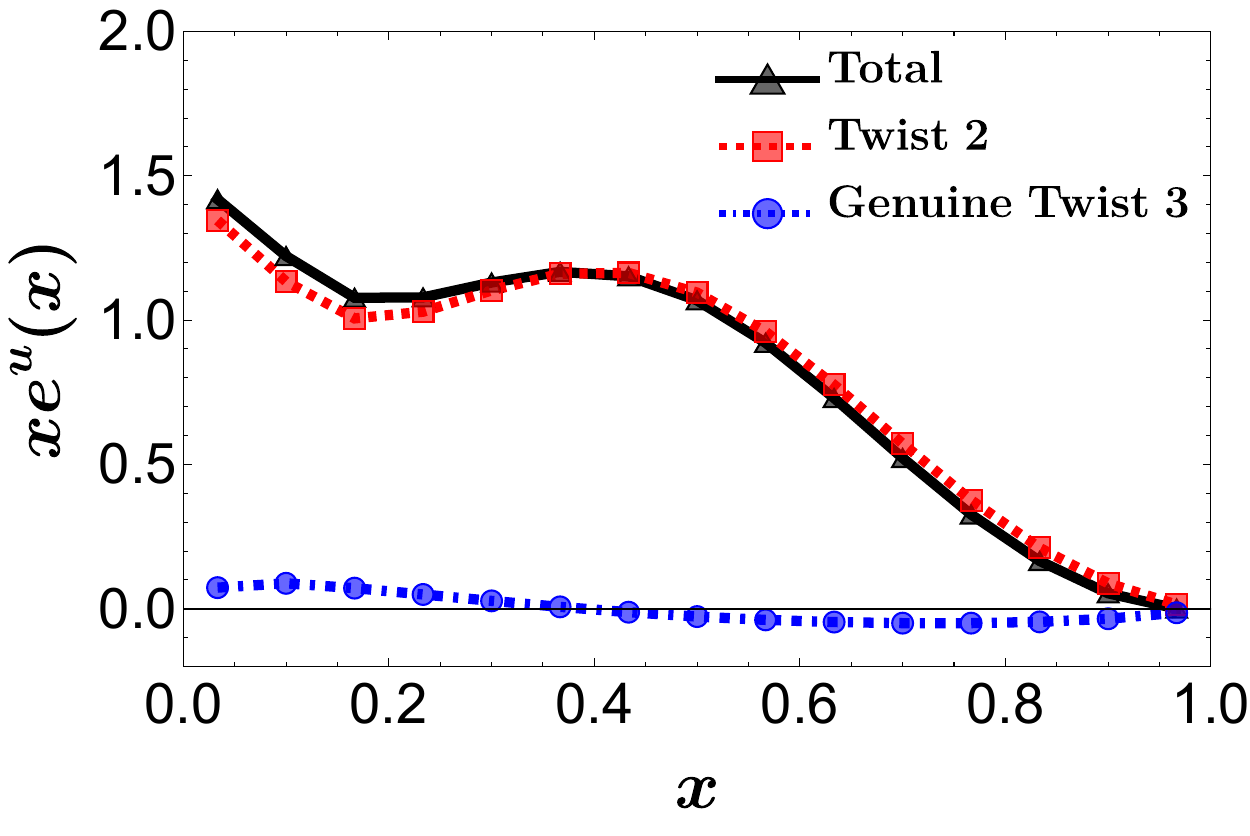}
    \includegraphics[width=0.4\textwidth]{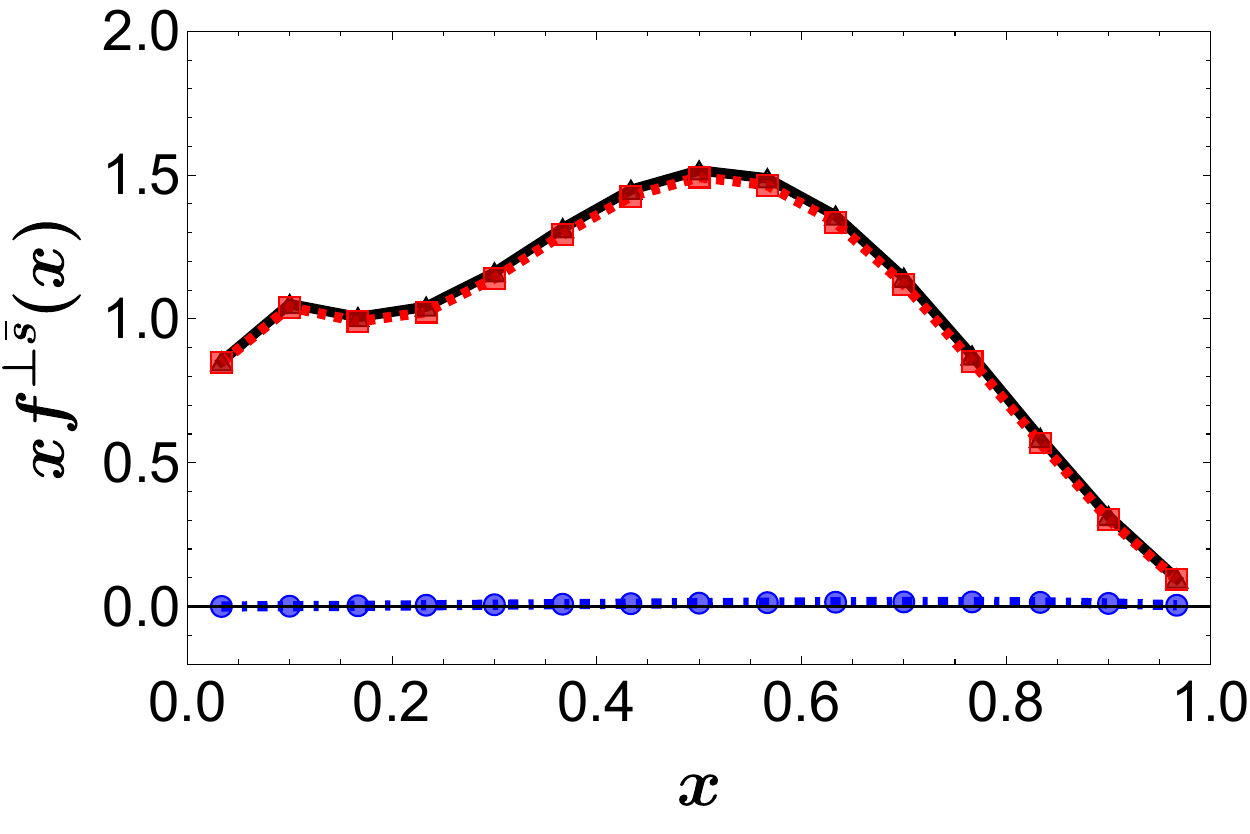}
    \includegraphics[width=0.4\textwidth]{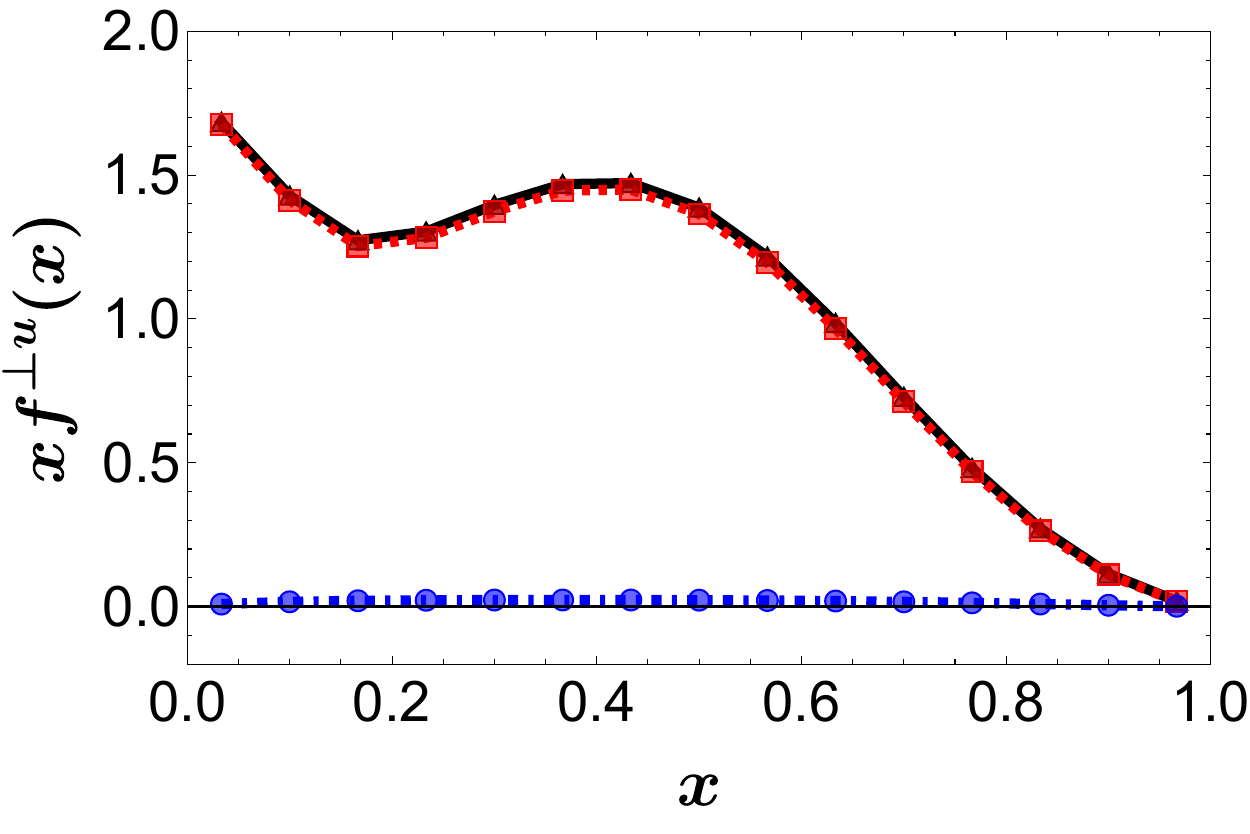}
    \caption{The total twist-3 PDFs (black curves), the genuine twist-3 contributions (blue curves), and the twist-2 contributions (red curves). The upper and lower panels correspond to the PDFs $e(x)$ and $f^{\perp}(x)$, respectively. The results are obtained in the BLFQ framework by averaging over truncation parameters $N_{\mathrm{max}}=\{12,14,16\}$ with $K=15$.}
    \label{Fig:t3_pdf}
\end{figure*}

\begin{figure*}[htp]
    \centering
    \includegraphics[width=0.4\textwidth]{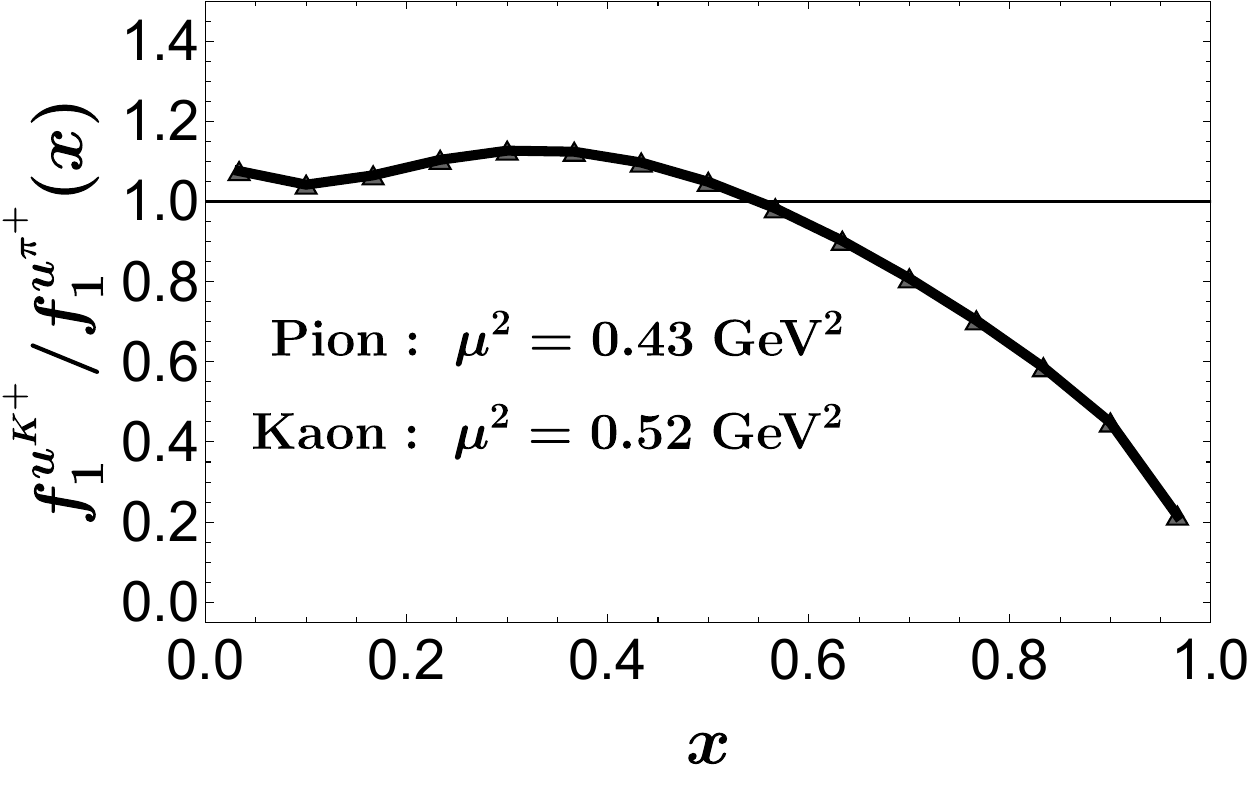}
    \includegraphics[width=0.4\textwidth]{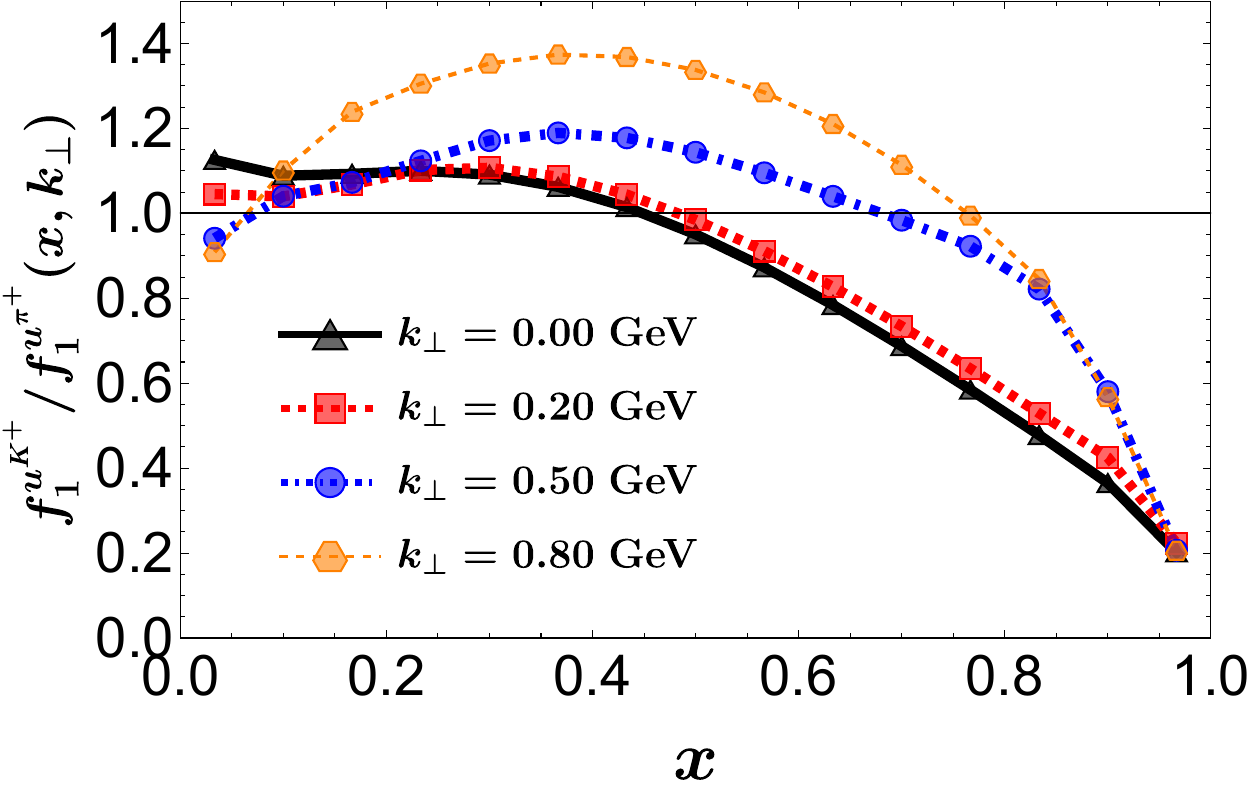}
    \includegraphics[width=0.4\textwidth]{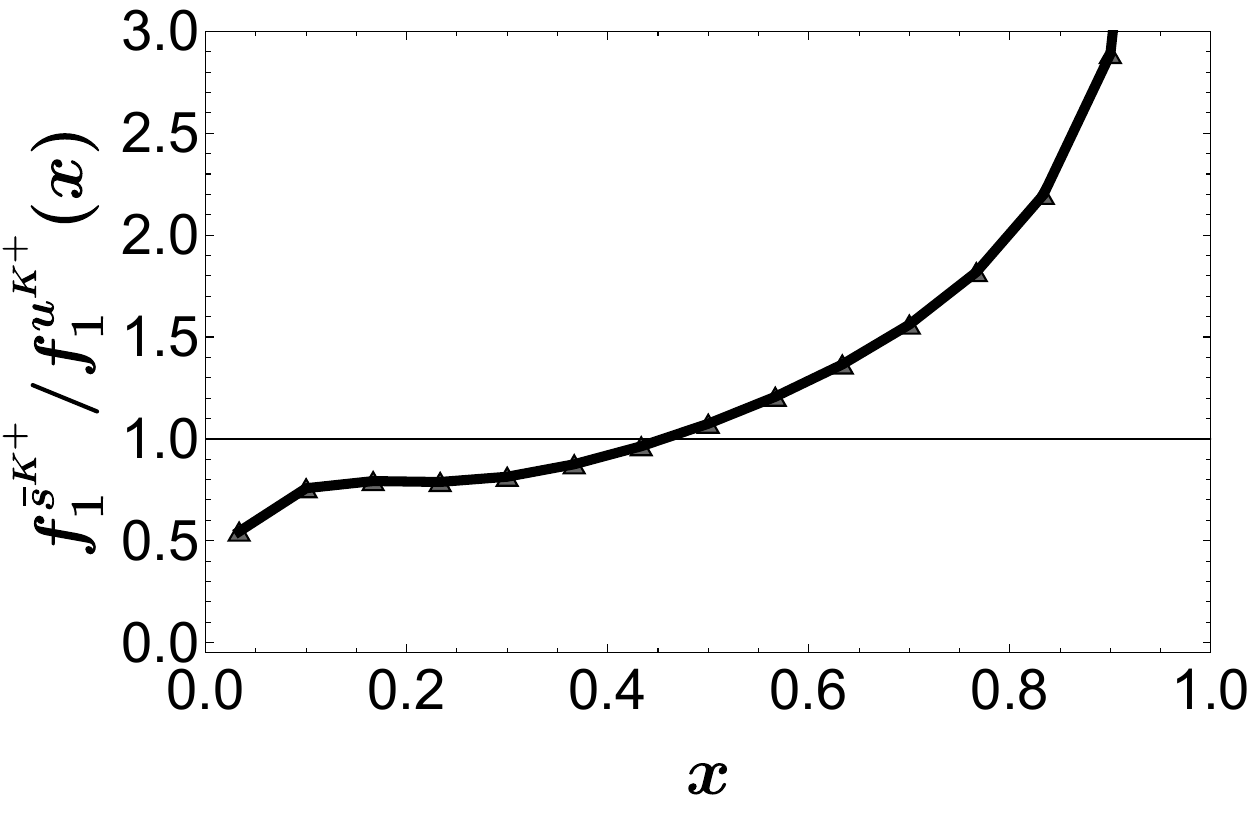}
    \includegraphics[width=0.4\textwidth]{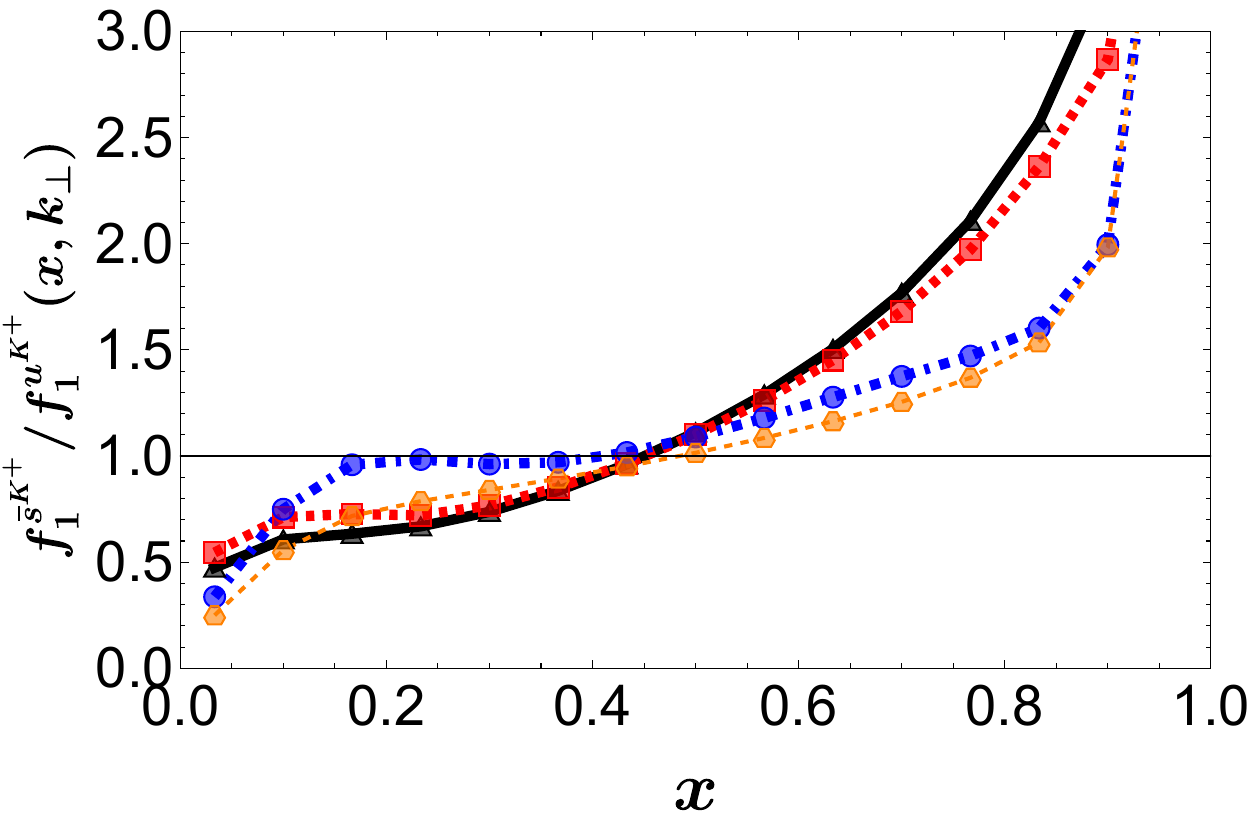}
    \includegraphics[width=0.4\textwidth]{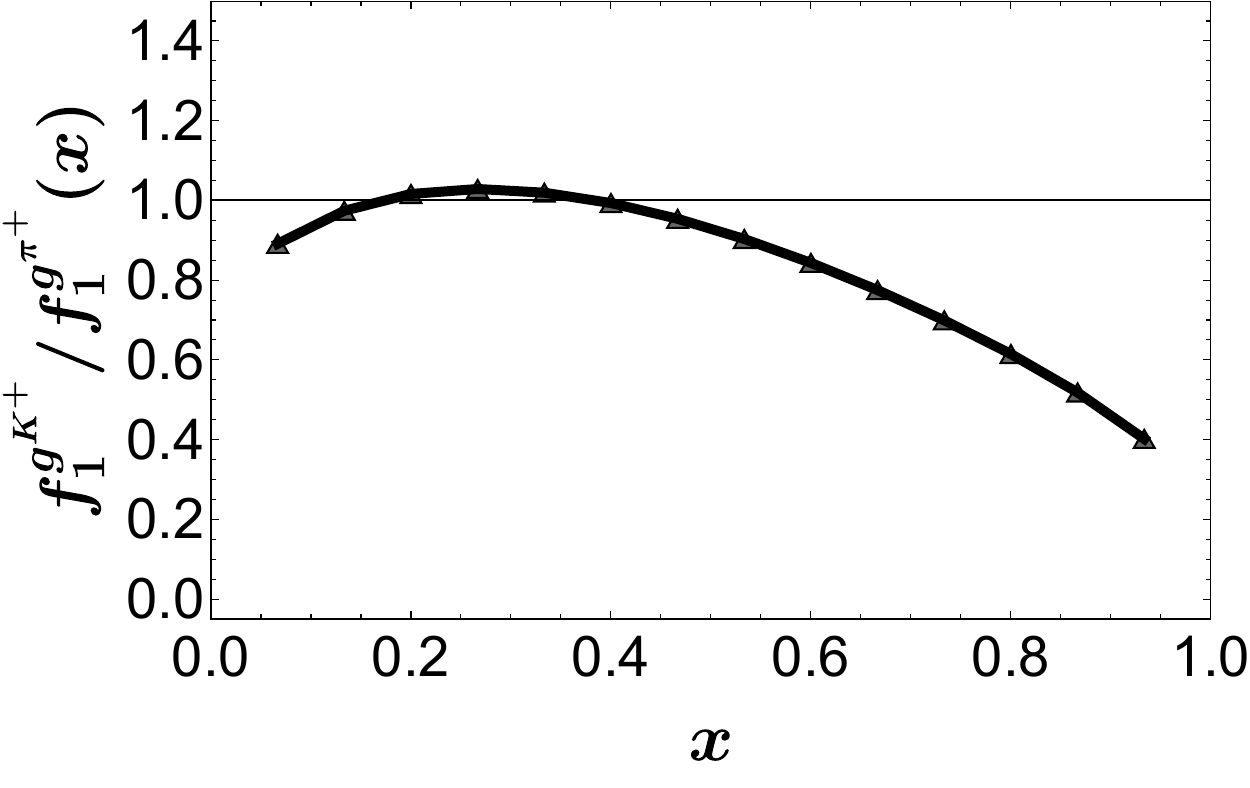}
    \includegraphics[width=0.4\textwidth]{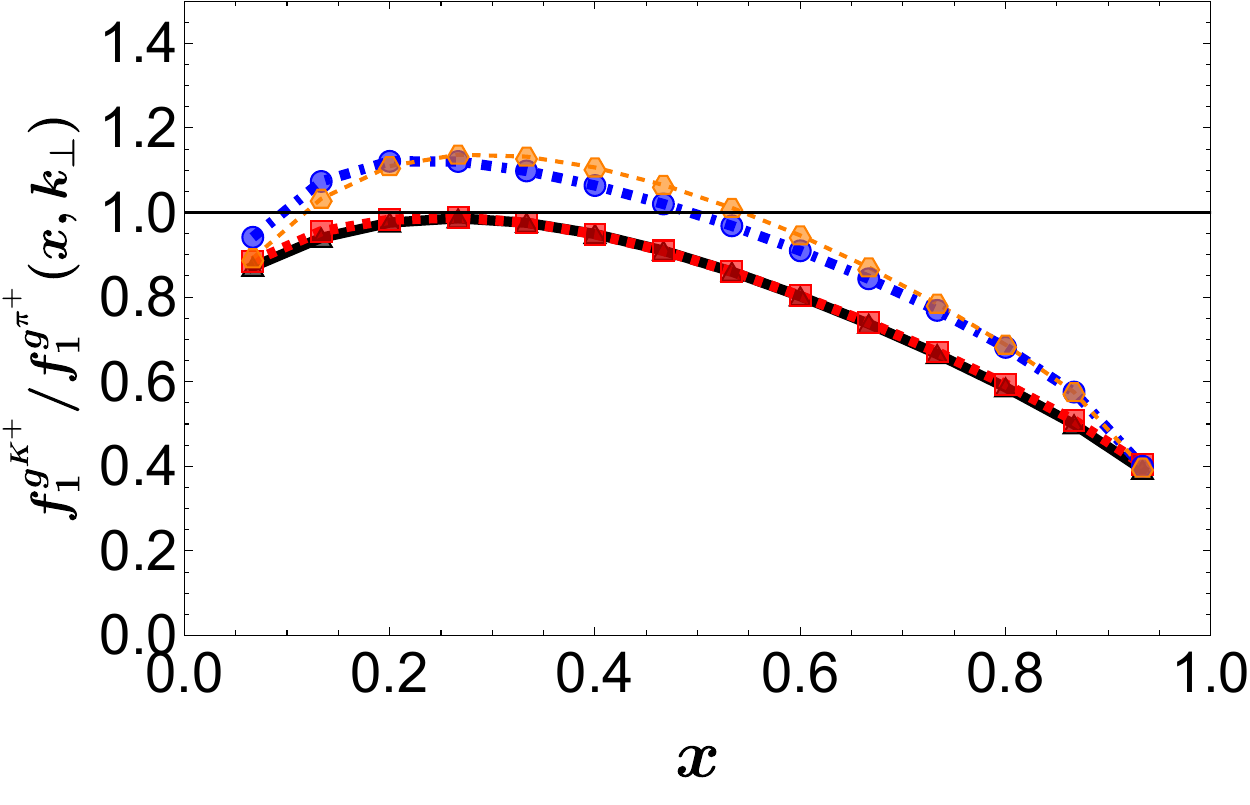}
\caption{Ratios of parton distributions. Upper panels: ratio of the up-quark distribution in the kaon to that in the pion. Middle panels: ratio of the strange-quark distribution to the up-quark distribution in the kaon. Lower panels: ratio of the gluon distribution in the kaon to that in the pion. The right and left panels show the TMDs and collinear PDFs, respectively. The TMD ratios are presented for different values of $k_\perp$.}
    \label{Fig:ratio}
\end{figure*}

\begin{figure*}[htp]
    \centering
    \includegraphics[width=0.4\textwidth]{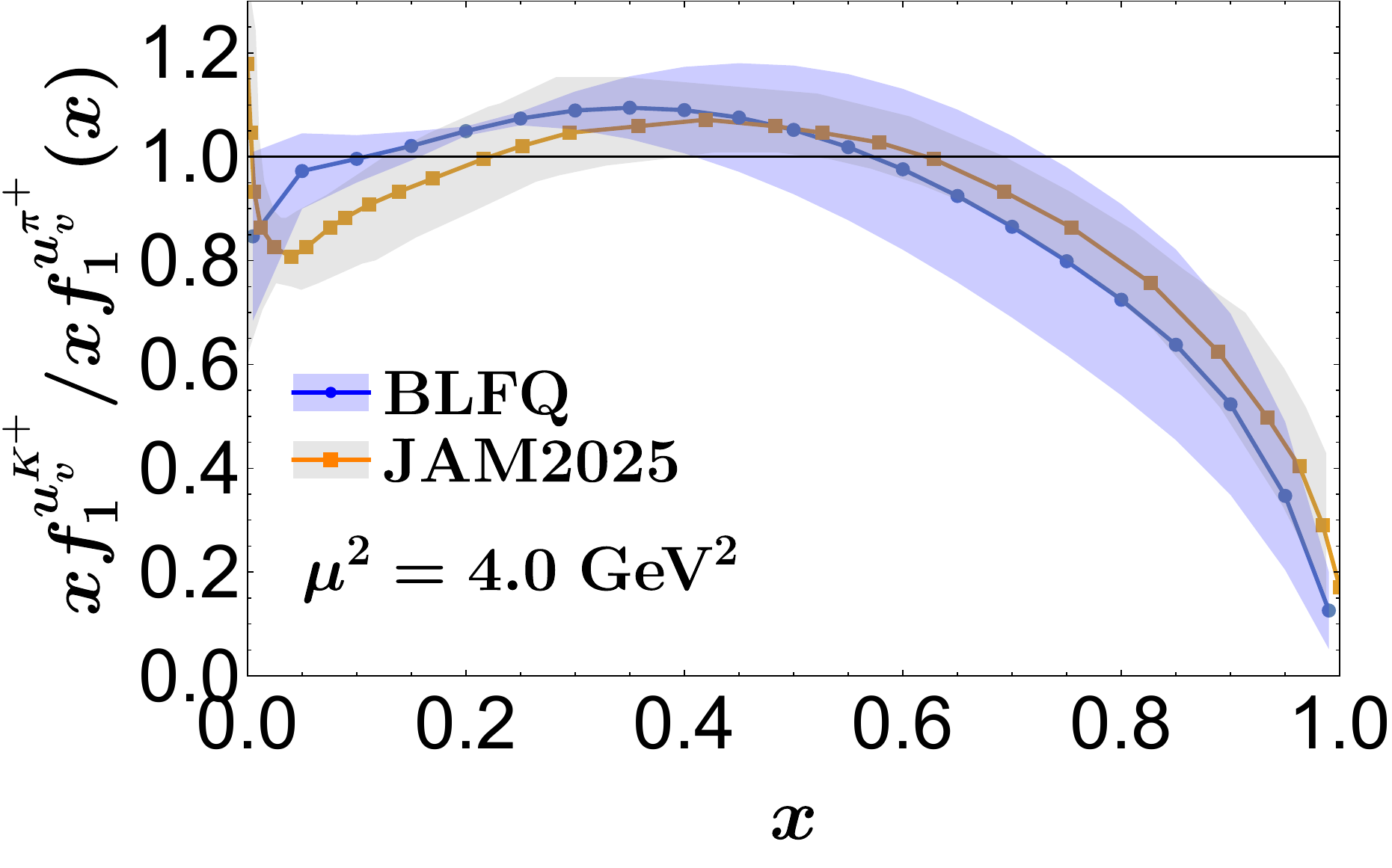}
    \includegraphics[width=0.4\textwidth]{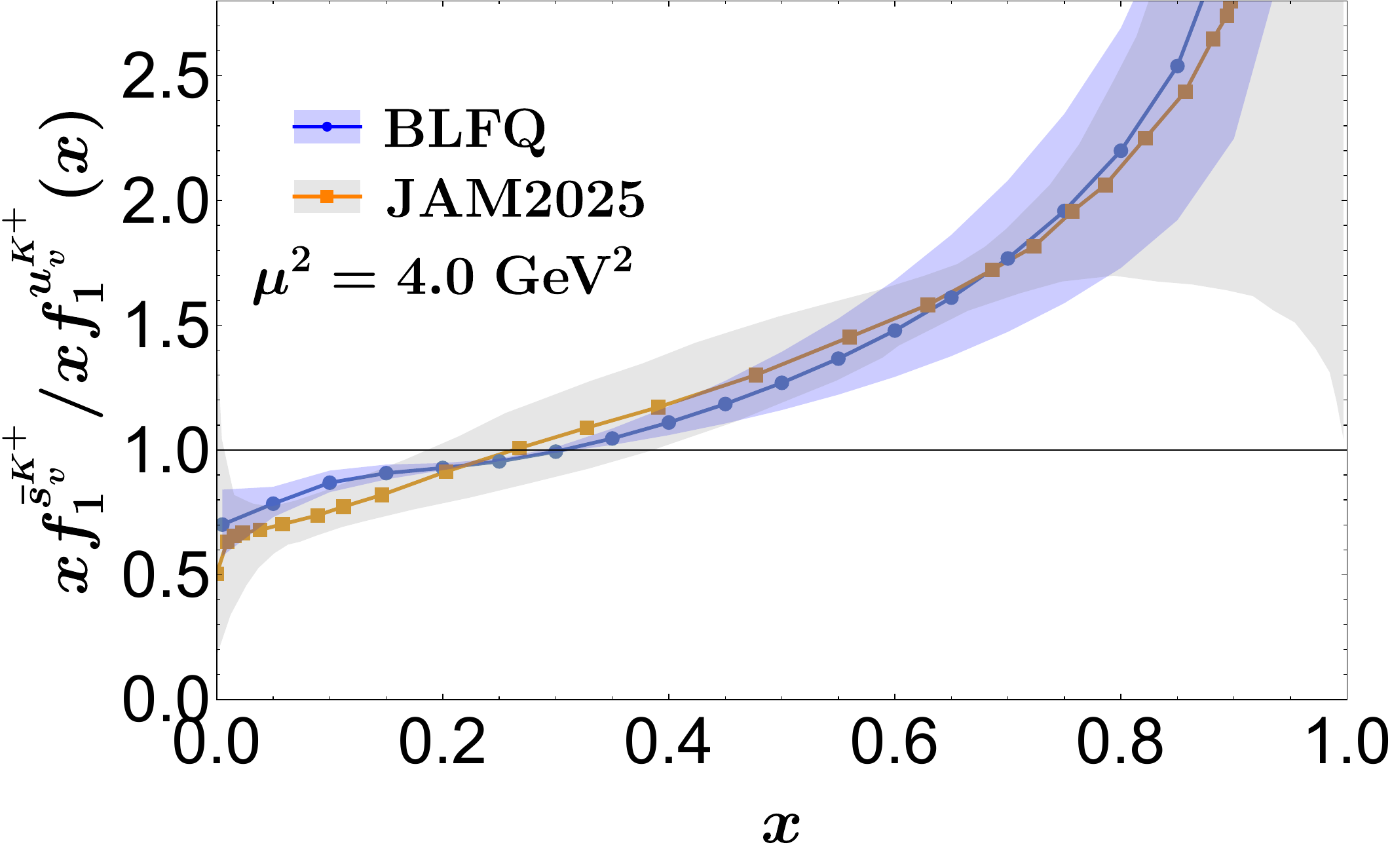}
    \includegraphics[width=0.4\textwidth]{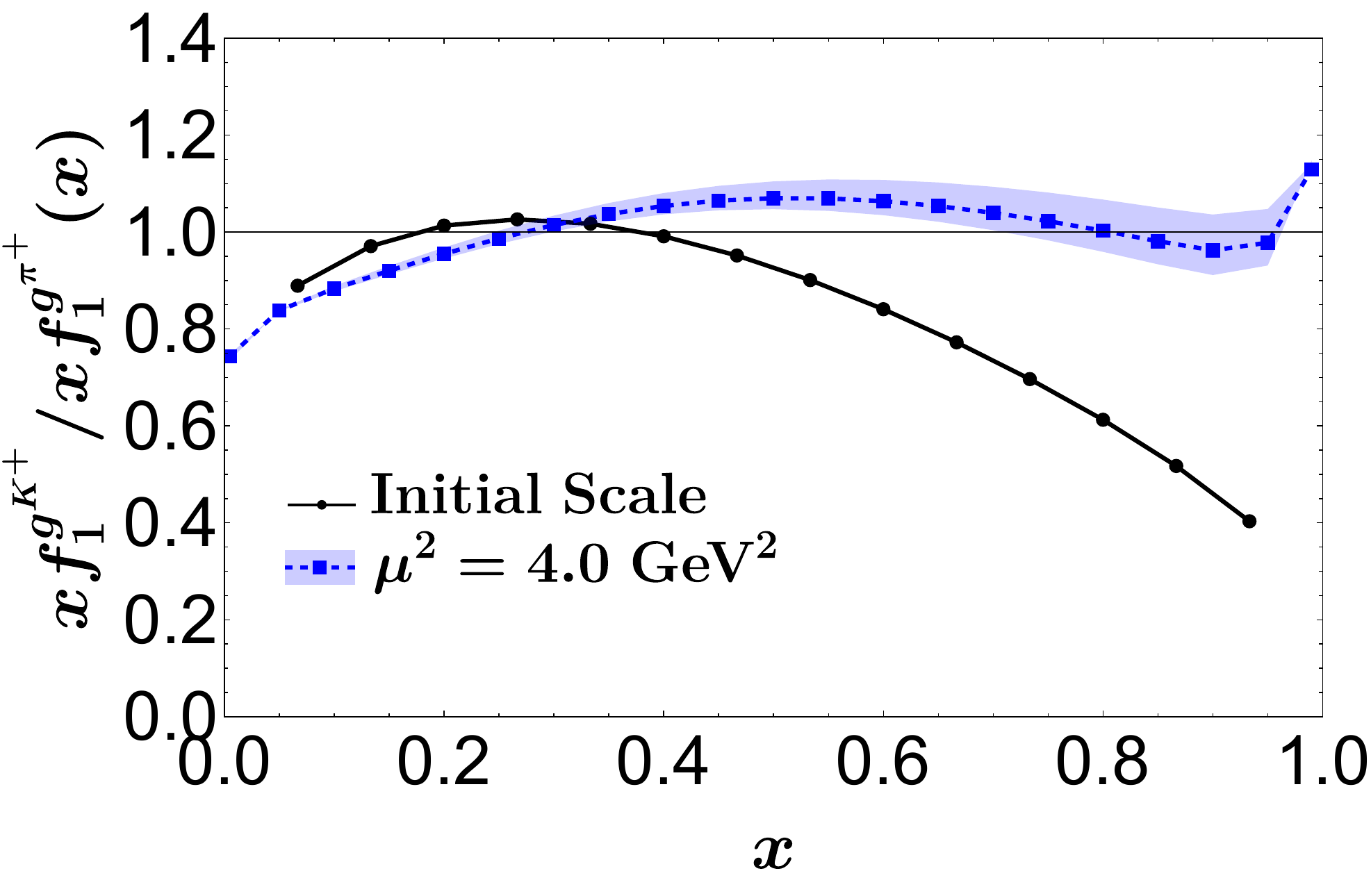}
    \includegraphics[width=0.4\textwidth]{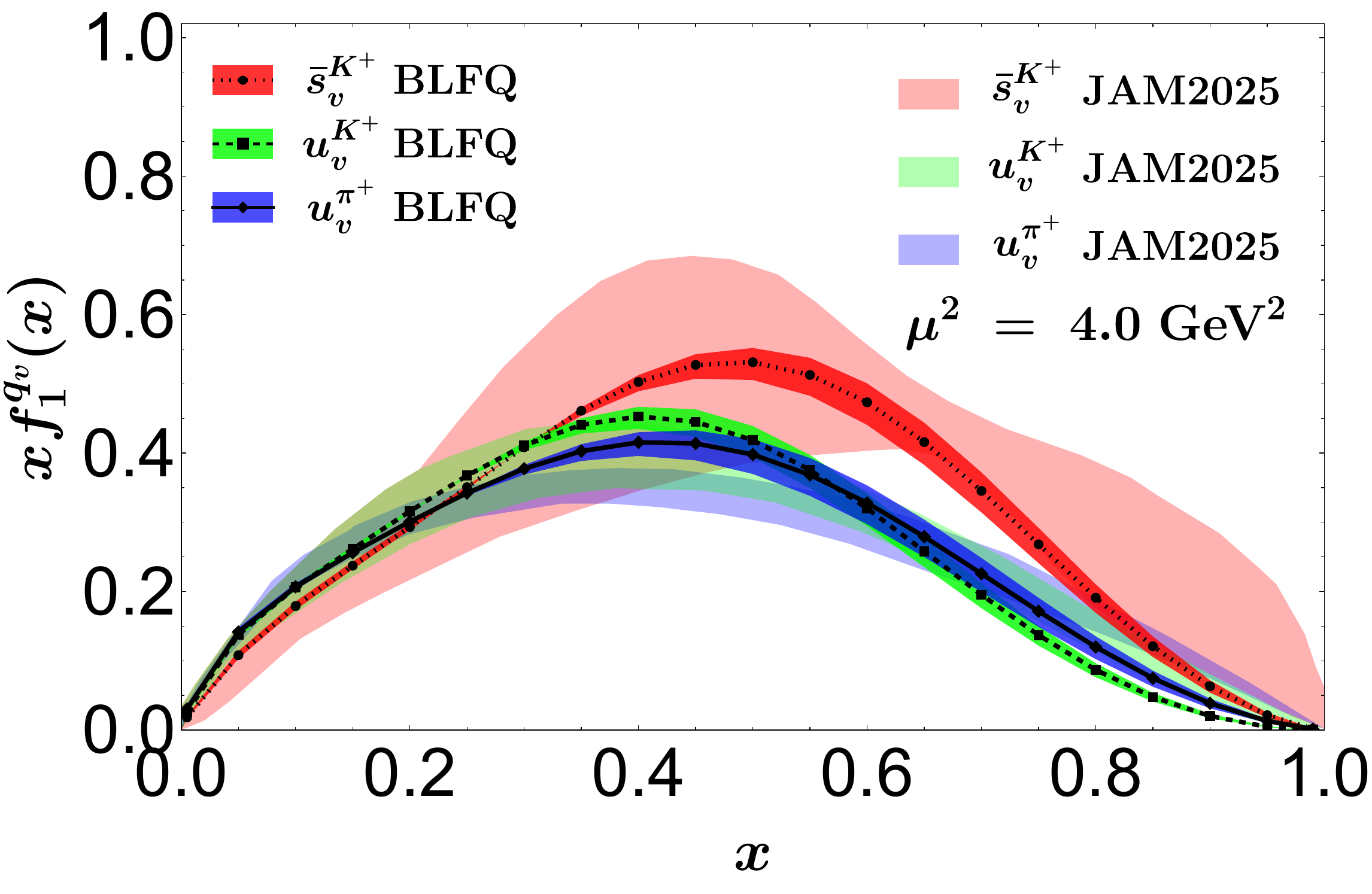}
\caption{Kaon PDFs after QCD evolution to $\mu^2 = 4~\mathrm{GeV}^2$. Upper-left panel: ratio of the up-quark distribution in the kaon to that in the pion. Upper-right panel: ratio of the strange-quark distribution to the up-quark distribution in the kaon. Lower-left panel: ratio of the gluon distribution in the kaon to that in the pion. Lower-right panel: valence-quark distributions in the kaon and pion. Our quark distributions and their ratios are compared with the recent global analysis by the JAM Collaboration~\cite{Barry:2025wjx}.}
    \label{Fig:ratioofevolution}
\end{figure*}

Figure~\ref{tw-3TMDs} shows the total twist-3 TMDs, $e(x,k_\perp)$ and $f^\perp(x,k_\perp)$, obtained using the EOM relations in Eq.~(\ref{16}).
Both distributions exhibit similar qualitative behavior, and their magnitudes are larger than that of the leading-twist TMD $f_1(x,k_\perp)$. 
Their shape is primarily driven by the function $f_1(x,k_\perp)/x$, which is singular at $x=0$. 
Nevertheless, the contributions of twist-3 TMDs to physical cross sections are suppressed by a factor of $M/P^+$~\cite{Bacchetta:2019qkv}, and thus higher-twist effects are reduced in high-energy processes. 
In addition, the twist-3 TMDs decrease rapidly with increasing transverse momentum. As noted earlier, twist-3 TMDs do not admit a probabilistic interpretation. 
These distributions are inherently multi-parton in nature, arising from the interference between different Fock sectors and encoding the dynamics of quark–quark–gluon correlations.

Figure~\ref{genuine:tw-3TMDs} presents the genuine twist-3 TMDs, $\tilde{e}(x,k_\perp)$ and $\tilde{f}^\perp(x,k_\perp)$. 
These distributions also display similar qualitative behavior, and their magnitudes are comparable to that of the leading-twist TMD $f_1(x,k_\perp)$. Both the valence quarks $\bar{s}$ and $u$ in the kaon show similar qualitative patterns, although the magnitude of the $u$-quark distribution is larger than that of the $\bar{s}$-quark. This is further illustrated in Fig.~\ref{genuine:tw-3TMDs_2d}, where the two-dimensional slices of the genuine twist-3 TMDs are presented.
An analysis of the operator structures in the twist-2 correlators indicates that twist-2 TMDs admit a probabilistic interpretation~\cite{Jaffe:1983hp,Barone:2001sp,Collins:2011zzd,Tangerman:1994eh,Zhu:2023nhl}. In contrast, genuine twist-3 TMDs do not have such an interpretation; rather, they correspond to unpolarized interference distributions~\cite{Mulders:1995dh}.

The total twist-3 TMDs, $e(x,k_\perp)$ and $f^\perp(x,k_\perp)$, which include contributions from both the twist-2 and genuine twist-3 terms, are shown in Fig.~\ref{tw-3TMDs_2D} as functions of $x$ for three different values of $k_\perp$. 
The same figure also displays the individual contributions from the twist-2 and genuine twist-3 components. 
We observe that the twist-2 contribution strongly dominates over the genuine twist-3 contribution. 
As a result, the overall behavior of the total twist-3 TMDs closely follows that of the twist-2 contribution, as evident from Fig.~\ref{tw-3TMDs_2D}.

\subsection{$k_{\perp}$ moments of the TMDs}
We define the $x$-dependent mean squared transverse momentum of a parton as~\cite{Hu:2022ctr,Zhu:2023lst}
\begin{equation}
    \langle k^2_{\perp}\rangle_{f_1}(x)=\frac{\int \mathrm{d}^2k_{\perp}k^2_{\perp}f_1(x,k_{\perp})}{\int \mathrm{d}^2k_{\perp}f_1(x,k_{\perp})}.
\end{equation}
In Fig.~\ref{Fig:moments}, we present the $x$-dependence of the mean squared transverse momentum, $\langle k^2_{\perp}\rangle_{f_1}(x)$, for the valence quarks ($u$ and $\bar{s}$) and the gluon in the kaon, computed within the Fock components $\vert q \bar{q}\rangle+\vert q \bar{q}g\rangle$. 
Within the BLFQ framework, $\langle k^2_{\perp}\rangle_{f_1}(x)$ exhibits a visible flavor and $x$ dependence. 
Due to flavor symmetry breaking arising from the mass difference between the $u$ and $\bar{s}$ quarks, the peak of $\langle k^2_{\perp}\rangle_{f_1}(x)$ is shifted away from $x=0.5$. 
The shift occurs towards higher $x$ for the $u$ quark and lower $x$ for the $\bar{s}$ quark.
Qualitatively, our results are consistent with experimental extractions that account for the $x$-dependence of $\langle k^2_{\perp} \rangle$, such as those reported in Refs.~\cite{Signori2013,Bacchetta2017}. The correlation between the gluon’s transverse and longitudinal momenta in the kaon is shown in the right panel of Fig.~\ref{Fig:moments}. 
The $\langle k^2_{\perp}\rangle_{f_1}(x)$ for the gluon is approximately symmetric about $x=0.5$, exhibiting a relatively larger value in the intermediate-$x$ region and decreasing rapidly at both small and large $x$. The results indicate that the ${k}_\perp$ moments of quarks and gluons are very similar across the entire range of the momentum fraction $x$.


Our calculations do not support the commonly used $x$–$k_\perp$ factorization ansatz often adopted in preliminary phenomenological studies~\cite{Anselmino2014,Bhattacharya2021,Anselmino2015,DAlesio2020,Cammarota2020,Lefky2015a}. 
In particular, the nontrivial $x$ dependence of $\langle k^2_{\perp}\rangle_{f_1}$ prevents the factorization of the TMD in the form
\begin{gather}
    f^q_1(x,k_\perp) = f^q_1(x)\frac{\hat{f}^q_1(k_\perp)}{\mathcal{N}} \, ,
\end{gather}
where $\mathcal{N}=\int \mathrm{d}^2 k_\perp\, \hat{f}^q_1(k_\perp)$.


\begin{table}[h]
  \caption{
 The first moment $\langle {k}_\perp \rangle$ (in GeV) and the square root of second moment $\langle k^2_\perp \rangle^{1/2}$ (in GeV) of the quark (twist-2 and twist-3) and gluon (twist-2) TMDs in the kaon are obtained by averaging over the basis truncation parameters $N_{\text{max}}=\{12,14,16\}$ truncation and $K=15$ as described in the text. 
  }
    \vspace{0.15cm}
  \centering
  \begin{tabular}{c|cc|cc|cc}
  \hline\hline 
       TMDs   & $\langle {k}_\perp\rangle_u$ & $\langle {k}_\perp\rangle_{\bar{s}}$ & $\langle k^2_\perp\rangle_u^{1/2}$ & $\langle k^2_\perp\rangle_{\bar{s}}^{1/2}$ & $\langle {k}_\perp\rangle_g$ &$\langle k^2_\perp\rangle_g^{1/2}$ \\        \hline 
     $f_1$ & 0.28 & 0.29 & 0.31 &  0.32 & 0.30 & 0.34 \\
     $e$ & 0.20 & 0.23 & 0.25 & 0.27 & - & - \\
     $f^\perp$ & 0.20 & 0.23 & 0.24 & 0.27 & - & - \\
   \hline \hline
  \end{tabular}
  \label{Table:moments}
\end{table}

Although twist-3 TMDs $e$ and $f^\perp$ do not admit a probabilistic interpretation, we compute their first and second transverse-momentum moments, defined as~\cite{Lorce:2014hxa}
\begin{equation}\label{moments} \langle k_\perp^n\rangle_{\rm TMD}=\frac{\int\mathrm{d}x\int\mathrm{d}^2{k}_\perp k_\perp^n {\rm TMD}}{\int\mathrm{d}x\int\mathrm{d}^2{k}_\perp \mathrm{TMD}}. \end{equation}
The corresponding results are summarized in Table~\ref{Table:moments}.


\subsection{Twist-2 and twist-3 PDFs}
In this subsection, we focus on the integrated TMDs over the transverse momentum $k_{\perp}$, namely the PDFs.  
The leading-twist unpolarized PDFs $f_1(x)$ for the valence quarks, including contributions from both the leading and next-to-leading Fock sectors, are presented in the left ($\bar{s}$ quark) and middle ($u$ quark) panels of Fig.~\ref{Fig:t2_pdf}.  
The separate contributions from the $|q\bar{q}\rangle$ and $|q\bar{q}g\rangle$ Fock components are also shown in the same plots.  
As seen in Fig.~\ref{Fig:t2_pdf}, the inclusion of a dynamical gluon leads to a noticeable enhancement of the quark PDFs in the small-$x$ region, reflecting the natural onset of gluon effects through higher Fock components in our framework.  
At small $x$, the $u$ distribution is slightly larger than that of the $\bar{s}$ quark, whereas in the large-$x$ region it becomes smaller.  
Within the valence Fock sector $|q\bar{q}\rangle$, the PDFs exhibit features consistent with previous model calculations and phenomenological analyses.

The gluon PDF, shown in the right panel of Fig.~\ref{Fig:t2_pdf}, is narrower in $x$ compared to the quark PDFs and therefore exhibits a more pronounced peak.  
This narrower shape arises from the larger effective gluon mass in our model, which suppresses contributions at small and large $x$ and concentrates the distribution around intermediate $x$.
 

As shown in Fig.~\ref{Fig:t4_pdf}, the genuine twist-3 function $x\tilde{e}(x)$ is positive for $x < 0.5$ and negative for $x > 0.5$, while $x\tilde{f}^{\perp}(x)$ remains positive over the entire $x$ range.
Figure~\ref{Fig:t3_pdf} shows the total twist-3 PDFs along with their decomposition into the twist-2 contribution and the genuine twist-3 parts. 
We note that the magnitude of the twist-2 contribution is significantly larger than that of the genuine twist-3 components.

We also find that the integrated value of $x\tilde{e}(x)$ over $x$ vanishes, demonstrating that our BLFQ results correctly satisfy the sum rule defined in Eq.~\eqref{23}. 
This sum rule originates from the EOM relations and is therefore expected to hold universally. 
Our findings are consistent with previous BLFQ studies of $\tilde{e}(x)$ for the pion~\cite{Zhu:2023lst} and proton~\cite{Zhu:2024awq}, both of which satisfy the same sum rule. 
For comparison, the calculation of $\tilde{e}(x)$ for the nucleon in Ref.~\cite{Pasquini:2018oyz}, which includes $|qqq\rangle$ and $|qqqg\rangle$ Fock components, does not fulfill the sum rule in Eq.~\eqref{23}. This may arise from neglecting zero modes and truncating the Fock sectors~\cite{Pasquini:2018oyz}.
In contrast, Ref.~\cite{Mukherjee:2009uy} reports a dressed-quark calculation of $e(x)$ and $\tilde{e}(x)$ at one loop, where $\tilde{e}(x)$ does satisfy the first-moment sum rule~(\ref{23}).

Finally, using the sum rule defined in Eq.~(\ref{scalarFF}), we compute the kaon scalar form factor at zero momentum transfer from our BLFQ results, obtaining $\sigma_K = 7.5~\text{GeV}$.

Figure~\ref{Fig:ratio} (upper panel) shows the ratios of the kaon to pion distributions for both the PDFs and TMDs of the $u$ quark under the initial scale, $\mu^2_{0\pi}=0.43\pm0.04\;\rm GeV^2$ for the pion and $\mu^2_{0K}=0.52\pm0.05\;\rm GeV^2$ for the kaon ~\cite{Lan:2021wok,Lan:2025qio,Lan:2025fia}\footnote{We note here a correction for the model scale of the kaon which is $\mu_{0,K}^2 = 0.52~\mathrm{GeV}^2$ for Phys.Lett.B 868 (2025) 139654~\cite{Lan:2025fia}.}.
These ratios provide indirect experimental constraints on the partonic structure of the kaon. For the pion distributions, we use our results obtained from the same light-front QCD Hamiltonian including the $|q\bar{q}\rangle$ and $|q\bar{q}g\rangle$ Fock sectors~\cite{Lan:2025qio}.
For the PDFs, the $u$-quark distribution in the kaon is nearly identical to that in the pion around $x \approx 0.1$.
In the intermediate region $0.1 < x < 0.5$, the kaon distribution is larger, whereas it becomes smaller for $x > 0.5$.
This pattern indicates that the $u$ quark in the kaon predominantly populates lower-$x$ regions, while in the pion it tends to carry a larger fraction of the longitudinal momentum. The middle panel shows the ratio of the $\bar{s}$ quark to the $u$ quark in the kaon. From these two figures, the strange quark has a larger longitudinal momentum fraction than the light quark in the larger $x$ region.

In contrast, the TMD ratios show that as $k_\perp$ increases, the momentum fraction $x$ carried by the $u$ quark in the kaon shifts toward the larger-$x$ region. 
This reflects the nontrivial interplay between the longitudinal and transverse momentum distributions.

The lower panel of Fig.~\ref{Fig:ratio} displays the ratios of the kaon to pion distributions for the gluon PDFs and TMDs. 
Except in the region $0.2 < x < 0.4$, where the gluon distributions in the kaon and pion are nearly identical, the pion gluon distribution dominates in both the small- and large-$x$ regions. 
Meanwhile, the TMD ratios reveal that as $k_\perp$ increases, the gluon TMD in the kaon becomes larger in the intermediate-$x$ range ($0.1 < x < 0.5$).

Figure~\ref{Fig:ratioofevolution} presents our results for kaon PDFs after QCD evolution at NNLO~\cite{Golec-Biernat:2025hwa,Hampson:2025pvi,Collins:2011zzd}. In the upper panel, we compare the ratios \(u_v^K/u_v^{\pi}\) and \(\bar{s}_v^K/u_v^{\pi}\) with the recent global analysis by the JAM Collaboration~\cite{Barry:2025wjx} at the scale \(\mu^2 = 4~\mathrm{GeV}^2\), which incorporates both experimental data and lattice QCD constraints. The uncertainty bands in our evolved distributions reflect a \(\pm 10\%\) uncertainty in the model input scales for the kaon and pion.
These initial scales were fixed by matching the evolved first moments of the valence quark and antiquark distributions to global QCD analyses.

Our results show good agreement with the JAM analysis, reproducing the characteristic decrease of the \(u_v^K/u_v^{\pi}\) ratio and the increase of the \(\bar{s}_v^K/u_v^{\pi}\) ratio with increasing \(x\). The lower-left panel of Fig.~\ref{Fig:ratioofevolution} compares the ratio of the evolved gluon distributions in the kaon and pion with the corresponding ratio at the initial scale. We observe that, although at the initial scale the kaon’s gluon distribution is smaller than that of the pion at both small and large $x$, QCD evolution drives the two distributions toward very similar behavior.  Finally, the lower-right panel presents a direct comparison of our BLFQ results with the JAM 2025 analysis~\cite{Barry:2025wjx} for the quark flavor distributions in the kaon and pion at \(\mu^2 = 4~\mathrm{GeV}^2\). Overall, our findings are in good agreement with the global QCD analysis by the JAM Collaboration.


\section{Conclusion and Outlook}
In this work, we have investigated the partonic structure of the kaon by computing its leading-twist TMD $f_1(x,k_\perp)$ for both quarks and gluons, and, for the first time, the genuine twist-3 TMDs $\tilde{e}(x,k_\perp)$ and $\tilde{f}^\perp(x,k_\perp)$ for quarks, using light-front wave functions obtained within the Basis Light-Front Quantization (BLFQ) framework. 
These wave functions are derived from the eigenvectors of the light-front QCD Hamiltonian in the light-cone gauge, incorporating both the $|q\bar{q}\rangle$ and $|q\bar{q}g\rangle$ Fock sectors, together with a three-dimensional confining potential in the leading Fock sector.


Using model-independent equations-of-motion (EOM) relations, we constructed the total twist-3 transverse-momentum--dependent distributions (TMDs) \(e(x,k_\perp)\) and \(f^{\perp}(x,k_\perp)\). In this study, the gauge link was set to unity, allowing us to focus exclusively on time-reversal--even (T-even) TMDs. To reduce artifacts arising from basis truncation, an averaging procedure was applied to the BLFQ results. The integrated TMDs exhibit good numerical stability with respect to basis truncation, and the resulting twist-3 PDFs \(e(x)\) satisfy the expected momentum sum rule.

We find that the genuine twist-3 contributions display a distinct \(x\)-dependence compared with the twist-2 terms, while their overall magnitude remains subdominant. The twist-2 PDFs, $f_1^q(x)$ obtained in this work are in good agreement with the recent global analysis by the JAM Collaboration~\cite{Barry:2025wjx}, which incorporates both experimental data and lattice QCD constraints.

Our results also highlight the important role of dynamical gluons, particularly at small $x$, where they significantly enhance the quark distributions. 
We have computed the first moments of the kaon PDFs, providing a clear decomposition of the longitudinal momentum among valence quarks and gluons. 
Moreover, a comparison of the kaon and pion TMDs reveals characteristic differences in the momentum distribution of the $u$ quark and gluon, offering valuable insights into flavor symmetry breaking in light mesons.

The present work focuses on the T-even distributions. Since the genuine twist-3 and T-odd TMDs share similar operator structures, our results pave the way for future studies of T-odd observables, such as the Boer–Mulders function $h_1^\perp$ in the kaon and its impact on spin asymmetries in Drell–Yan processes. 
In addition, a detailed investigation of the gluon TMDs and their evolution properties within the BLFQ framework will be presented in future work.

We believe that these results, including the twist-3 TMDs, provide valuable predictions for future experimental programs, such as precise measurements of kaon-induced Drell–Yan processes by COMPASS++/AMBER at CERN and forthcoming studies of the 3D structure of the kaon at facilities like the EIC and EicC~\cite{Chavez:2021koz}. 
Such measurements will offer critical opportunities to refine our understanding of the kaon’s three-dimensional partonic structure and its internal QCD correlations.

\section*{Acknowledgements}
We thank Zhi Hu, Jiatong Wu, and Jialin Chen for many helpful discussions.
This work is supported by National Natural Science Foundation of China, Grant No. 12375143 and 12305095.
Z. Z. is supported by China Association for Science and Technology. C. M. is supported by new faculty start up funding by the Institute of Modern Physics,
Chinese Academy of Sciences, Grant No. E129952YR0. J.
L. is supported by the Special Research Assistant Funding Project, Chinese Academy of Sciences, by the National
Science Foundation of Gansu Province, China, Grant No.
23JRRA631. X. Z. is supported by National Natural Science Foundation of China, Grant No. 12375143,
by new faculty startup funding by the Institute of Modern Physics, Chinese Academy of Sciences, by Key Research Program of Frontier Sciences, Chinese Academy of
Sciences, Grant No. ZDBS-LY-7020, by the Foundation
for Key Talents of Gansu Province, by the Central Funds
Guiding the Local Science and Technology Development of
Gansu Province, Grant No. 22ZY1QA006, by Gansu International Collaboration and Talents Recruitment Base of
Particle Physics (2023-2027), by International Partnership
Program of the Chinese Academy of Sciences, Grant No.
016GJHZ2022103FN,  by National Key
R$\And$D Program of China, Grant No. 2023YFA1606903 and
by the Strategic Priority Research Program of the Chinese
Academy of Sciences, Grant No. XDB34000000. J. P. V.
is supported by the Department of Energy under Grant
No. DE-SC0023692. A portion of the computational resources were also provided by Advanced Computing Center in Taiyuan and by Sugon Computing Center in Xi’an.

\newpage
\remove{\section{Reference}}
\bibliography{bibliography1}

\end{document}